\documentclass[twocolumn,10pt,aps,prl,amsmath,amssymb,showpacs,superscriptaddress]{revtex4-2}
\usepackage[utf8]{inputenc}

\usepackage{amsmath} 
\usepackage{graphicx}
\usepackage{dcolumn}
\usepackage{bm}
\usepackage{epsfig} 
\usepackage{hyperref} 
\usepackage{color} 
\usepackage[normalem]{ulem} 
\usepackage{cancel} 
\usepackage{ stmaryrd }
\usepackage{comment}
\usepackage[capitalize]{cleveref}
\usepackage{enumitem}

\newcommand{\ve}[1]{\boldsymbol{#1}}
\newcommand{\mf}{\text{MF}}
\newcommand{\eF}{\varepsilon_F}
\newcommand{\khs}{\ve{k}_{\text{hs}}}

\begin{document}
\title{The fate of the Fermi surface coupled to a single-wave-vector cavity mode}
\author{Bernhard Frank}
\affiliation{Institut f\"{u}r Theoretische Physik and W\"{u}rzburg-Dresden Cluster of Excellence ct.qmat, Technische Universit\"{a}t Dresden, 01062 Dresden, Germany}
\author{Michele Pini}
\email{michele.pini@uni-a.de}
\affiliation{Theoretical Physics III, Center for Electronic Correlations and Magnetism, Institute of Physics, University of Augsburg, 86135 Augsburg, Germany}
\affiliation{Max-Planck-Institut f\"{u}r Physik komplexer Systeme, 01187 Dresden, Germany}
\author{Johannes Lang}
\affiliation{Institut f\"{u}r Theoretische Physik, Universit\"{a}t zu K\"{o}ln, 50937 Cologne, Germany}
\author{Francesco Piazza}
\affiliation{Theoretical Physics III, Center for Electronic Correlations and Magnetism, Institute of Physics, University of Augsburg, 86135 Augsburg, Germany}
\affiliation{Max-Planck-Institut f\"{u}r Physik komplexer Systeme, 01187 Dresden, Germany}
\date{\today}

\begin{abstract}
The electromagnetic field of standing-wave or ring cavities induces a spatially modulated, infinite-range interaction between atoms in an ultracold Fermi gas, with a single wavelength comparable to the Fermi length. This interaction has no analog in other systems of itinerant particles and has so far been studied only in the regime where it is attractive at zero distance. Here, we fully solve the problem of competing instabilities of the Fermi surface induced by single-wavelength interactions. We find that while the density-wave (superradiant) instability dominates on the attractive side, it is absent for repulsive interactions, where the competition is instead won by non-superradiant superfluid phases at low temperatures, with Fermion pairs forming at both vanishing and finite center-of-mass momentum. Moreover, even in the absence of such symmetry-breaking instabilities, the Fermi surface exhibits a peculiar anisotropic deformation. We estimate this full phenomenology to be within reach of dedicated state-of-the-art experimental setups.
\end{abstract}
\maketitle

\paragraph{Introduction ---} The problem of an electron gas coupled to transverse photons is a longstanding one in many-body physics~\cite{reizer1989relativistic}, showing an intriguing phenomenology deviating from standard metallic behavior. Recent developments in the field of cavity quantum materials~\cite{Schlawin2022} promise to bring this rich physics within experimental reach~\cite{rokaj2022free}, owing to the enhancement of the light-matter coupling via optical confinement. The latter introduces a tunable gap; it closes at the superradiant phase transition, where non-Fermi-liquid behavior has indeed been predicted~\cite{Rao2023}.

Light confinement via mirrors enables strong coupling of matter to a single electromagnetic mode. In solids, the wavelength of this mode is orders of magnitude larger than the Fermi length~\cite{Gao2020,Chakraborty2021}. By contrast, in ultracold atomic systems, where electrons are replaced by neutral fermionic atoms, these two scales can be comparable~\cite{Mivehar2021rev}. This gives access to a new regime where the photon-mediated interaction transfers a single, but finite momentum between Fermions, set by the cavity-photon wave vector. This situation, absent in other itinerant systems, becomes particularly interesting in the presence of a Fermi surface (FS). 

To date, both theory~\cite{Piazza2013,Piazza2014,Piazza2014PRL,Keeling2014,Chen2014,Colella2018,Colella2019,Schlawin2019atoms,Zheng2020} 
and experiment~\cite{ZhangWu2021,HelsonBrantut2023,ZwettlerBrantut2026} have mainly focused on the attractive regime, where the superradiant (SR) (density wave) phase dominates due to a collective enhancement caused by the single-wave-vector nature of the photon-mediated interaction. 
\begin{figure}[t]    
\includegraphics[width=0.9\columnwidth]{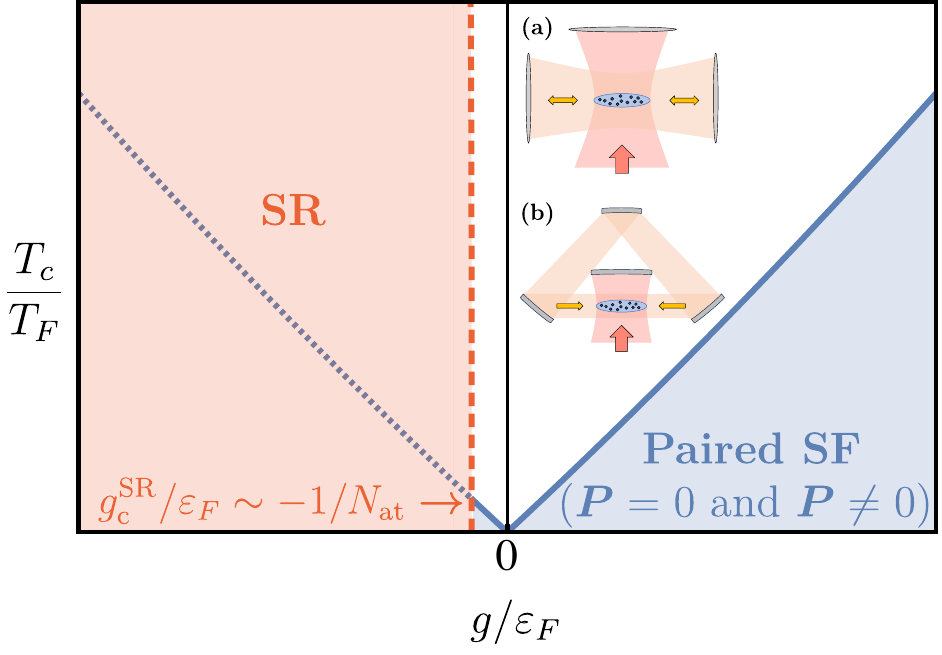} \\
\caption{
Phase diagram for a transversely driven ultracold polarized Fermi gas in a cavity. For attractive interactions ($g<0$), below a critical coupling scaling as the inverse of the number of atoms, the system is dominated by a superradiant (SR) phase with modulated density ordering in the atom cloud. For repulsive interactions ($g>0$), the SR phase is suppressed, and a paired superfluid (SF) phase emerges with a superposition of pairs with zero and finite c.o.m.~momentum $\ve P$. Inset: Sketch of two possible cavity geometries, (a) a standing-wave cavity and (b) a ring cavity.}
\label{fig:PD-sketch}
\end{figure}
In this work, we leverage the latter feature to fully solve the competition problem via a mean-field treatment, which becomes exact in the thermodynamic limit \footnote{Here we mean the thermodynamic limit relevant for ultracold-atom-cavity experiments, where the trap size within the cavity is large enough to be well described by a momentum continuum, while the size of the cavity itself is such that the relevant single mode is still well separated in energy from all other electromagnetic modes.}. 

\begin{figure}[t]    
\includegraphics[width=0.99\columnwidth]{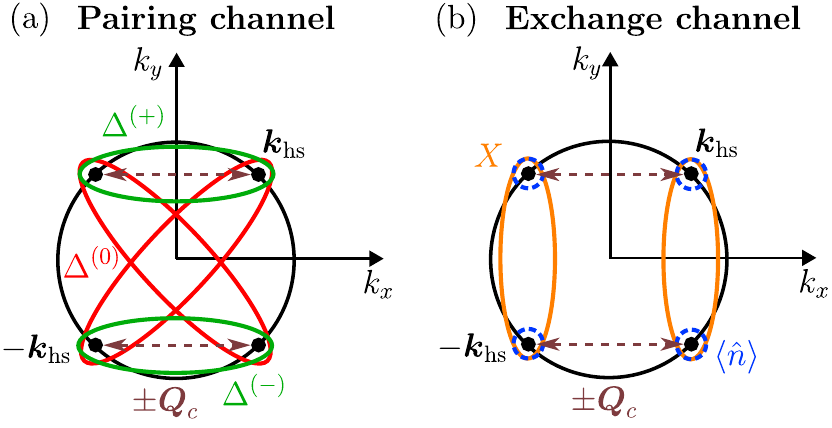}
\caption{
Sketch of the mean-field decouplings evaluated at the hot spots on the FS. (a) Cooper order parameters $\Delta^{(0)}_{\khs} = g \langle \hat{c}_{-\khs} \hat{c}_{\khs}  \rangle $ and $\Delta^{(0)}_{\khs-\ve{Q}_c} = g \langle \hat{c}_{-\khs+\ve{Q_c}} \hat{c}_{\khs-\ve{Q}_c}  \rangle $  (red), together with PDW order parameters $\Delta^{(\pm)}_{\pm(\khs-\ve Q_c)} = \langle  \hat{c}_{\pm \khs} \hat{c}_{\pm(\khs -\ve Q_c)}\rangle$ (green). (b)  Exciton order parameters  $X_{\pm \khs} = g \langle \hat{c}^\dagger_{\pm \khs} \hat{c}_{\mp(\khs - \ve Q_c)} \rangle$ (orange), together with momentum occupation numbers $\langle \hat n_{\pm\khs}\rangle = \langle \hat{c}^\dagger_{\pm\khs} \hat{c}_{\pm\khs}\rangle$ and $\langle \hat n_{\pm(\khs-\ve{Q}_c)}\rangle = \langle \hat{c}^\dagger_{\pm(\khs-\ve{Q}_c)} \hat{c}_{\pm(\khs-\ve{Q}_c)}\rangle$ (blue dashed). The latter are at the origin of the FS deformation (see \cref{fig:n_delta_composite_plot}), whereas the others are associated with instabilities due to spontaneous symmetry breaking.
} 
\label{fig:MF-decoupling}
\end{figure}

As shown in Fig.~\ref{fig:PD-sketch}, on the repulsive side of the photon-mediated interaction, the SR phase is absent, giving way to a non-trivial competition between non-superradiant phases, governed by the hot spot points on the FS which are connected by a photon wave vector (see~\cref{fig:MF-decoupling}). 
At low temperatures, this competition is won by superfluid phases, which exhibit degenerate pairing channels with either vanishing or finite center-of-mass (c.o.m.) momentum. Remarkably, independent of the sign of the interaction and even in the absence of symmetry breaking, we find that the Fermi surface is always anisotropically deformed. This deformation is a true many-body effect due to the anisotropy of photon-mediated interactions, similarly to what happens in dipolar Fermi gases \cite{Miyakawa2008,Fregoso2009,Aikawa2014}. Contrary to the latter case, however, the FS is not an ellipsoid, but rather shows a new type of deformation geometry due to the single-wave-vector nature of the interaction.

We fully characterize the behavior of the pairing gap, which strongly depends on momentum, and estimate that the required temperature and light-matter couplings lie within reach of dedicated, state-of-the-art experimental setups.

\paragraph*{Model ---}
We study a fully spin-polarized ultracold gas of fermionic atoms, at temperature $T$, confined by an external potential to two spatial dimensions~\cite{levinsen2015strongly} in a homogeneous trap~\cite{hueck2018,Navon2021}.
Due to spin polarization, intrinsic short-range interactions are absent~\cite{zwerger2011bcs, zwer14varenna, levinsen2015strongly}, allowing us to focus exclusively on photon-mediated interactions.

The dispersion relation is given by the standard form $ \xi_{\ve k} = \ve k^2/(2m)- \varepsilon_F$
(we work in units with $\hbar=k_B=1$), where the Fermi energy $\varepsilon_F$ determines the particle density $n = k_F^2/(4\pi)$ via the Fermi wave vector $k_{F} = (2m \varepsilon_F)^{1/2}$, which sets the radius of the Fermi disk in the absence of interactions at zero temperature. 
The Fermi gas is located within an optical ring cavity [Fig.~\ref{fig:PD-sketch}(b)] that supports two counterpropagating cavity modes $\exp(\pm i \ve Q_c \cdot \ve r)$ with cavity momentum $\ve Q_c$ and corresponding frequency $\omega_c$, and transversally driven by a laser pump of frequency $\omega_p$, with an adjustable detuning $\Delta_c=\omega_c-\omega_p$~\cite{Bux2011,Bux2013}. 
Typical cavity experiments with ultracold Fermi gases~\cite{ZhangWu2021,HelsonBrantut2023,ZwettlerBrantut2026} operate at densities where $k_F$ and $Q_c$ are of comparable magnitude.
Here we focus on a ring cavity, being translationally invariant and thus the simplest case. However, we find similar results for the experimentally more common standing-wave cavity [see Fig.~\ref{fig:PD-sketch}(a)]; a summary of this generalization is provided below and discussed in full in the Supplemental Material \cite{SM}.
In the dispersive regime, the system is described by the Hamiltonian~\cite{Mivehar2021rev}
\begin{align}\label{eq:model}
  \hat H =\sum_{\ve k} \xi_{\ve k} \hat c^\dagger_{\ve k} \hat{c}_{\ve k}   +\frac{g}{2} \sum_{\substack{\ve p, \ve k\\ s= \pm 1}}  \hat{c}^\dagger_{\ve p+s\ve Q_c} \hat{c}^\dagger_{\ve k-s\ve Q_c}  \hat{c}_{\ve k} \hat{c}_{\ve p}\, ,
\end{align}
where the geometry of the ring cavity implies momentum conservation in the interaction term.
In the above, the operators $\hat c_{\ve{k}}$ and $\hat c^\dagger_{\ve{k}}$ annihilate and create a Fermion with momentum $\ve k$. Assuming $|\Delta_c|$ much larger than the cavity loss rate $\kappa$, the tunable coupling strength is given by $g\approx U_0 V_0/\Delta_c$, where $U_0$ is the depth of the cavity potential per photon and $V_0$ corresponds to the light shift induced by the pump, proportional to the pump intensity
\cite{ritsch2013review,Mivehar2021rev,HelsonBrantut2023}.

Physically, these cavity-mediated interactions transfer only fixed momenta $\pm \ve{Q}_c$ and are therefore of infinite range in real space, with a periodic modulation with wave vector $\ve{Q}_c$. The sign of $g$, which is tunable via the cavity detuning $\Delta_c$, determines the sign of the interaction at zero distance. 
Because of the long-range nature of the interactions, mean-field theory becomes exact in the thermodynamic limit~\cite{Mivehar2021rev}, which forms the basis of the analysis of the phase diagram below.

\paragraph*{Mean-field analysis ---}  
We first identify the most relevant low-energy processes in the pairing and exchange channel~\cite{Altland&Simons}, since the direct channel is dominated by the density wave (superradiance). 
For a given $\ve Q_c$ with $Q_c \sim k_F$ and $Q_c < 2k_F$ (necessary for hot spots to exist), the Hamiltonian $\hat H$ singles out special interaction processes where all incoming \emph{and} outgoing momentum states reside on the FS, which, therefore, correspond to the primary candidates for instabilities. As depicted in Fig.~\ref{fig:MF-decoupling}, these occur if and only if the four momentum hot spots $\pm \khs$ and $\pm(\khs - \ve Q_c)$ are involved. 

In $\hat H$, these two processes are given by the contributions 
\begin{align}
    &g \hat{c}^\dagger_{\khs - \ve Q_c} \hat{c}^\dagger_{-\khs+\ve Q_c} \hat{c}_{-\khs } \hat{c}_{\khs} \label{eq:H_contr1} \\
    &g \hat{c}^\dagger_{\pm(\khs - \ve Q_c)} \hat{c}^\dagger_{\pm\khs} \hat{c}_{\pm(\khs-\ve Q_c)} \hat{c}_{\pm\khs} \label{eq:H_contr2}
\end{align}
A standard mean-field decoupling of \eqref{eq:H_contr1} in both the pairing and the exchange channel yields $g \hat{c}^\dagger_{\khs - \ve Q_c} \hat{c}^\dagger_{-\khs+\ve Q_c} \hat{c}_{-\khs } \hat{c}_{\khs} \simeq \bar\Delta^{(0)}_{\khs-\ve Q_c} \hat{c}_{-\khs} \hat{c}_{\khs } - \bar X_{-\khs} \hat{c}^\dagger_{-\khs +\ve Q_c} \hat{c}_{\khs} + \text{h.c.} + \text{const}$.
 Here, a finite value of $\Delta^{(0)}_{\khs} = g \langle \hat{c}_{-\khs} \hat{c}_{\khs}  \rangle $ [red in Fig.~\ref{fig:MF-decoupling}(a)] describes a state with pairs of vanishing  c.o.m.~momentum, which is referred to as Cooper pairing in the following. 
On the other hand, for the exchange channel, the particle-hole expectation value $X_{-\khs} \equiv g \langle c^\dagger_{-\khs} c_{\khs - \ve Q_c} \rangle$  [orange in Fig.~\ref{fig:MF-decoupling}(b)] describes exciton condensation (XCON) with finite relative momentum $\ve K = 2 \khs - \ve Q_c$. 
Applying the same decoupling to process~\eqref{eq:H_contr2} leads to $g \hat{c}^\dagger_{\pm(\khs - \ve Q_c)} \hat{c}^\dagger_{\pm\khs} \hat{c}_{\pm(\khs-\ve Q_c)} \hat{c}_{\pm\khs} \simeq  \bar\Delta^{(\pm)}_{\pm(\khs-\ve Q_c)} \hat{c}_{\pm(\khs-\ve Q_c)} \hat{c}_{\pm \khs }- g \langle \hat n_{\pm(\khs-\ve Q_c)} \rangle \hat{c}^\dagger_{\pm\khs } \hat{c}_{\pm\khs} + \text{h.c.} + \text{const}$. 
The order parameter $\Delta^{(\pm)}_{\pm(\khs-\ve Q_c)} = \langle  \hat{c}_{\pm \khs} \hat{c}_{\pm(\khs -\ve Q_c)}\rangle$ [green in Fig.~\ref{fig:MF-decoupling}(a)] describes pair-density-wave (PDW) pairing with fixed c.o.m.~momentum $\pm \ve P = \pm(2 \khs -\ve Q_c)$. In the exchange channel, momentum occupation numbers $\langle \hat n_{\khs}\rangle = \langle \hat{c}^\dagger_{\khs} \hat{c}_{\khs}\rangle$ [blue dashed in Fig.~\ref{fig:MF-decoupling}(b)] emerge, which do not give rise to instabilities but instead lead to a reshaping of the FS. Moreover, as we shall see below, the pairing instabilities $\Delta^{(0,\pm)}$ are always dominant over the XCON instability.

To extend the mean-field approach beyond the hot spots, we introduce the momentum deviations  $\ve \delta \ve k^{\pm}=(\pm \delta k_x,\delta k_y)$
and include shifted copies of the hot spots, displaced by multiples of $\ve Q_c$. 
The generalized pairing order parameters then become ($l \in \mathbb Z)$
\begin{align}\label{eq:main_def_Delta}
\begin{split}
    \Delta^{(0)}_{\khs + \ve \delta \ve k^{\pm}+l \ve Q_c} & \equiv g \langle \hat{c}_{-(\khs + \ve \delta \ve k^{\pm}+l\ve Q_c)} \hat{c}_{\khs + \ve \delta \ve k^{\pm} + l \ve Q_c}\rangle \\
    \Delta^{(+)}_{\khs + \ve \delta \ve k^+ +l \ve Q_c} & \equiv g \langle \hat{c}_{\khs+\ve \delta \ve k^- -(l+1)\ve Q_c} \hat{c}_{\khs + \ve \delta \ve k^++ l \ve Q_c}\rangle \\
\Delta^{(-)}_{-\khs -\ve \delta \ve k^+ +l \ve Q_c} & \equiv g \langle \hat{c}_{-\khs - \ve \delta \ve k^+ -(l-1)\ve Q_c} \hat{c}_{-\khs - \ve \delta \ve k^- + l \ve Q_c}\rangle \, .
\end{split}
\end{align}
Further details--including the analogous expressions for the exchange channels $X$, $Y$ [The latter becomes distinct only beyond the hot spots, see Eq.~\eqref{eq:Y}], and $\langle \hat{n} \rangle$--are presented in the End Matter (EM), along with the full mean-field Hamiltonian. Moreover, the Hamiltonian~\eqref{eq:model} exhibits a special $\mathrm{SO}(2)$ symmetry in momentum space
originating from the fact that the $k_y$ components ($\hat{\ve e}_y\perp \ve Q_c$) are untouched by the interaction, such that any Fermion may be replaced by a superposition with the Fermion with flipped $k_y$ momentum. This leads to an exact degeneracy between Cooper and PDW pairing, as is detailed in the EM, too. Upon entering the paired phase, the system spontaneously breaks this symmetry by selecting a specific superposition of Cooper and PDW states.

Given the full mean-field decoupled Hamiltonian, one can derive (see EM) a closed set of coupled equations for all the above order parameters, which we use to solve the full problem of competing instabilities.

\begin{figure}[tb]    
\includegraphics[width=0.99\columnwidth]{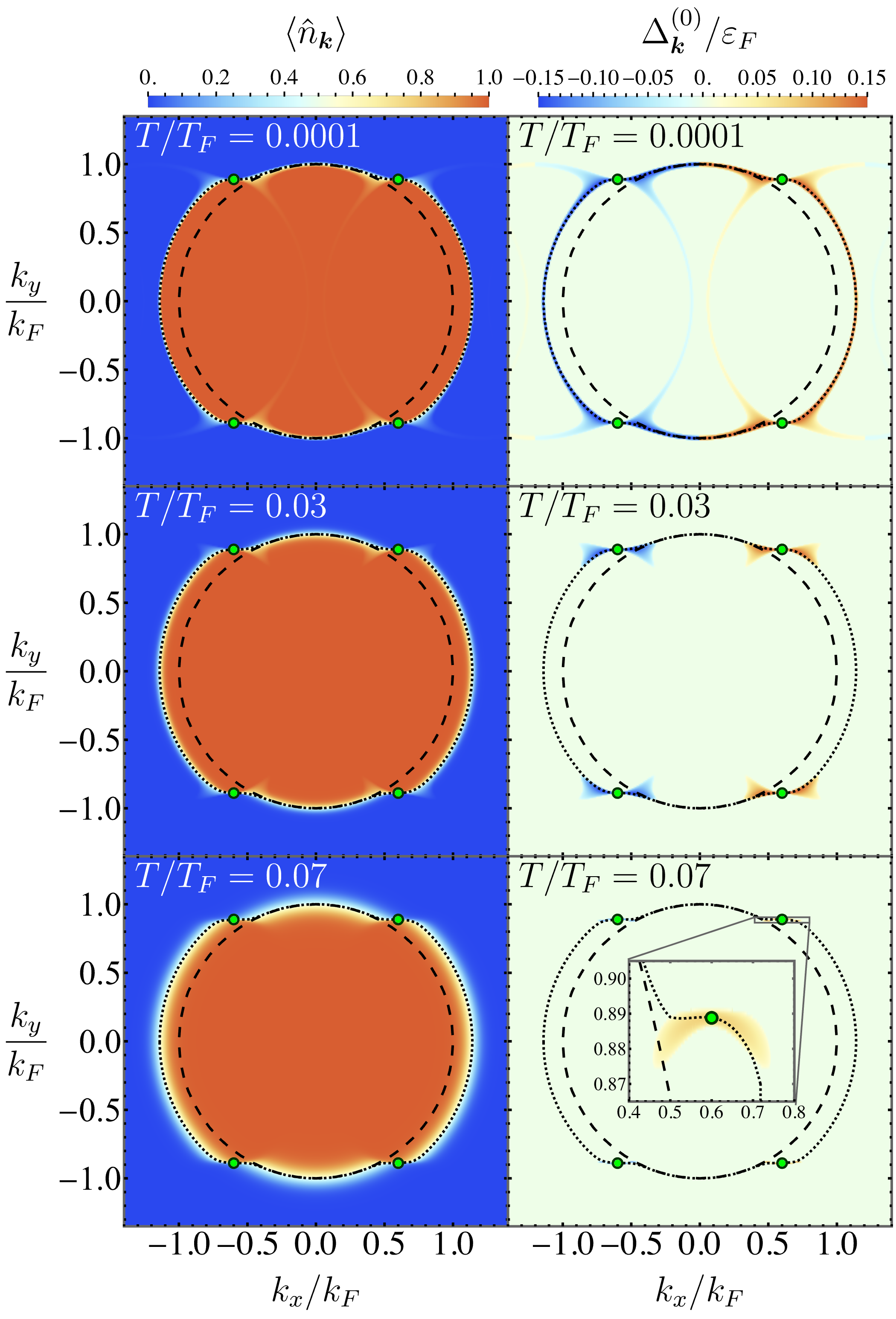}
\caption{Momentum distribution $\langle \hat n_{\ve{k}} \rangle$ (left) and Cooper gap $\Delta^{(0)}_{\ve{k}}$ (right) for different temperatures below $T_c$. 
Dotted line: FS reshaped by interactions. The dotted arches on the left and right side correspond to the analytic result $\xi_{\ve k} = g$. Dashed line: bare FS. Green dots: renormalized hot spots.
Inset in the lower-right panel: zoom on the region around a hot spot for a temperature just below $T_c$.}
\label{fig:n_delta_composite_plot}
\end{figure}

\begin{figure}[tb]    
\includegraphics[width=0.9\columnwidth]{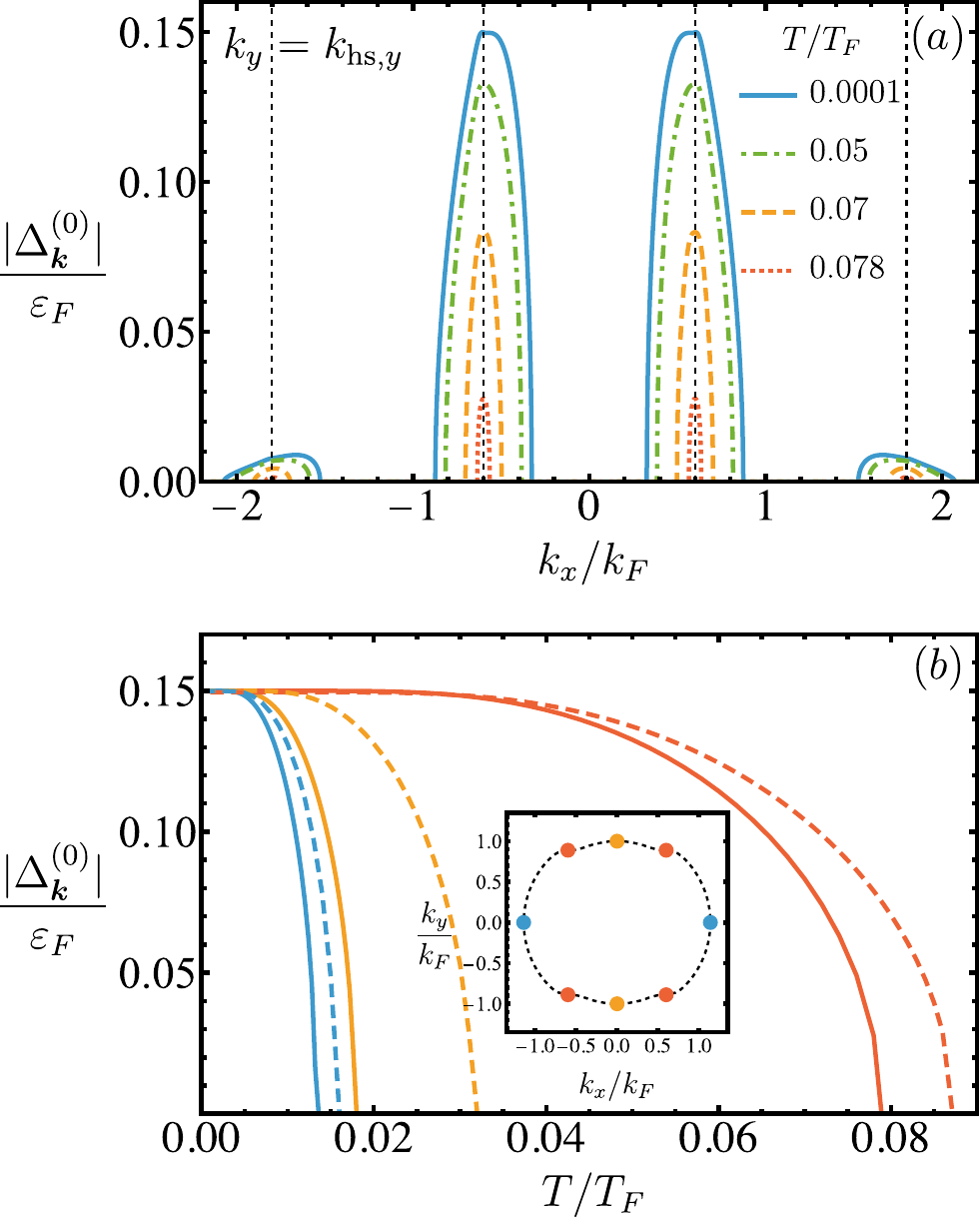}
\caption{(a) $|\Delta^{(0)}_{\ve{k}}|$ as a function of $k_x$ along a horizontal momentum cut passing through the hot spots ($k_y=k_{\mathrm{hs},y}$) at different temperatures below $T_c$. The dashed vertical lines correspond to the $k_x$ of the hot spots and their first shifted copies. (b) $|\Delta^{(0)}_{\ve{k}}|$ as a function of temperature $T$ for ring (solid lines) and standing-wave (dashed lines) cavities at specific $\ve{k}$-points on the FS. For the ring cavity, these are indicated with dots of the corresponding color in the inset (hot spots shown in red). The standing-wave cavity calculation has been performed on the equivalent points of the circular FS.}
\label{fig:results_momenta}
\end{figure}

\paragraph*{Results ---} We focus on repulsive interactions, $g>0$, such that the superradiant density wave instability (direct channel) is absent (see EM for details).
To fully solve the coupled mean-field equations~\eqref{eq:MF_equations}, we resort to numerics on a truncated momentum grid containing the vicinity of the Fermi sea and the corresponding momenta shifted by $\pm \ve Q_c$. Including further shifted copies amounts to a perturbative expansion in $g/\epsilon_F$, yielding small corrections in the regime $|g|/\eF \ll 1$. The results in Figs.~\ref{fig:n_delta_composite_plot} and \ref{fig:results_momenta} are shown for $g=0.3 \eF$ and $Q_c = 1.2 k_F$.

To gain a deeper understanding of the underlying physics, we first analyze the mean-field equations in the absence of pairing and exchange. The Hamiltonian then becomes diagonal with a renormalized dispersion, and the self-consistency equation reduces to
\begin{align}\label{eq:disp_ren}
    \epsilon^{(\hat n)}_{\ve k} = \xi_{\ve k} - g \sum_{s=\pm 1} \langle \hat n_{\ve k + s \ve Q_c}\rangle   \qquad 
    \langle \hat n_{\ve k}\rangle = n_F(\epsilon^{(\hat n)}_{\ve k})\, ,
\end{align}
with the Fermi-Dirac distribution $n_F$. This results in a reshaped FS: From a grand-canonical perspective, the ground-state corresponds to a Fermi sea that includes all single-particle states with $\epsilon^{(\hat{n})}_{\ve k} < 0$.
For $g>0$ (see \cite{SM} for the case $g<0$), Eq.~\eqref{eq:disp_ren} implies that not only the non-interacting FS $\xi_{\ve k} < 0$ is occupied, but also momentum states with $\xi_{\ve k} - g <0$, provided that simultaneously $\xi_{\ve k \pm \ve Q_c} - g <0$. The left column of Fig.~\ref{fig:n_delta_composite_plot} shows the reshaped FS with renormalized hot spots (green dots) for three different temperatures below $T_c$. The new $k_{\text{hs},y}$ hot spot coordinates are determined by the equation $\xi_{\ve k_{\text{hs}}} =g$ (hot spots on reshaped FS), where $k_{\text{hs},x} = \pm Q_c/2$ because of the symmetry $k_x \leftrightarrow - k_x$. For the parameters used, this predicts the hot spots at $(\pm 0.6 k_F, \pm 0.889 k_F)$, in perfect agreement with the numerics.

We can now use the mean-field equations (restricted to the renormalized hot spots) to separately determine the critical temperatures for Cooper and PDW instabilities, yielding
\begin{align}\label{eq:main_TCD}
     T_c^{(0)}=T_c^{(\pm)} = \frac{|g|}{4} + \frac{g^2}{8 \epsilon^{(\hat n)}_{\khs + \ve Q_c}} + \mathcal{O}(|g|^3/\eF^2) \, ,
\end{align}
whereas for the XCON sector one obtains \begin{align}\label{eq:main_TCX}
    T_c^{(X)} = \frac{|g|}{4} + |g| e^{-4\epsilon^{(\hat n)}_{\khs + \ve Q_c}/|g|}+ ...\, ,
\end{align}
where we give the leading exponential correction. 
The linear dependence on $g$ also appears in other models with fixed momentum transfer~\cite{Rademaker2016,Gao2020,Chakraborty2021}. However, in our case where this fixed momentum is also strictly non-zero, $T_c$  becomes independent of the sign of $g$, since any initial state has to undergo an even number of scattering events to return to itself: e.g.~in the Cooper channel $(\khs, -\khs) \xrightarrow{g}(\khs-\ve Q_c, -\khs +\ve Q_c) \xrightarrow{g}(\khs, -\khs)$. As a consequence, all the non-superradiant instabilities occur for both attractive and repulsive interactions, as shown in Fig.~\ref{fig:PD-sketch}. Due to the  symmetry~\eqref{eq:alpha_rot}, Cooper pairing and PDW states are degenerate, as the results for $T_c$ confirm. Since the degeneracy follows from a symmetry of the microscopic Hamiltonian, it is exact and will not be lifted upon the inclusion of higher orders or fluctuations. Comparison of Eqs.~\eqref{eq:main_TCD} and~\eqref{eq:main_TCX} shows that pairing always dominates in the relevant experimental regime $\epsilon_{\khs + \ve Q_c} \sim \eF $ and $|g|/\eF\ll 1$ due to the next-to-leading order terms. These arise from the first shifted copies of the hot spots and explain the suppression of XCON: To create a particle-particle pair, only the necessary energy to scatter from the FS to the shifted copies is required. In contrast, the particle-hole pair in the case of XCON needs, in addition, a thermally excited particle at the shifted copy that turns into the hole. At low $T$, such an excited state has an exponentially small probability. This has also been confirmed in the numerics: By initializing the system in a generic superposition of Cooper, PDW and XCON order parameters, the iterative calculation always converges to a state where the XCON is suppressed. For this reason, in the following, we focus only on the dominant pairing channels.

The right column of Fig.~\ref{fig:n_delta_composite_plot} depicts the pairing gap 
$\Delta^{(0)}_{\ve{k}}=g\langle\hat{c}_{-\ve k}\hat{c}_{\ve k}\rangle$ for three different temperatures. Due to the symmetry-induced degeneracy, $\Delta^{(\pm)}_{\ve{k}}$ is identical. The critical temperature corresponds to the instability at the hot spots, and pairing progressively gaps the FS as the temperature is lowered.
The sign structure of $\Delta^{(0)}_{\ve k}$ is due to the form of the mean-field equations (see EM) and the triplet nature of the pairing: under $\ve k \to - \ve k$, the pairing amplitude $\langle \hat{c}_{-\ve k} \hat{c}_{\ve k} \rangle$ changes sign due to the fermionic anticommutation.

In Fig.~\ref{fig:results_momenta}(a) we plot $\Delta^{(0)}_{\ve{k}}$ at fixed $k_y = k_{\text{hs},y}$: The maxima of the order parameter occur at the hot spots while side peaks are found in the vicinity of the shifted copies. Both the variation of $\Delta^{(0)}$ around $\khs$, as well as the reduction of the maximum from $\khs$ to the shifted copy depend algebraically on $g$ and $\xi_{\khs + \ve Q_c}$ (see EM). Furthermore, analytics at $T=0$ provide the gap value $|\Delta^{(0,\pm)}_{\ve k}(T\to 0)| = |g|/2$ everywhere on the reshaped FS, in agreement with Figs.~\ref{fig:n_delta_composite_plot} and \ref{fig:results_momenta}(a). 
Fig.~\ref{fig:results_momenta}(b) shows the dependence of $\Delta_{\ve k}^{(0)}$ on the temperature for different momenta on the FS. The blue dotted line corresponds to temperatures where the full FS is gapped. The evaluation of Eq.~\eqref{eq:main_TCD} for our parameters yields $T^{(0)}_c = 0.0789 T_F$ again in perfect agreement with Fig.~\ref{fig:results_momenta}(b).

\paragraph*{Standing-wave cavity ---}
On top of the momentum-conserving interactions in Eq.~\eqref{eq:model}, the standing-wave-cavity setup exhibits additional interactions that break momentum conservation and change the c.o.m.~momentum of any pair by $\pm 2 \ve Q_c$. Nevertheless, analogous derivations lead to similar conclusions (see \cite{SM} for details).
Again, pairing dominates over the exchange channel. However, now a third type of leading  pairing instability occurs, which arises from pairs of the form $\Delta^{(\parallel)}_{\khs} = g \langle \hat{c}_{-\khs+\ve Q_c} \hat{c}_{\khs} \rangle$. It is related to $\Delta^{(0)}$ pairs by an additional $\text{SO}(2)$ symmetry of a form analogous to Eq.~\eqref{eq:alpha_rot} with different momentum indices, while the latter symmetry also remains intact. As a result, $\Delta^{(0)}$, $\Delta^{(\pm)}$, and $\Delta^{(\parallel)}$ pairs are all exactly degenerate, independently of the mean-field approximation. The critical temperature is derived as $T_c = |g|/4 + 3 g^2/(8  \epsilon^{(\hat n)}_{\khs + \ve Q_c} ) + \mathcal{O}(|g|^3/\eF^2) $, where the $g^2$ correction is enhanced by a factor of 3 with respect to  Eq.~\eqref{eq:main_TCD}, owing to pairs connected to the hot spots via non-momentum conserving interactions.  
This result, as well as the symmetry-induced degeneracy, is confirmed numerically [see Fig.~\ref{fig:results_momenta}(b)]. Consequently, the phase diagram of Fig.~\ref{fig:PD-sketch} applies to both cavity geometries.

\paragraph*{Experimental parameters considerations ---}
Our predictions are general and do not rely on a specific atomic species; multiple choices of Fermions and cavity configurations may be explored to optimize parameters. To compare with the present state of the art, we focus on the commonly used choice of $^6\mathrm{Li}$~\cite{HelsonBrantut2023,Buehler2025,ZwettlerBrantut2026,ZhangWu2021}. The key requirement for observing the predicted phase transition, based on the leading order of Eq.~\eqref{eq:main_TCD} for $T_c$, is the dimensionless ratio $|g|/4T > 1$, where $g \approx U_0 V_0/\Delta_c$ [see below Eq.~\eqref{eq:model}]. Present experiments typically operate with $U_0/h\approx 10\text{--}20\;\mathrm{Hz}$~\cite{HelsonBrantut2023,Buehler2025,ZwettlerBrantut2026}, while the pump-induced light shift $V_0$ can reach values up to $100$ times the lithium recoil energy, i.e., $(V_0)_{\max}/h\approx 7.4\;\mathrm{MHz}$. Since avoiding dynamical instabilities demands $\Delta_c>\kappa$, where $\kappa/h \approx 77 \text{ kHz}$ is the cavity linewidth, these values, together with a typical Fermi energy of $\varepsilon_F/h\approx 10 \text{ kHz}$, imply a maximal coupling of $g_{\max}/\varepsilon_F = U_0 V_0/\kappa \varepsilon_F \approx 0.2$.
Using the lowest temperatures achieved in ultracold Fermi gas experiments $T/T_F\approx 0.03$~\cite{Navon2010,Sobirey2021}, one obtains
$|g_{\max}|/4T \approx 1.7$,
showing that the required regime is in principle accessible with current state-of-the-art setups.
In addition, further improvements are feasible: employing cavities with reduced loss rate $\kappa$ allows one to work at smaller $\Delta_c$, and tighter mode confinement at the cavity waist increases $U_0$~\cite{Bolognini2025}.
Finally, observing the predicted ordering requires nesting between the cavity wave vector and the FS. Because the cavity-mediated interaction is of global range, such nesting requires a globally uniform density, realized in homogeneous (box-like) traps~\cite{hueck2018,Navon2021}.
To summarize, the pairing phase can be realized by combining state-of-the-art technologies for both cavity setups and box-like traps.

\paragraph*{Conclusions ---} Standing-wave or ring cavities mediate a unique type of interaction between ultracold Fermions, involving the transfer of a single momentum, which can be comparable with the Fermi momentum. This property enabled us to fully solve the problem of competing Fermi-surface instabilities in such systems. We found that the superradiant density-wave instability dominates for attractive interactions but is absent in the repulsive case. There, the Fermi surface is instead gapped in a strongly momentum-dependent way by the formation of a spontaneously chosen superposition of zero- and finite-momentum Fermion-pairs. The corresponding temperature scales are estimated to be accessible to state-of-the-art experiments.

\emph{Note added ---} While preparing this manuscript we became aware of the work~\cite{OrtunoGonzalez2025}, focusing on the superradiant density wave order and the resulting Fermi-surface reshaping, and its interplay with the Pauli crystal characterizing the Fermi gas.

\paragraph*{Acknowledgments ---}
B.F. acknowledges support by the DFG through the W\"urzburg-Dresden Cluster of Excellence on Complexity and Topology in Quantum Matter – ct.qmat (EXC 2147, project id 390858490).
The work of J.L. was supported by the Deutsche Forschungsgemeinschaft (DFG, German Research Foundation) under Germany’s Excellence Strategy Cluster of Excellence Matter and Light for Quantum Computing (ML4Q) EXC 2004/1 390534769, and by the DFG Collaborative Research Center (CRC) 183 Project No. 277101999.
M.P. and F.P. acknowledge valuable discussions with J.P.~Brantut and T.~B\"uhler. 

F.P.~and B.F.~conceived the study. M.P.~and F.P.~supervised the work.
B.F.~developed the theoretical framework, performed the majority of the analytic calculations, implemented and tested the core code. M.P.~generated numerical data, performed some analytic calculations, and investigated the experimental feasibility. J.L.~contributed to the initial development of the code. B.F., M.P.~and F.P.~wrote the original draft.
All authors contributed to the interpretation of the results and to the review and editing of the manuscript.

\bibliography{Comp_inst_bibliography}

\clearpage

\appendix*
\setcounter{equation}{0}
\section{End Matter}\label{sec:EM}
\emph{Superradiant transition ---} At mean-field level, the superradiant phase transition condition can be found by imposing the closing of the gap in the cavity photon spectrum~\cite{Piazza2013,Piazza2014PRL,Keeling2014}, which for $\Delta_c\ll \kappa$ corresponds to
\begin{equation}
\Delta_c= \mathrm{Re}[ \Pi^R(|\ve{Q}|= Q_c, \Omega=0)] \, ,
\label{eq:SR_conditon}
\end{equation}
where $\Pi^R(\mathbf{Q},\Omega)$  
is the density response of the gas given by the Lindhard function
\begin{equation}
\Pi^R(\mathbf{Q},\Omega)=  \frac{2  g \Delta_c N_\mathrm{at}}{n} \int \! \frac{d\ve{k}}{(2\pi)^2} \frac{n_F(\xi_{\ve k})-n_F(\xi_{\ve k + \ve Q})}{ \Omega +\xi_{\ve k}-\xi_{\ve k + \ve Q}+i0^+} \,,
\label{eq:Lindhard}
\end{equation}
where $N_\mathrm{at}$ is the total number of atoms, $n$ is the atomic density, and $n_F$ is the Fermi-Dirac distribution. Performing the angular integral in Eq.~\eqref{eq:Lindhard} and expressing the relation~\eqref{eq:SR_conditon} in dimensionless units using $\eF=2\pi n/m$ and $k_F=\sqrt{2m\eF}$, one obtains the condition
\begin{equation}
1=-8 N_\mathrm{at} \frac{g}{\eF} \int_0^{1/2} \!\! d\tilde{k}  \frac{\tilde{k} }{\sqrt{1-4\tilde{k}^2}} n_F(\xi_{\tilde{\ve{k}}})\,,
\end{equation}
where $\tilde{k}=k/k_F$.
All the terms on the r.h.s.~are positive except for $g$, thus the condition is satisfied only for $g<0$. At $T=0$, we find that $g_c$ scales as $1/N_\mathrm{at}$ (see Fig.~\ref{fig:PD-sketch}).

\begin{figure}[t]    
\includegraphics[width=0.9\columnwidth]{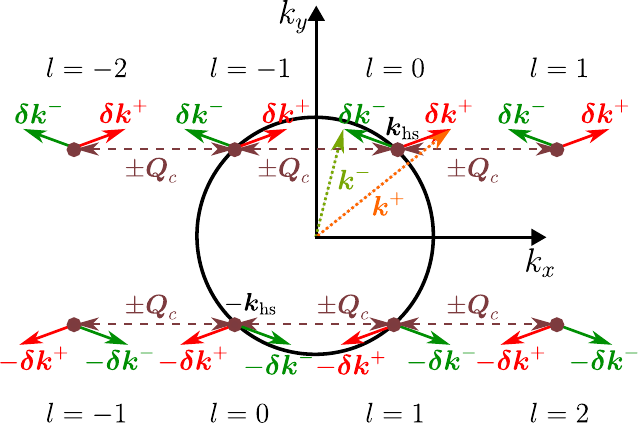}
\caption{
Schematic representation of the momentum structure used for the mean-field decoupling. The hot spots and their mirror images are indicated by brown dots, while $l$ gives the corresponding index. The momentum deviations $\ve \delta \ve k^+$ and $\ve \delta \ve k^-$ are depicted by red and green arrows.}
\label{fig:MF-momentum-sketch}
\end{figure}

\emph{Details on mean-field approach ---} This section provides further details on the extension of the mean-field approach to the entire momentum space, introduced in Eq.~\eqref{eq:main_def_Delta}. 
A graphical representation of the momentum description is given in Fig.~\ref{fig:MF-momentum-sketch}.
To simplify the notation we define $\ve k^\pm  = \khs + \ve{\delta k}^\pm$,
with $\ve \delta \ve k^{\pm}=(\pm \delta k_x,\delta k_y)$.
As can be seen in Fig.~\ref{fig:MF-momentum-sketch}, this choice of momenta together with their mirror images conserves the $k_{x,y} \to - k_{x,y}$ mirror symmetry of the system. 
Moreover, we note that the full Hamiltonian $\hat{H}$ from Eq.~\eqref{eq:model} possesses a special continuous rotation symmetry in momentum space
\begin{align}\label{eq:alpha_rot}
\begin{split}
  \left(\!\!
    \begin{array}{c}
         \hat{c}_{\ve k^-+l \ve Q_c}  \\
         \hat{c}_{-\ve k^+ +(l+1) \ve Q_c}
    \end{array}
    \!\!\right) 
    \! \to \!
    \underbrace{
    \left(
    \begin{array}{cc}
        \cos(\alpha) & - \sin(\alpha)  \\
         \sin(\alpha) & \cos(\alpha) 
    \end{array}
    \right)}_{\equiv \underline R(\alpha)}
    \left(\!\!
    \begin{array}{c}
         \hat{c}_{\ve k^-+l \ve Q_c}  \\
         \hat{c}_{-\ve k^++(l+1)\ve Q_c}
    \end{array}
    \!\!\right)\!,
    \end{split}
\end{align}
as is evidenced in the SM. Physically, this symmetry arises from the fact that any scattering process $(\ve q_1^{(\text{in})}, \ve q_2^{(\text{in})}) \to (\ve q_1^{(\text{out})}, \ve q_2^{(\text{out})})$ has an equivalent counterpart obtained by flipping the sign of the $q_y$ components of either $\ve q_{1}^{(\text{in, out})}$ or $\ve q_{2}^{(\text{in, out})}$ since $\xi_{q_x, -q_y} = \xi_{q_x, q_y}$. Since these processes are independent, they can occur in arbitrary superpositions.

Next, we construct the mean-field Hamiltonian $\hat H_\mf$ for the full momentum space under the condition that it is invariant under the action of $\underline R(\alpha)$. 
With the $\ve k^\pm$ notation, the generalization of Cooper pairs of Eq.~\eqref{eq:main_def_Delta} then becomes
\begin{align}
\Delta^{(0)}_{\ve k^{\pm}+l \ve Q_c} & \equiv g \langle \hat{c}_{-(\ve k^{\pm}+l\ve Q_c)} \hat{c}_{\ve k^{\pm} + l \ve Q_c}\rangle \, .
\end{align}
Setting $l=0$ and taking the limit $\ve \delta \ve k \to 0$ returns the hot spots. The corresponding mean-field decoupling of the interaction Hamiltonian is given by (see \cite{SM} for details):
\begin{align}
\begin{split}
\hat H_{\text{Cooper}} = & \!\!\sum_{\substack{\delta k_x, \delta k_y\\ s = \pm 1,l}}\! \big[ (
    \bar \Delta^{(0)}_{\ve k^+ + (l+s)\ve Q_c} \hat{c}_{-(\ve k^+ + l \ve Q_c)} \hat{c}_{\ve k^+ + l\ve Q_c} + \text{h.c.})\\ & \quad  + g^{-1} \bar\Delta^{(0)}_{\ve k^+ + (l+s) \ve Q_c} \Delta^{(0)}_{\ve k^+ + l \ve Q_c} + (\ve k^+ \leftrightarrow \ve k^-)\big] .
\end{split}
\end{align}

Regarding PDW states, arbitrary c.o.m.~momenta $\ve P$ are, in principle, possible. In the ring-cavity setup, however, the Hamiltonian~\eqref{eq:model} conserves momentum such that PDW states with different $\ve P$ are decoupled. This is not valid in the case of a Fabry-Pérot cavity, where pairing sectors of different c.o.m.~momenta mix such that all possibilities have to be taken into account, which changes the results quantitatively but leaves the overall physical picture untouched. Here, we focus on the specific generalization of the PDW expectation values [equivalent to Eq.~\eqref{eq:main_def_Delta}]
\begin{align}\label{eq:PDW_Delta}
\begin{split}
\Delta^{(+)}_{\ve k^++l \ve Q_c} & \equiv g \langle \hat{c}_{\ve k^- -(l+1)\ve Q_c} \hat{c}_{\ve k^++ l \ve Q_c}\rangle \, , \\
\Delta^{(-)}_{-\ve k^- +l \ve Q_c} & \equiv g \langle \hat{c}_{-\ve k^+-(l-1)\ve Q_c} \hat{c}_{-\ve k^- + l \ve Q_c}\rangle \, ,
\end{split}
\end{align}
since this choice preserves the continuous rotation symmetry~\eqref{eq:alpha_rot}. In particular, the corresponding decoupled Hamiltonian reads
\begin{align}\label{eq:main_HPDW}
\begin{split}
&\hat H_{\text{PDW}} = \sum_{\substack{\delta k_x, \delta k_y \\ s = \pm 1, l}} \big[(
    \bar \Delta^{(+)}_{\ve k^+ + (l+s)\ve Q_c} \hat{c}_{\ve k^- - (l+1) \ve Q_c} \hat{c}_{\ve k^+ + l\ve Q_c}\\
    & \qquad \qquad +\bar \Delta^{(-)}_{-\ve k^- + (l+s)\ve Q_c} \hat{c}_{-\ve k^+ - (l-1) \ve Q_c} \hat{c}_{-\ve k^- + l\ve Q_c} + \text{h.c.}) \\
    & + g^{-1}(\bar\Delta^{(+)}_{\ve k^+ + (l+s) \ve Q_c} \Delta^{(+)}_{\ve k^+ + l \ve Q_c}+ \bar\Delta^{(-)}_{-\ve k^- + (l+s) \ve Q_c} \Delta^{(-)}_{-\ve k^- + l \ve Q_c}) \big]\, .
\end{split}
\end{align}
while $\hat H_{\text{Cooper}} + \hat H_{\text{PDW}}$ remains invariant under $\underline R{(\alpha)}$ (see \cite{SM} for a proof). However, the order parameters in the pairing sector transform as follows: $(\Delta^{(0)}_{\ve k^+ + l \ve Q_c}, \Delta^{(+)}_{\ve k^+ + l \ve Q_c}) \to \underline R^{-1}(\alpha) (\Delta^{(0)}_{\ve k^+ + l \ve Q_c}, \Delta^{(+)}_{\ve k^+ + l \ve Q_c})$, and 
$(\Delta^{(0)}_{-\ve k^- + l \ve Q_c}, \Delta^{(-)}_{-\ve k^- + l \ve Q_c}) \to \underline R(\alpha) (\Delta^{(0)}_{-\ve k^- + l \ve Q_c}, \Delta^{(-)}_{-\ve k^- + l \ve Q_c})$. As a consequence, the Cooper pairing and PDW states do not compete in breaking the global U(1) symmetry of $\hat H$ associated with the particle number conservation. Instead, the pairing instability occurs simultaneously for $\Delta^{(0,\pm)}$ by also spontaneously breaking the $\mathrm{SO}(2)$ symmetry of Eq.~\eqref{eq:alpha_rot}. The attained value of $\alpha$ describes the relative admixture of Cooper and PDW pairing. This behavior is indeed observed both analytically and numerically.
Finally, Eq.~\eqref{eq:PDW_Delta} implies $\pm\ve P(\ve \delta \ve k) = \pm (2 \khs -\ve Q_c + 2 \delta k_y \hat{\ve e}_{k_y})$, such that the hot spots are again recovered for $l=0$ and $\ve\delta \ve k \to 0$.

In the exchange channel, the generalization of the XCON away from the hot spot is defined as
\begin{align}
\begin{split}
    X_{\ve k^+ + l \ve Q_c} & \equiv g \langle \hat c^\dagger_{-\ve k^+ + (l+1) \ve Q_c} \hat c^{}_{\ve k^+ + l \ve Q_c}\rangle \\
    X_{\ve k^- + l \ve Q_c} & \equiv g \langle \hat c^\dagger_{-\ve k^- + l \ve Q_c} \hat c^{}_{\ve k^- + (l-1)\ve Q_c} \rangle\, ,
\end{split}
\end{align}
since the other option $g\langle \hat c^\dagger_{-\ve k^{\pm} + (l+1) \ve Q_c} \hat c^{}_{\ve k^{\mp} + l \ve Q_c} \rangle$ requires the excitation of two particles or two holes.
The corresponding relative momenta become $\ve K (\ve \delta \ve k^{\pm}) = 2(\khs + \ve \delta \ve k^{\pm}) - \ve Q_c$ and we have 
\begin{align}
\begin{split}
    \hat H^{(X)}_{\text{XCON}}\! = & \!\!\!\sum_{\substack{\delta k_x, \delta k_y \\ s = \pm 1, l}}\! \!\!\big[ (
    \bar X_{\ve k^+ + (l+s)\ve Q_c} \hat c^\dagger_{-(\ve k^+ + (l+1) \ve Q_c)} \hat{c}_{\ve k^+ + l\ve Q_c}+  \text{h.c.})\\ &\quad \quad + g^{-1}\bar X_{\ve k^+ + (l+s) \ve Q_c} X_{\ve k^+ + l \ve Q_c} + (\ve k^+ \leftrightarrow \ve k^-)\big] \, . 
\end{split}
\end{align}
In order to comply with the rotation symmetry~\eqref{eq:alpha_rot},
we additionally have to take
\begin{align}\label{eq:Y}
    Y_{\pm \ve k^+ + l \ve Q_c} \equiv g \langle \hat c^\dagger_{\pm\ve k^- + l \ve Q_c} \hat c^{}_{\pm \ve k^+ + l \ve Q_c} \rangle
\end{align}
into account. These expectation values describe exciton condensation around the same hot spot, which entails the smaller relative momenta $\ve K(\ve \delta \ve k) = 2 \delta k_x \hat{\ve e}_{k_x}$ and the mean-field interaction
\begin{align}\label{eq:main_HYXCON}
\begin{split}
    \hat H^{(Y)}_{\text{XCON}} = & \sum_{\substack{\delta k_x, \delta k_y \\ s = \pm 1, l}}  \big[ (
    \bar Y_{\ve k^+ + (l+s)\ve Q_c} \hat c^\dagger_{\ve k^- + l \ve Q_c} \hat{c}_{\ve k^+ + l\ve Q_c}  + \text{h.c.}) \\ & \quad  + g^{-1} \bar Y_{\ve k^+ + (l+s) \ve Q_c} Y_{\ve k^+ + l \ve Q_c} + (\ve k^{\pm} \leftrightarrow - \ve k^{\pm})\big]\, . 
\end{split}
\end{align}
In this case, one finds that $\hat H^{(X)}_{\text{XCON}} + \hat H^{(Y)}_{\text{XCON}}$ is invariant under $\underline R(\alpha)$, while the order parameters transform like $(Y_{\ve k^+ + l \ve Q_c}, X_{\ve k^+ + l \ve Q_c}) \to \underline R(\alpha)(Y_{\ve k^+ + l \ve Q_c}, X_{\ve k^+ + l \ve Q_c})$ and $(Y_{-\ve k^+ + l \ve Q_c}, X_{\ve k^- + l \ve Q_c}) \to \underline R^{-1}(\alpha) (Y_{-\ve k^+ + l \ve Q_c}, X_{\ve k^- + l \ve Q_c})$. Like in the pairing sector, this means that $X$ and $Y$ occur in a superposition and thus $T_c^{(Y)} = T_c^{(X)}$ from Eq.~\eqref{eq:main_TCX}.

Finally, we also include the momentum occupation numbers, which in fact do not give rise to an instability but affect the shape of the interacting FS. The corresponding Hamiltonian contribution reads:
\begin{align}\label{eq:main_Hn}
\begin{split}
    \hat H_{\hat n} = -g & \sum_{\substack{\delta k_x, \delta k_y \\ u, s = \pm 1, l}}\big[\langle \hat n_{u  \ve k^+ + (l +s) \ve Q_c} \rangle \hat c^\dagger_{u \ve k^+ + l \ve Q_c}\hat c^{}_{u \ve k^+ + l \ve Q_c} \\ & - \frac{1}{2} \langle \hat n_{u \ve k^+ + (l +s) \ve Q_c} \rangle \langle \hat n_{u \ve k^+ + l \ve Q_c} \rangle + (\ve k^+ \leftrightarrow \ve k^-)\big] \, .
\end{split}
\end{align}
The total mean-field Hamiltonian $\hat H_\mf  = \hat H_0 + \hat H_{\text{Cooper}} + \hat H_\text{PDW} + \hat H_{\hat n} + \hat H^{(X)}_{\text{XCON}}+\hat H^{(Y)}_{\text{XCON}}$,
where $\hat H_0 =  \sum_{\delta k_x,\delta k_y, u = \pm, l } \left[ \xi_{u(\ve k^+ + l \ve Q_c)} \hat n_{u(\ve k^+ + l \ve Q_c)} + (\ve k^+ \leftrightarrow \ve k^-) \right]$ denotes the kinetic energy,
can be diagonalized via a Bogoliubov transformation~\cite{BlaizotBook}
\begin{align}
    \hat H_\mf = \sum_{\alpha} \epsilon_\alpha \hat \gamma^\dagger_\alpha \hat \gamma^{}_\alpha + E_0\, ,
\end{align}
to fermionic excitations $\hat \gamma_{\alpha}$ with positive energies $\epsilon_\alpha$ and the mean-field ground-state energy $E_0$ (see \cite{SM} for details), which both depend on the order parameters. From the corresponding free energy density $\mathcal F_\mf$, one obtains via the thermodynamic conditions 
\begin{align}
    \begin{split}
        \frac{\delta \mathcal F_\mf}{\delta \bar \Delta^{(0)}_{\ve k^{\pm} + l \ve Q_c}} = \frac{\delta \mathcal F_\mf}{\delta\bar\Delta^{(+)}_{\ve k^+ + l \ve Q_c}} = \frac{ \delta \mathcal F_\mf}{\delta \bar \Delta ^{(-)}_{\ve - \ve k^- + l \ve Q_c}} & = 0 \, , \\
        \frac{\delta \mathcal F_\mf}{\delta \bar X_{\ve k^{\pm} + l \ve Q_c}} = \frac{\delta \mathcal F_\mf}{\delta \bar Y_{\pm \ve k^+ + l \ve Q_c}} & = 0 \, , \\
        \frac{\delta \mathcal F_\mf}{\delta \bar \langle \hat n_{\pm \ve k^{\pm} + l \ve Q_c} \rangle} & = 0 \, ,
    \end{split}
\end{align}
the set of self-consistent mean-field equations: 
\begin{align} \label{eq:MF_equations}
    \begin{split}
        \Delta^{(r)}_{\ve q+l \ve Q_c} & = - \frac{g}{2} \sum_{\alpha,l'} \left(\underline M\right)^{-1}_{l l'}  \text{th}\left(\frac{\beta\epsilon_\alpha}{2}\right)\frac{\delta \epsilon_\alpha}{\delta \bar\Delta^{(r)}_{\ve q + l' \ve Q_c}} \, , \\
        A^{(+)}_{\ve q + l \ve Q_c} & = \frac{g}{2} \sum_{\alpha,l'}  \left(\underline M\right)^{-1}_{l l'}  \text{th}\left(\frac{\beta\epsilon_\alpha}{2}\right)\frac{\delta \epsilon_\alpha}{\delta \bar A^{(+)}_{\ve q+l' \ve Q_c}} \, ,\\
        \langle \hat n_{\pm \ve k^\pm+l\ve Q_c} \rangle -\frac{1}{2} & = \frac{1}{2} \sum_{\alpha,l'}  \left(\underline M\right)^{-1}_{l l'}  \text{th}\left(\!\frac{\beta\epsilon_\alpha}{2}\!\right)\frac{\delta (\epsilon_\alpha/g)}{\delta \langle \hat n_{\pm \ve k^\pm+l' \ve Q_c} \rangle}.
    \end{split}
\end{align}
Here we define the indices $(r,\ve q) = (0,\ve k^{\pm}),(+,\ve k^+),(-,-\ve k^-)$ for the first line and $(A,\ve q) = (X,\ve k^{\pm}),(Y,\pm \ve k^+)$ for the second line. 
The matrix
$\underline M_{l l'} = \delta_{l, l'+1}+\delta_{l, l'-1},$
which takes the fixed momentum transfer into account, makes the mean-field equations resemble a nearest-neighbor tight-binding model in momentum space. This gives rise, for instance, to the sign change of $\Delta^{(0,\pm)}_{\ve k}$ upon shifting $\ve k \to \ve k + \ve Q_c$ as is depicted Fig.~\ref{fig:n_delta_composite_plot}. 

\end{document}


\title{Supplemental Material for ``The fate of the Fermi surface coupled to a single-wave-vector cavity mode"}
\author{Bernhard Frank}
\affiliation{Institut f\"{u}r Theoretische Physik and W\"{u}rzburg-Dresden Cluster of Excellence ct.qmat, Technische Universit\"{a}t Dresden, 01062 Dresden, Germany}
\author{Michele Pini}
\affiliation{Theoretical Physics III, Center for Electronic Correlations and Magnetism, Institute of Physics, University of Augsburg, 86135 Augsburg, Germany}
\affiliation{Max-Planck-Institut f\"{u}r Physik komplexer Systeme, 01187 Dresden, Germany}
\author{Johannes Lang}
\affiliation{Institut f\"{u}r Theoretische Physik, Universit\"{a}t zu K\"{o}ln, 50937 Cologne, Germany} 
\author{Francesco Piazza}
\affiliation{Theoretical Physics III, Center for Electronic Correlations and Magnetism, Institute of Physics, University of Augsburg, 86135 Augsburg, Germany}
\affiliation{Max-Planck-Institut f\"{u}r Physik komplexer Systeme, 01187 Dresden, Germany}
\date{\today}

\begin{abstract}
\centering
The Supplemental Material contains details on the calculations described in the main text and also provides an extension of the procedure to the standing-wave cavity.
\end{abstract}
\maketitle
\section{Overview}
The Supplemental Material is organized as follows: First we define the momentum labeling that is used throughout and give the full Hamiltonian in our formulation in Sec.~\ref{sec:Hamiltonian_mom_struc}. Then we perform the mean-field analysis based on the arguments in the main text for the most important contributions. In Sec.~\ref{sec:MF-analysis} we give details on the construction of the mean-field Hamiltonian and derive the mean-field equations. Next, we focus on the reshaping of the Fermi Surface in Sec.~\ref{sec:MF_Ana_reshape} before we turn to the instabilities in Sec.~\ref{sec:MF_Ana_inst}. In particular, we derive the analytic statements from the main text and discuss the $\mathrm{SO}(2)$ symmetry in momentum space. Finally, in \cref{sec:instabilities_SWcavity} we discuss how the instabilities are modified in the case of a standing-wave cavity.

\section{Structure of momentum space}\label{sec:Hamiltonian_mom_struc}
The model Hamiltonian introduced in the main text to describe the Fermi gas with cavity-mediated interaction reads as:
\begin{align}\label{eq:model}
  \hat H =\sum_{\ve k} \xi_{\ve k}
  \hat c^\dagger_{\ve k} \hat c_{\ve k}   +\frac{g}{2} \sum_{\substack{\ve p, \ve k\\ s= \pm 1}}  \hat{c}^\dagger_{\ve p+s\ve Q_c} \hat{c}^\dagger_{\ve k-s\ve Q_c}  \hat{c}_{\ve k} \hat{c}_{\ve p}\, ,
\end{align}
where we use the standard short-hand notation $\xi_{\ve k} = {\ve k}^2/2m - \eF$. As discussed in the main text, it is essential to consider the structure imprinted by the cavity-mediated interactions on momentum space in order to understand the physics of the system. To this end, we first introduce a convenient momentum representation, which is then used to express the Hamiltonian in an ideally suited form for the mean-field analysis.
\subsection{Description of momenta }
\begin{figure}[ht]
    \centering
    \includegraphics[width = .6 \columnwidth]{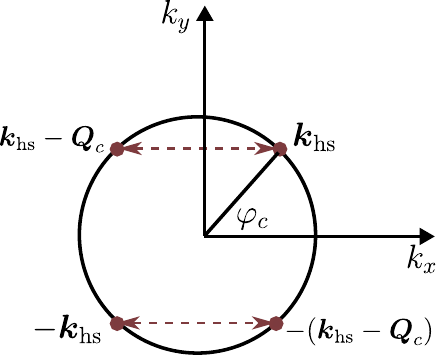}
    \caption{Geometry and labeling of the hot spots on the FS.}
    \label{fig:hotspots}
\end{figure}

\begin{figure}[ht]
\includegraphics[width=0.9\columnwidth]{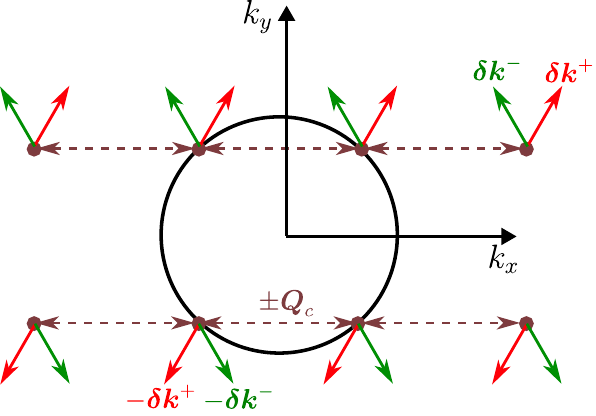}
\caption{Sketch of the momentum structure including the hot spots and their nearest shifted copies away from the FS and a given pair $\ve \delta\ve k^+ = (\delta k_x, \delta k_y)$ (red) $\ve \delta \ve k^- = (-\delta k_x, \delta k_y)$ (green) and their negative counterparts. 
The red (green) shaded area shows as an example the region of momentum space covered by $\ve{k}^{(i)}_{-2}(\ve{\delta k}^+) \quad (\boldsymbol{k}^{(ii)}_{-2}(\ve{\delta k}^-)) $ as function of $\ve{\delta k}^+ \quad (\ve{\delta k}^-)$. The truncation to 16 momenta for each $\boldsymbol \delta\ve k^+$, $\ve \delta \ve k^-$ forms the basis of the mean-field decoupling presented here.}
\label{fig:dkStructure}
\end{figure}

The physical peculiarities of the system arise from the interplay of the interactions with fixed momentum transfer $\pm \ve Q_c$ and the existence of a Fermi surface (FS). 
In the main text four hot spots $\pm \khs$ and $\pm(\khs - \ve Q_c)$, where the FS shows nesting with respect to translations by the cavity momentum, were identified as most prominent candidates for instabilities of the FS. These are depicted in Fig.~\ref{fig:hotspots}. Equivalently, the hot spots can be parameterized as $k_F (\pm \cos\varphi_c, \pm \sin \varphi_c) $ in terms of the angle $\cos(\varphi_c) = Q_c/(2 k_F)$. 
As we will see in the following, a complete description of the interaction effects requires to consider, in addition to the hot spots, any momenta obtained by repeated scattering events that add arbitrary multiples of $\ve Q_c$ to the hot spots, i.e.~$\pm \khs + l \ve Q_c$ and $\pm(\khs - \ve Q_c) + l \ve Q_c$, with $l \in \mathds Z$. This gives rise to a periodic structure along the direction of the cavity momentum formed by the hot spots and their repeated images away from the FS in momentum space.
To eventually cover momentum space completely we consider, in addition, finite deviations $\ve \delta \ve k^+= (\delta k_x, \delta k_y)$ and $\ve \delta \ve k^- = (-\delta k_x, \delta k_y)$ away from the hot spots and their repeated images. With the restriction $\delta k_x \in [- Q_c/2, Q_c/2)$ and $\delta k_y \in [-k_F \sin \varphi_c, \infty)$ every momentum $\ve k$ is uniquely defined as $\ve k = \pm \khs + l \ve Q_c \pm \ve \delta \ve k^{\pm}$ where always the minimal $|l|$ is chosen, see also Fig.~\ref{fig:dkStructure}.

For given $\delta k_x, \delta k_y$, we have four different groups of momenta:
\begin{align}\label{eq:defMomGroups}
\begin{array}{r c}
   (i):   & \ve k_l^{(i)}(\ve{\delta k}^+)  = \khs + l \ve Q_c + \ve \delta \ve k^+ \\
  (ii):  & \ve k_l^{(ii)}(\ve{\delta k}^-)   = \khs + l \ve Q_c + \ve \delta  \ve k^- \\ 
  (iii): &  \ve k_l^{(iii)}(\ve{\delta k}^-)   = - \khs + l \ve Q_c - \ve \delta \ve k^- \\
  (iv):  & \ve k_l^{(iv)}(\ve{\delta k}^+)  = -\khs + l \ve Q_c -\ve \delta  \ve k^+
\end{array}
\end{align} 
that represent identical single-particle energies in the sense $\xi_{\ve k^{(i)}_l} = \xi_{\ve k^{(ii)}_{-l-1}} = \xi_{\ve k^{(iii)}_{l+1}} = \xi_{\ve k^{(iv)}_{-l}}$.
In Fig.~\ref{fig:dkStructure} these groups correspond to the red momenta in the upper half $(i)$, the green momenta in the upper half  
$(ii) $, the green momenta in the lower half $(iii)$ and the red momenta in the lower half $(iv) $. We note that a single application of the cavity-mediated interaction leads to the change $l \to l \pm 1$ but does not cause a change of the momentum group. 
As Fig.~\ref{fig:dkStructure} also reveals, the momentum inversion $k_x \to - k_x$ relates $(i) \leftrightarrow (ii)$ via $\ve k^{(i)}_l \leftrightarrow \ve k^{(ii)}_{-l-1}$ and $(iii) \leftrightarrow (iv)$ via $ \ve k^{(iii)}_{l} \leftrightarrow \ve k^{(iv)}_{-l+1}$ whereas the inversion $k_y \to - k_y$ leads to $(i) \leftrightarrow (iii)$ via $\ve k^{(i)}_l \leftrightarrow \ve k^{(iii)}_{l+1}$ and $(ii) \leftrightarrow (iv)$ via $\ve k^{(ii)}_l \leftrightarrow \ve k^{(iv)}_{l+1}$. We emphasize again that these operations do not affect the isotropic dispersion relation $\xi_{\ve k} =\ve k^2/(2m) - \eF$. 

\subsection{Representation of the Hamiltonian within the momentum structure}
Expressing the Hamiltonian Eq.~\eqref{eq:model} in this basis yields
\begin{align}\label{eq:modelII}
\begin{split}
    H = & \sum_{\substack{\nu = i,...,iv\\
    l \in \mathds Z}} \sum_{\delta k_x,\delta k_y}  \xi_{\ve{k^{(\nu)}_l}}
    \hat{c}^\dagger_{k^{(\nu)}_l} \hat{c}_{k^{(\nu)}_l} \\
    & + \frac{g}{2} \sum_{\substack{\nu,\nu' = i,...,iv\\
    l,l' \in \mathds Z; s= \pm 1}}
    \sum_{\substack{\delta k_x,\delta k_y \\ \delta k'_x,\delta k'_y}} \hat{c}^\dagger_{k^{(\nu)}_{l+s}}\hat{c}^\dagger_{k^{(\nu')}_{l'-s}} \hat{c}_{k^{(\nu')}_{l'}} \hat{c}_{k^{(\nu)}_l} \, ,
\end{split}
\end{align}
with the definition~\eqref{eq:defMomGroups} and the ranges of $\delta k_{x,y}$ from the previous paragraph. As this representation is just an exact rewriting of the original Hamiltonian, it does not really simplify the problem. However, this form is well-suited to capture the relevant channels in the mean-field decoupling introduced below. 
We point out that the full Hamiltonian~\eqref{eq:model} is invariant under the operations $k_x \to -k_x$ or $k_y \to - k_y$ such that Eq.~\eqref{eq:modelII} is, too, which can be easily proven on the basis of Eq.~\eqref{eq:defMomGroups}.
These symmetries will have implications on the different types of order-parameters. 

\section{Mean-field analysis}\label{sec:MF-analysis}
\subsection{Strategy of the MF decoupling from the perspective of the hot spots}
\begin{figure}[t]
\includegraphics[width=0.9\columnwidth]{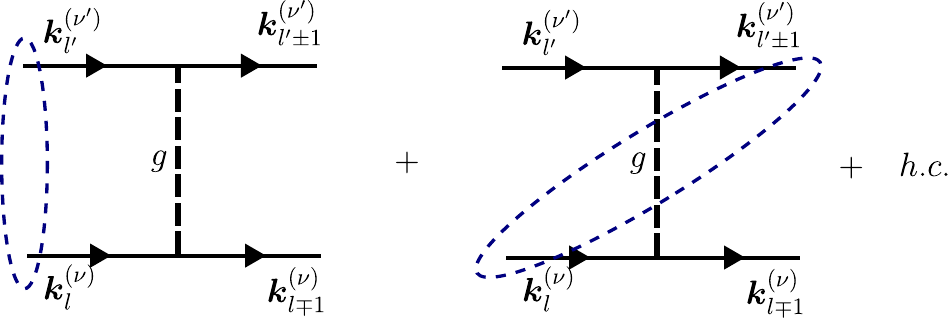}
\caption{Schematic representation of the MF decoupling of the interaction with momentum labels in the parameterization of Eq.~\eqref{eq:defMomGroups}. Left: Pairing channel. Right: Exchange channel. The hermitean conjugation includes two diagrams with the other pair of legs paired. }
\label{fig:decoup_channels}
\end{figure}

In the following, we perform a mean-field analysis, which becomes exact in the presence of the infinite-ranged cavity-mediated interactions. As the direct channel has already been discussed previously, we focus now on the pairing and exchange channels, shown in Fig.~\ref{fig:decoup_channels}. 

\begin{figure}[tb]
    \centering
    \includegraphics[width = 0.8 \columnwidth]{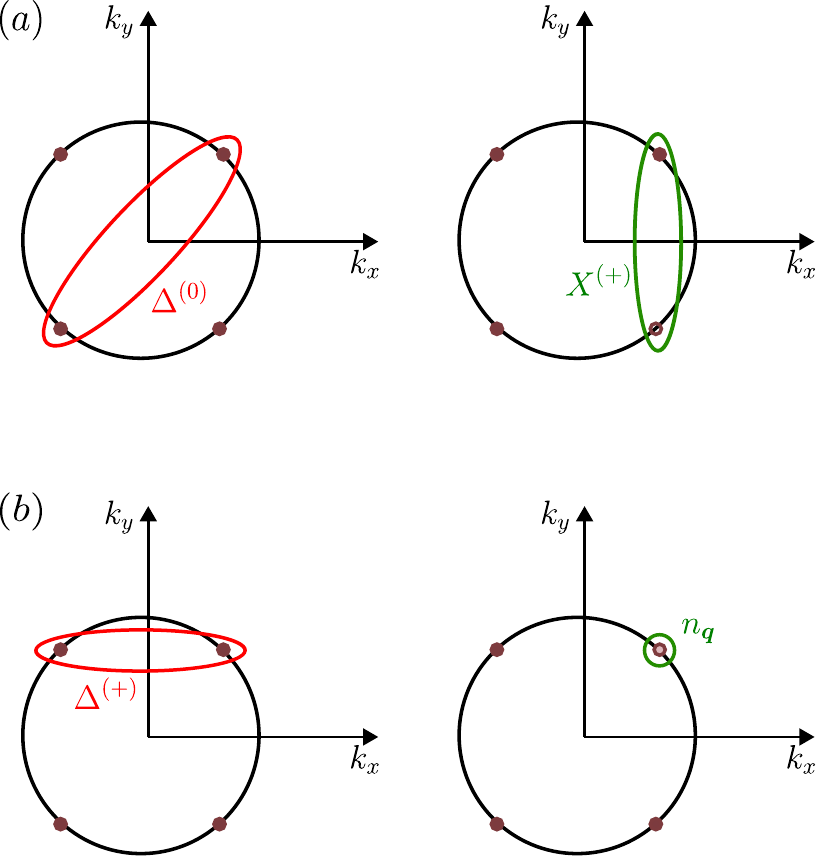}
    \caption{Instabilities arising on the hot spots in the two cases outlined in the text. The left (red) side represents the pairing channel that considers correlations between two particles (solid dots) and the right one the exchange channel with interactions between particles (solid dot) and holes (empty dot). Cooper pairs in scenario (a) with vanishing center-of-mass momentum correspond to the order parameter $\Delta^{(0)}$ while (b) has a pair-density-wave state with order parameter $\Delta^{(+)}$, which carries the total momentum $\ve P = 2\khs-\ve Q_c$. Analogously, the pairing may occur in the lower part with pairing momentum $-\ve P$ and order parameter $\Delta^{(-)}$ (not shown). The exchange channel in scenario (a) describes exciton condensation $X^{(+)}$ with relative momentum $\ve K = \ve P$, while for (b) the defining operator is given by momentum occupation at the hot spots as indicated for $\khs$.} 
    \label{fig:inst_hot_spots}
\end{figure}

On physical grounds, one expects the most dominant instabilities if all of the fermions undergoing the interaction are on the FS. This special case can be realized in two ways: (a) both incoming fermions have opposite momenta like $\pm \khs$ or $\pm (\khs - \ve Q_c)$ or (b) the momenta of the two incoming fermions correspond to two hot spots with the same component perpendicular to the cavity momentum: $\khs$ and $\khs -\ve Q_c$ or $-\khs$ and $-\khs +\ve Q_c$. These cases are depicted in Fig.~\ref{fig:inst_hot_spots}. 
The situation with incoming momenta $\khs$ and $-\khs + \ve Q_c$ that have the same component parallel to $\ve Q_c$, in contrast, does not lead to scattering both the outgoing momenta back on the FS due to momentum conservation. Therefore, instabilities arising from this scenario are suppressed. Finite expectation values in the situations (a) and (b) correspond to different instabilities. 
We illustrate this first with examples before we consider the full theory. In situation (a) let us assume that the incoming momenta are given by $\pm \khs$, which leads to following contribution to the mean-field Hamiltonian:
\begin{align}\label{eq:HS_a}
\begin{split}
& g \hat{c}^\dagger_{\khs - \ve Q_c} \hat{c}^\dagger_{-\khs + \ve Q_c} \hat{c}_{-\khs} \hat{c}_{\khs} \\
& \simeq   \Delta^{(0)}_{\khs} \hat{c}^\dagger_{\khs -\ve Q_c} \hat{c}^\dagger_{-\khs + \ve Q_c} + \bar\Delta^{(0)}_{\khs-\ve Q_c} \hat{c}_{-\khs} \hat{c}_{\khs } \\
& \quad -\frac{\bar\Delta^{(0)}_{\khs-\ve Q_c} \Delta^{(0)}_{\khs}}{g} \\
& \quad - X^{(+)}_{\khs} \hat{c}^\dagger_{\khs-\ve Q_c} \hat{c}_{-\khs} - X^{(-)}_{-\khs} \hat{c}^\dagger_{-\khs +\ve Q_c} \hat{c}_{\khs} \\
&\quad + \frac{X^{(+)}_{\khs} X^{(-)}_{-\khs}}{g}
\end{split}
\end{align}
The central line describes the pairing channel where an instability caused by Cooper pairs with vanishing center-of-mass momentum occurs. This is indicated by the special instances of the order parameter
\begin{align}\label{eq:defDelta0_HS}
\Delta^{(0)}_{\ve k} \equiv g \langle \hat{c}_{-\ve k} \hat{c}_{\ve k}  \rangle \, ,
\end{align}
where the bar denotes complex conjugation.

In the exchange channel, the above gives rise to a condensation of excitons (XCON) or, equivalently, particle-hole pairs with finite relative momentum $\pm \ve K  = \pm( 2 \khs - \ve Q_c)$ as special cases of the general order parameter
\begin{align}\label{eq:defXK_HS}
    X^{(\ve q)}_{\ve k} \equiv g \langle \hat{c}^\dagger_{\ve k} \hat{c}_{\ve k + \ve q} \rangle \, , 
\end{align}
characterized by the relative momentum $\ve q$.
In the above, we have used the convenient short-hand notation $X^{(\pm)}_{\pm \khs} = X^{(\pm \ve K)}_{\pm \khs}$. 

To elucidate situation (b) we consider $\khs$ and $ \khs -\ve Q_c$ as the two incoming momenta, which implies the following contribution to the mean-field Hamiltonian:
\begin{align}\label{eq:HS_b}
\begin{split}
& g \hat{c}^\dagger_{\khs - \ve Q_c} \hat{c}^\dagger_{\khs} \hat{c}_{\khs-\ve Q_c} \hat{c}_{\khs} \\
& \simeq   \Delta^{(+)}_{\khs} \hat{c}^\dagger_{\khs -\ve Q_c} \hat{c}^\dagger_{\khs} \!\! + \bar\Delta^{(+)}_{\khs-\ve Q_c} \hat{c}_{\khs-\ve Q_c} \hat{c}_{\khs }\!\! -\frac{\bar\Delta^{(+)}_{\khs-\ve Q_c} \Delta^{(+)}_{\khs}}{g} \\
& \quad -g \langle \hat n_{\khs} \rangle \hat{c}^\dagger_{\khs-\ve Q_c} \hat{c}_{\khs - \ve Q_c} - g \langle \hat n_{\khs-\ve Q_c} \rangle \hat{c}^\dagger_{\khs} \hat{c}_{\khs} \\
& \quad + g\langle \hat n_{\khs} \rangle \langle \hat n_{\khs-\ve Q_c}\rangle    
\end{split}
\end{align}

In the pairing channel, this leads to a pair density wave (PDW) state with total momentum $\ve P = 2\khs- \ve Q_c = \ve K$. Here, we introduce
\begin{align}\label{eq:defDeltaP_HS}
    \Delta_{\ve k}^{(\ve p)} \equiv g \langle \hat{c}_{\ve p - \ve k} \hat{c}_{\ve k}\rangle \, ,
\end{align}
and write $\Delta^{(\pm)}_{\pm \khs} = \Delta^{(\pm \ve P)}_{\pm \khs}$ or $\Delta^{(\pm)}_{\pm (\khs-\ve Q_c)} = \Delta^{(\pm \ve P)}_{\pm (\khs - \ve Q_c)}$ for the two most relevant center-of-mass momenta. Note that pairing with $-\ve P$ occurs by considering the two hot spots $-\khs$ and $-\khs +\ve Q_c$. Regarding the exchange channel, the above actually does not directly correspond to an instability since it rather contains the momentum occupation numbers $\left \langle \hat n_{\ve q} \right \rangle = \left \langle \hat{c}^\dagger_{\ve q} \hat{c}_{\ve q} \right \rangle $ at the hot spots. 

A graphic representation of these decoupling channels in cases (a) and (b) can be found again in Fig.~\ref{fig:inst_hot_spots}. 
\subsection{Construction of the mean-field Hamiltonian}
For a full picture of the mean-field theory, however, it is important to include finite momentum deviations from the hot spots.
Regarding possible instabilities, we have at finite $\ve \delta \ve k^\pm$ and arbitrary values of $l \in \mathds Z$, in principle, many more options to generate order parameters, as compared to the pure hot spot theory outlined in the previous paragraph. 
To progress, we follow the physical principle to take only those types of instabilities into account that are characterized by  order parameters which for $\ve \delta \ve k^\pm \to 0$ reproduce the dominant instabilities obtained at the hot spots. 
Let us first consider pairing: Cooper pairs can be formed by combining any fermion from group $(i)$ with the fermion of opposite momentum that is contained in group $(iv)$, see Fig.~\ref{fig:dkStructure}. Analogously, Cooper pairs can be obtained from momenta chosen appropriately out of $(ii)/(iii)$.
More precisely, the corresponding order parameters read
\begin{align}\label{eq:defDelta0gen}
    \Delta^{(0)}_{\ve k_l^{(i)}} \equiv g\langle \hat{c}_{\ve k_{-l}^{(iv)}} \hat{c}_{\ve k_l^{(i)}} \rangle\, , \quad \Delta^{(0)}_{\ve k_l^{(iii)}} \equiv  g\langle \hat{c}_{\ve k_{-l}^{(ii)}} \hat{c}_{\ve k_l^{(iii)}} \rangle\, ,
\end{align}
which generalizes~\eqref{eq:defDelta0_HS}.
The corresponding contribution to the mean-field Hamiltonian $\hat H_\mf$ is obtained by considering those interaction terms of~\eqref{eq:modelII} that have a vanishing total momentum of the incoming fermions and decoupling them in the pairing channel:
\begin{align}\label{eq:H_Cooper}
\begin{split}
& g \sum_{\substack{\delta k_x, \delta k_y \\ l \in \mathds{Z} \\ s = \pm 1}} \sum_{\substack{(\nu, \nu') = (i,iv) \\
(\nu,\nu') = (iii,ii)}} \hat{c}^\dagger_{\ve k^{(\nu)}_{l+s}} \hat{c}^\dagger_{\ve k^{(\nu')}_{-l-s}} \hat{c}_{\ve k^{(\nu')}_{-l}} \hat{c}_{\ve k^{(\nu)}_{l}} \longrightarrow \\
&\hat H_{\text{Cooper}} = \sum_{\substack{\delta k_x, \delta k_y \\ l \in \mathds{Z} \\ s = \pm 1}} \sum_{\substack{(\nu, \nu') = (i,iv) \\
(\nu,\nu') = (iii,ii)}} \bigg[
    \bar \Delta^{(0)}_{\ve k^{(\nu)}_{l+s}} \hat{c}_{\ve k^{(\nu')}_{-l} } \hat{c}_{\ve k^{(\nu )}_l} \\
    & \qquad \qquad \qquad+  \Delta^{(0)}_{\ve k^{(\nu)}_{l}} \hat{c}^\dagger_{\ve k^{(\nu)}_{l+s} } \hat{c}^\dagger_{\ve k^{(\nu')}_{-l-s}} - \frac{\bar \Delta^{(0)}_{\ve k^{(\nu)}_{l+s}} \Delta^{(0)}_{\ve k^{(\nu)}_{l}}}{g} \bigg] 
\end{split}
\end{align}
We note that the factor of $1/2$ has been absorbed by fixing the order of the group pairs $(\nu,\nu')$. Furthermore, $\hat H_{\text{Cooper}}$ is hermitean as can be shown by the concatenating the index shifts $l \to l -s$ and $s\to -s$. 

PDW states with energetically equivalent single-particle content as compared to the Cooper case can be formed by fermions from $(i)/(ii)$ or $(iii)/(iv)$ provided that the center-of-mass momentum is chosen as $\ve P(\delta \ve k^+) = \pm (2\khs - \ve Q_c + 2 \delta k_y \ve e_y)$. Here, the positive (negative) sign refers to the first (second) combination. The condition $\ve P(\ve{\delta k}^+\to \ve 0) = \pm \ve P$ is satisfied in both cases. In addition, the symmetry of the FS chooses $\ve P(\ve{\delta k}^+)$ perpendicular to $\ve Q_c$ for all momentum shifts $\ve \delta \ve k^+$. 
The resulting PDW order parameters are given by
\begin{align}\label{eq:defDeltaPDW}
\Delta^{(+)}_{\ve k_l^{(i)}} \equiv  g \langle \hat{c}_{\ve k_{-l-1}^{(ii)}} \hat{c}_{\ve k_l^{(i)}} \rangle\, ,\quad   \Delta^{(-)}_{\ve k_{l}^{(iii)}} \equiv g \langle \hat{c}_{\ve k_{-l+1}^{(iv)}} \hat{c}_{\ve k_l^{(iii)}} \rangle\, ,
\end{align} 
in agreement with Eq.~\eqref{eq:defDeltaP_HS}, where the super indices refer to the special cases $\pm \ve P( \ve \delta \ve k^+)$ considered here.
Now, the contribution to $\hat H_\mf$ can be obtained from those interaction terms with center-of-mass momentum $\pm\ve P(\ve \delta \ve k^+)$. Decoupling again in the pairing channel leads to:
\begin{align}\label{eq:H_PDW}
\begin{split}
& g\!\! \sum_{\substack{\delta k_x, \delta k_y \\ l \in \mathds{Z} \\ s = \pm 1}} \sum_{\substack{[(\nu, \nu'),r] = [(i,ii),+] \\
[(\nu,\nu'),r] = [(iii,iv),-]}}\!\!\! \hat{c}^\dagger_{\ve k^{(\nu)}_{l+s}} \hat{c}^\dagger_{\ve k^{(\nu')}_{-l-r-s}} \hat{c}_{\ve k^{(\nu')}_{-l-r}} \hat{c}_{\ve k^{(\nu)}_{l}} \! \longrightarrow \\
&\hat H_{\text{PDW}} = \sum_{\substack{\delta k_x,\delta k_y \\ l \in \mathds{Z} \\ s = \pm 1}} \sum_{\substack{[(\nu, \nu'),r] = [(i,ii),+] \\
[(\nu,\nu'),r] = [(iii,iv),-]}} \bigg[
    \bar \Delta^{(r)}_{\ve k^{(\nu)}_{l+s}} \hat{c}_{\ve k^{(\nu')}_{-l-r} } \hat{c}_{\ve k^{(\nu )}_l} \\
    & \qquad \qquad \qquad+  \Delta^{(r)}_{\ve k^{(\nu)}_{l}} \hat{c}^\dagger_{\ve k^{(\nu)}_{l+s} } \hat{c}^\dagger_{\ve k^{(\nu')}_{-l-r-s}} - \frac{\bar \Delta^{(r)}_{\ve k^{(\nu)}_{l+s}} \Delta^{(r)}_{\ve k^{(\nu)}_{l}}}{g} \bigg] 
\end{split}
\end{align}
Again, the factor $1/2$ has been absorbed into fixing the momentum groups $\nu$ and $\nu'$ while the identity $\hat H_{\text{PDW}}^\dagger = \hat H_{\text{PDW}}$ follows via the index shift $l \to l-s$ followed by $s \to -s$.

In the pure hot spot theory from the previous section the two interaction contributions (a) with vanishing center-of-mass momentum~\eqref{eq:HS_a} and (b) with the special total momentum $\pm \ve P$~\eqref{eq:HS_b} are responsible for the most important contributions in the exchange channel, the XCON with relative momentum $\ve K$ and the effects from the momentum occupations of the FS, too. For finite $\delta k_x,\delta k_y$, this is no longer the case and decoupling Eqs.~\eqref{eq:H_Cooper} and~\eqref{eq:H_PDW} in the exchange channel does not give rise to expectation values that are characterized by a unique relative momentum. In order to capture the exchange processes that reflect the results from the hotpots one has to take different total momenta into account. We consider first the terms that contain the momentum occupation numbers $\langle \hat n_{\ve k}\rangle$. They arise from:
\begin{align}\label{eq:H_mom_occ}
\begin{split}
& \frac{g}{2} \sum_{\substack{\delta k_x, \delta k_y \\ l \in \mathds{Z} \\ s = \pm 1}} \sum_{\substack{\nu = i,ii,\\iii,iv}} \hat{c}^\dagger_{\ve k^{(\nu)}_{l+s}} \hat{c}^\dagger_{\ve k^{(\nu)}_{l}} \hat{c}_{\ve k^{(\nu)}_{l+s}} \hat{c}_{\ve k^{(\nu)}_{l}} \longrightarrow \\
&\hat H_{\hat n} = -\frac{g}{2}\sum_{\substack{\delta k_x,\delta k_y \\ l \in \mathds{Z} \\ s = \pm 1}} \sum_{\substack{\nu= i,ii,\\ iii,iv}} \bigg[
    \langle \hat{n}_{\ve k^{(\nu)}_{l+s}}\rangle \hat{c}^\dagger_{\ve k^{(\nu)}_{l} } \hat{c}_{\ve k^{(\nu )}_l} \\
    & \qquad \qquad \qquad+  \langle \hat n_{\ve k^{(\nu)}_{l}}\rangle \hat{c}^\dagger_{\ve k^{(\nu)}_{l+s} } \hat{c}_{\ve k^{(\nu)}_{l+s}} - \langle \hat n_{\ve k^{(\nu)}_{l+s}}\rangle \langle \hat n_{\ve k^{(\nu)}_{l}}\rangle \bigg] \,.
\end{split}
\end{align}
Here, the factor $1/2$ cannot be included into the sum but since all four momentum groups are summed over independently $\hat H_{\hat n}$ represents as many terms as $\hat H_{\text{Cooper}}$ and $\hat {H}_{\text{PDW}}$.
As already mentioned above, these expectation values do not directly represent an instability but are finite for certain momenta at any finite density even in the absence of interactions. As we will see below, this entails a renormalization of the momentum-distribution of the ground state and thereby affects the momentum structure of the order parameters associated with all other occurring instabilities.

Finally, we turn to the generalization of the exciton condensation.We have identified the special relative momentum $\ve K = 2\khs -\ve Q_c$ in the case of hot spots. With finite momentum deviations $\delta k_x, \delta k_y$ we focus on generalizations to the hot spot scenario obtained by the specific combinations of groups
$(i)/(iv)$ and $(iii)/(ii)$~\footnote{Very similar particle-hole expectation values $g \langle \hat{c}^\dagger_{\ve k^{(iii)}_{l+1}} \hat{c}_{\ve k^{(i)}_l}\rangle$ or $ g \langle \hat{c}^\dagger_{\ve k^{(iv)}_{l+1}} \hat{c}_{\ve k^{(ii)}_l}\rangle $ can be formed from $(i)/(iii)$ and $(ii)/(iv)$. Both are characterized by $\ve K(\ve \delta \ve k^+) = 2 \khs -\ve Q_c +2 \delta k_y \hat{\boldsymbol e}_y = \ve P(\ve \delta \ve k^+)$.   However, only the choice from the main text, allows for the familiar particle-hole excitations with a hole inside the FS and particle outside of the FS when a hot spot and  small enough values of $|\delta k_y|$ are considered.  As a consequence, also the bare particle-hole susceptibility is larger for the choice in the main text, which therefore represents the dominant instability.}.: 
\begin{align}\label{eq:def:Xldelta}
    X^{(+)}_{\ve k_l^{(i)}}  \equiv g \langle \hat{c}^\dagger_{\ve k^{(iv)}_{l+1}} \hat{c}_{\ve k^{(i)}_l} \rangle\, , \quad X^{(+)}_{\ve k_l^{(ii)}}  \equiv g \langle \hat{c}^\dagger_{\ve k^{(iii)}_{l}} \hat{c}_{\ve k^{(ii)}_{l-1}} \rangle \, .
\end{align}
This gives rise to particle-hole pairs carrying the
relative momenta $\ve K(\ve \delta \ve k^+) =2 \khs - \ve Q_c + 2 \ve \delta \ve k^+  $ or $\ve K(\ve \delta \ve k^-) =2 \khs - \ve Q_c + 2 \ve \delta \ve k-$, respectively  according to the general definition~\eqref{eq:defXK_HS}. The superindex $\pm$ represents again our short-hand notation. Inverting the order of the momenta in Eq.~\eqref{eq:def:Xldelta} gives rise to the corresponding expectation values $X^{(-)}_{\ve k^{(iv)}_{l+1}} = \bar X^{(+)}_{\ve k_l^{(i)}}$ and $X^{(-)}_{\ve k^{(iii)}_{l+1}} = \bar X^{(+)}_{\ve k_l^{(ii)}}$ with relative momenta $-\ve K(\ve \delta \ve k^+)$ or $-\ve K(\ve \delta \ve k^-)$, respectively. 
The related term to the mean-field Hamiltonian is obtained through decoupling in the exchange channel of the following contribution to~\eqref{eq:modelII}:
\begin{align}\label{eq:H_XCON_X}
\begin{split}
& g \sum_{\substack{\delta k_x, \delta k_y \\ l \in \mathds{Z} \\ s = \pm 1}} \sum_{\substack{(\nu, \nu') = (i,iv) \\
(\nu,\nu') = (ii,iii)}} \hat{c}^\dagger_{\ve k^{(\nu)}_{l+s}} \hat{c}^\dagger_{\ve k^{(\nu')}_{l+1}} \hat{c}_{\ve k^{(\nu')}_{l+1+s}} \hat{c}_{\ve k^{(\nu)}_{l}} \longrightarrow \\
&\hat H_{\text{XCON}}^{(X)} = - \sum_{\substack{\delta k_x,\delta k_y \\ l \in \mathds{Z} \\ s = \pm 1}} \sum_{\substack{(\nu, \nu') = (i,iv) \\
(\nu,\nu') = (ii,iii)}} \bigg[
    \bar{X}^{(+)}_{\ve k^{(\nu)}_{l+s}} \hat{c}^\dagger_{\ve k^{(\nu')}_{l+1}} \hat{c}_{\ve k^{(\nu )}_l} \\
    & \qquad \qquad \qquad+  X^{(+)}_{\ve k^{(\nu)}_{l}} \hat{c}^\dagger_{\ve k^{(\nu)}_{l+s} } \hat{c}_{\ve k^{(\nu')}_{l+1+s}} - \frac{\bar{X}^{(+)}_{\ve k^{(\nu)}_{l+s}} X^{(+)}_{\ve k^{(\nu)}_{l}}}{g} \bigg] \, .
\end{split}
\end{align}
The factor $1/2$ has again been absorbed in the usual way, while the identity $[\hat H_{\text{XCON}}^{(X)}]^\dagger = \hat H_{\text{XCON}}^{(X)}$ can again be shown by $l \to l-s$ and $s \to -s$. 

However, particle-hole pairs with almost identical energy content are formed also by the combinations
\begin{align}\label{eq:def:Yldelta}
    Y^{(+)}_{\ve k_l^{(i)}}  \equiv g \langle \hat{c}^\dagger_{\ve k^{(ii)}_{l}} \hat{c}_{\ve k^{(i)}_l} \rangle\, , \quad Y^{(+)}_{\ve k_l^{(iv)}}  \equiv g \langle \hat{c}^\dagger_{\ve k^{(iii)}_{l}} \hat{c}_{\ve k^{(iv)}_l} \rangle \, .
\end{align}
As is detailed in Sec.~\ref{sec:MF_Ana_symm}, these contribution turn also out to be relevant in order conserve the symmetries of the general Hamiltonian~\eqref{eq:modelII} when the mean-field theory is applied. 
Here, the relative momenta read $\pm  \ve q (\ve \delta \ve k^\pm) = \pm (\ve \delta \ve k^+ + \ve \delta \ve k^-) = \pm 2 \delta k_x \hat{\ve e}_x$, as follows from the definition~\eqref{eq:defXK_HS}.

From a physical perspective these terms can be interpreted as shifting the two operators forming $\hat n_{\ve q}$ in the hot spot case by the momentum deviations $\ve \delta \ve k^+$ and $\ve \delta \ve k^-$.
Exchanging the momenta in Eq.~\eqref{eq:def:Yldelta} yields
$Y^{(-)}_{\ve k^{(ii)}_{l}} = \bar Y^{(+)}_{\ve k_l^{(i)}}$ and $Y^{(-)}_{\ve k^{(iii)}_{l}} = \bar Y^{(+)}_{\ve k_l^{(iv)}}$ while mean-field decoupling leads to
\begin{align}\label{eq:H_XCON_Y}
\begin{split}
& g \sum_{\substack{\delta k_x, \delta k_y \\ l \in \mathds{Z} \\ s = \pm 1}} \sum_{\substack{(\nu, \nu') = (i,ii) \\
(\nu,\nu') = (iv,iii)}} \hat{c}^\dagger_{\ve k^{(\nu)}_{l+s}} \hat{c}^\dagger_{\ve k^{(\nu')}_{l}} \hat{c}_{\ve k^{(\nu')}_{l+s}} \hat{c}_{\ve k^{(\nu)}_{l}} \longrightarrow \\
&\hat H_{\text{XCON}}^{(Y)} = - \sum_{\substack{\delta k_x,\delta k_y \\ l \in \mathds{Z} \\ s = \pm 1}} \sum_{\substack{(\nu, \nu') = (i,ii) \\
(\nu,\nu') = (iv,iii)}} \bigg[
    \bar{Y}^{(+)}_{\ve k^{(\nu)}_{l+s}} \hat{c}^\dagger_{\ve k^{(\nu')}_{l}} \hat{c}_{\ve k^{(\nu )}_l} \\
    & \qquad \qquad \qquad+  Y^{(+)}_{\ve k^{(\nu)}_{l}} \hat{c}^\dagger_{\ve k^{(\nu)}_{l+s} } \hat{c}_{\ve k^{(\nu')}_{l+s}} - \frac{\bar{Y}^{(+)}_{\ve k^{(\nu)}_{l+s}} Y^{(+)}_{\ve k^{(\nu)}_{l}}}{g} \bigg] \, .
\end{split}
\end{align}
Taking Eqs.~\eqref{eq:H_XCON_X} and~\eqref{eq:H_XCON_Y} together, we find for the excition-condensation sector:
\begin{align}\label{eq:H_XCON}
    \hat H_{\text{XCON}} = \hat H_{\text{XCON}}^{(X)} +\hat H_{\text{XCON}}^{(Y)} \, .
\end{align}
As a result, the total mean-field Hamiltonian reads
\begin{align}\label{eq:HMFfull}
\begin{split}
    \hat H_\mf = & \sum_{\substack{\delta k_x,\delta k_y \\ l \in \mathds Z}} \sum_{\substack{\nu = i,ii,\\iii,iv}}   \xi_{\ve{k^{(\nu)}_l}}
    \hat{c}^\dagger_{\ve k^{(\nu)}_l} \hat{c}_{\ve k^{(\nu)}_l} \\
    & + \hat H_{\text{Cooper}} + \hat H_\text{PDW} + \hat H_{\hat n} + \hat H_{\text{XCON}}  \, ,
\end{split}
\end{align}
with the the kinetic part from Eq.~\eqref{eq:modelII} 
and the interaction terms of \cref{eq:H_Cooper,eq:H_PDW,eq:H_mom_occ,eq:H_XCON}.
Notice that because of the special form of the cavity-mediated interactions the momentum-structure of these mean-field interaction terms resemble a nearest-neighbor tight-binding model frequently encountered in real-space. 
 
 It turns out in the following that the instabilities introduced previously 
 occur below a certain critical temperature at renormalized hot spots. With decreasing temperature the momentum region where the FS becomes unstable spreads out from these renormalized hot spots. Since the interior of these regions has already developed a symmetry-breaking expectation value in any of these states at larger temperatures, the energetically favorable choices for the momenta of the PDW and exciton condensates do not change and are given by $\ve P(\boldsymbol \delta \ve k^+)$ and $\ve K(\boldsymbol \delta \ve k^+)$ or $\ve q (\ve \delta \ve k^+)$. This provides an a posteriori justification for the way how we constructed the mean field-Hamiltonian.

\subsection{Mean-field equations}
After constructing $\hat H_\mf$ in Eq.~\eqref{eq:HMFfull}, we next turn to its solution. 
We have already noted that $\hat H_\mf$ is hermitean. 
In addition, the anomalous expectation values are odd under exchange of the momentum indices in the sense
\begin{align}
\begin{split}
    \Delta^{(0)}_{\ve k^{(\nu)}_l} & = - \Delta^{(0)}_{\ve k^{(\nu')}_{-l}}\   \ \text{for } (\nu,\nu') = (i,iv),(iii,ii) \\
    \Delta^{(r)}_{\ve k^{(\nu)}_l} & = -  \Delta^{(r)}_{\ve k^{(\nu')}_{-l-r}} \   \ \;\text{for } (\nu,\nu',r) = (i,ii,+),(iii,iv,-) \, ,
\end{split}
\end{align}  
according to Eq.~\eqref{eq:defDelta0gen} or Eq.~\eqref{eq:defDeltaPDW}. Since this exchange of momentum indices is formally equivalent to taking the transpose, the mean-field Hamiltonian~\eqref{eq:HMFfull} is amenable to a general fermionic Bogoliubov transformation~\cite{BlaizotBook}. Therefore, it is equivalent to the diagonal form
\begin{align}\label{eq:Bogoliubov}
\begin{split}
     \hat H_\mf =
       \sum_{\alpha} \epsilon_\alpha \hat{\gamma}^\dagger_\alpha \hat{\gamma}^{}_\alpha + E_0 \, .
\end{split}
\end{align}
The fermionic operators $\hat{\gamma}^{(\dagger)}_\alpha$ annihilate (create) a Bogoliubov quasiparticle with excitation energy $\epsilon_\alpha > 0$ where $\alpha$ is just an arbitrary label to enumerate the eigenvalues. Below we relate $\alpha$ to more physical descriptions. The $\epsilon_\alpha$ are given 
by the positive eigenvalues of $2\hat H_\mf$, which generically appear in pairs $\pm \epsilon_\alpha$ due to the properties of fermionic Bogoliubov transformations. The factor two arises from doubling the dimensions of the Hamiltonian when introducing a Nambu structure to perform the Bogoliubov transformation~\cite{BlaizotBook}. 
$E_0$ denotes the (mean-field) ground-state energy and reads:
\begin{align}
\begin{split}
     & E_0  = \frac{1}{2}\!\sum_{\substack{\delta k_x,\delta k_y \\ l \in \mathds Z}}\sum_{\substack{\nu = i,ii,\\iii,iv}}   \!\left[ \xi_{\ve{k}^{(\nu)}_l}-\sum_{s=\pm 1} g\langle \hat{n}_{\ve k^{(\nu)}_{l+s}}\rangle\right]\!\! -\frac{1}{2} \sum_{\alpha} \epsilon_\alpha \\
    & - \sum_{\substack{\delta k_x,\delta k_y \\ l \in \mathds Z\\ s = \pm 1}}
    \left[\sum_{\substack{\nu = i,iii}}  \frac{\bar \Delta^{(0)}_{\ve k^{(\nu)}_{l+s}} \Delta^{(0)}_{\ve k^{(\nu)}_{l}}}{g} + \sum_{\substack{[\nu,r] = [i,+],\\ [iii,-]}}\frac{\bar\Delta^{(r)}_{\ve k^{(\nu)}_{l+s}} \Delta^{(r)}_{\ve k^{(\nu)}_{l}}}{g}\right]\\
    & +\sum_{\substack{\delta k_x,\delta k_y \\ l \in \mathds Z \\ s = \pm 1}}  \left[ \sum_{\substack{\nu = i,ii,\\iii,iv}} \frac{g}{2} \langle \hat n_{\ve k^{(\nu)}_{l+s}}\rangle \langle \hat n_{\ve k^{(\nu)}_{l}}   \rangle
    +\sum_{\substack{\nu = i,iii}} \frac{\bar{X}^{(+)}_{\ve k^{(\nu)}_{l+s}} X^{(+)}_{\ve k^{(\nu)}_{l}}}{g}\right] \\
    & +\sum_{\substack{\delta k_x,\delta k_y \\ l \in \mathds Z \\ s = \pm 1}} \sum_{\substack{\nu = i,iv}} \frac{\bar{Y}^{(+)}_{\ve k^{(\nu)}_{l+s}} Y^{(+)}_{\ve k^{(\nu)}_{l}}}{g}
    \, . 
\end{split}
\end{align}

Now, we can state the mean-field equations in general form. After diagonalizing $H_\mf$ like in Eq.~\eqref{eq:Bogoliubov}, the contribution to the free energy $\mathcal{F}_\mf$ is given by the standard expression for noninteracting fermions in addition to the constant ground-state energy
\begin{align}\label{eq:FreeEnMF}
\begin{split}
& \mathcal F_\mf =-T  \sum_{\alpha} \log(1+e^{- \beta \epsilon_\alpha}) +E_0 \, ,    
\end{split}
\end{align}
where the $\epsilon_\alpha$ denote again the positive eigenvalues of $\hat H_\mf$. Setting the variations of $\mathcal F_\mf$ as function of the mean-field parameters to zero yields the mean-field equations:
\begin{align}\label{eq:MF_equations}
    \begin{split}
        \Delta^{(r)}_{\ve k^{(\nu)}_l} & = - \frac{g}{2} \sum_{l'} \left(\underline M\right)^{-1}_{\ve k^{(\nu)}_l \ve k^{(\nu)}_{l'}} \sum_{\alpha} \text{th}\left(\frac{\beta\epsilon_\alpha}{2}\right)\frac{\delta \epsilon_\alpha}{\delta \bar\Delta^{(r)}_{\ve k^{(\nu)}_{l'}}} \\
        A^{(+)}_{\ve k^{(\nu)}_l} & = \frac{g}{2} \sum_{\ve k^{(\nu)}_{l'}} \left(\underline M\right)^{-1}_{\ve k^{(\nu)}_l \ve k^{(\nu)}_{l'}} \sum_{\alpha} \text{th}\left(\frac{\beta\epsilon_\alpha}{2}\right)\frac{\delta \epsilon_\alpha}{\delta \bar A^{(+)}_{\ve k^{(\nu)}_{l'}}} \\
        \langle \hat n_{\ve k^{(\nu)}_l} \rangle -\frac{1}{2} & = \frac{1}{2} \sum_{\ve k^{(\nu)}_{l'}} \left(\underline M\right)^{-1}_{\ve k^{(\nu)}_l \ve k^{(\nu)}_{l'}} \sum_{\alpha} \text{th}\left(\frac{\beta\epsilon_\alpha}{2}\right)\frac{\delta (\epsilon_\alpha/g)}{\delta \langle \hat n_{\ve k^{(\nu)}_{l'}} \rangle}\, ,
    \end{split}
\end{align}
with $r=0,\pm$ and $A^{(+)}_{\ve k^{(\nu)}_l}= X^{(+)}_{\ve k^{(\nu)}_l},Y^{(+)}_{\ve k^{(\nu)}_l}$. Furthermore, the matrix 
\begin{align} \label{eq:def_M}
\underline M_{\ve k^{(\nu)}_l \ve k^{(\nu)}_{l'}} = \delta_{\ve k^{(\nu)}_l, \ve k^{(\nu)}_{l'+1}}+\delta_{\ve k^{(\nu)}_l, \ve k^{(\nu)}_{l'-1}} \, ,
\end{align}
which resembles a tight-binding matrix with open boundary conditions, takes the fixed momentum transfer into account. 

As usual, these equations identify local extrema of the free energy. Determination of the thermodynamic state as global minimum requires to compare the values of $\mathcal F_\mf$ in case of several solutions. This allows to answer whether the instabilities correspond to first or second order phase transitions. 
In the presence of nonvanishing order parameters and finite renormalizations of the occupation numbers in all channels the eigenvalues $\epsilon_\alpha(\Delta^{(0)},\Delta^{(\pm)},X^{(\pm)},Y^{(\pm)}, \langle \hat n \rangle)$ and the corresponding operators cannot be determined analytically. However, certain limiting cases allow for analytic statements that will be discussed in the following paragraphs.
To proceed in the most general case one considers the mean-field equations as a recursive problem that is treated iteratively until convergence signals the self-consistent solution. This requires the truncation of the momentum grids. We will mostly consider grids derived from the four hot spots and their four closest shifted copies (cf. Fig.~\ref{fig:dkStructure}), such that $\hat H_\mf$ corresponds to a $32 \times 32$ matrix for each value of $\ve \delta \ve k^+$ (4 hot spots + four shifted copies, each with 2 values $\ve \delta \ve k^+$ and $\ve \delta \ve ^+k$ and finally the standard doubling of dimension by taking the anomalous terms into account within a Nambu structure). The generalization to more shifted copies or the reduction to a pure hot spot picture can be easily accomplished for the numerics but our choice allows to understand the essential physics to a large extent by analytic means.

\section{Reshaping of the bare Fermi surface}\label{sec:MF_Ana_reshape}
After having established the structure of the mean-field approach in the previous chapters we turn now to the solution of the corresponding mean-field equations and their physical implications. To this end, we consider first the effects of the momentum occupation terms~\eqref{eq:H_mom_occ}, which are present without any instability, while neglecting the order parameters in the Cooper and PDW channels as well as the XCON. They will be studied in detail below in Sec.~\ref{sec:MF_Ana_inst}. This stepwise approach helps to elucidate the full picture obtained from coupled mean-field equations.

More precisely, we consider in this section the restricted mean-field Hamiltonian (cf. Eq.~\eqref{eq:HMFfull})
\begin{align}
\begin{split}
& \sum_{\substack{\delta k_x,\delta k_y \\ l \in \mathds Z}} \sum_{\substack{\nu = i,ii,\\iii,iv}}   \xi_{\ve{k}^{(\nu)}_l}
    \hat{c}^\dagger_{\ve k^{(\nu)}_l} \hat{c}_{\ve k^{(\nu)}_l} +\hat H_{\hat n}\\
 & = \sum_{\substack{\delta k_x,\delta k_y \\ l \in \mathds Z}} \sum_{\substack{\nu = i,ii,\\iii,iv}} \left(\xi_{\ve{k^{(\nu)}_l}}
 - g \sum_{s=\pm1} \langle \hat n_{\ve k^{(\nu)}_{l+s}}\rangle 
 \right)
    \hat{c}^\dagger_{\ve k^{(\nu)}_l} \hat{c}_{\ve k^{(\nu)}_l}  \\
   &\quad - \sum_{\substack{\delta k_x,\delta k_y \\ l \in \mathds Z\\ s = \pm 1}} \sum_{\substack{\nu = i,ii,\\iii,iv}}  \langle \hat n_{\ve k^{(\nu)}_{l+s}}\rangle  \langle \hat n_{\ve k^{(\nu)}_l}\rangle \, .
\end{split}
\end{align}
With other OPs being absent, the Hamiltonian is already diagonal with renormalized dispersion 
\begin{align}\label{eq:disp_ren}
\epsilon^{(\hat n)}_{\ve k^{(\nu)}_l} =  \xi_{\ve{k^{(\nu)}_l}}
 - g \sum_{s=\pm1} \langle \hat n_{\ve k^{(\nu)}_{l+s}}\rangle  \, ,
\end{align}
while the occupation numbers of the modes obey Fermi-Dirac statistics
\begin{align}\label{eq:occ_ren}
  \langle \hat n_{\ve k^{(\nu)}_l}\rangle = n_F\left(\epsilon^{(\hat n)}_{\ve k^{(\nu)}_l}\right)\, .
\end{align}
The same is result is obtained after inserting the positive solution of the two eigenvalues $\pm \epsilon^{(\hat n)}_{\ve k^{(\nu)}_l}$ as Bogoliubov energy into the last mean-field equation from~\eqref{eq:MF_equations}. The other equations are trivially satisfied for vanishing order parameters. We note that the latter two equations represent a very simple variant of Landau Fermi Liquid theory, where the dispersions and the occupations of all quasiparticle excitations become renormalized by the presence of the other quasiparticles. Here, the special momentum structure of the interaction drastically  restricts these mutual modifications between the quasiparticles.

\begin{figure*}[htb]
  \includegraphics[width=\textwidth]{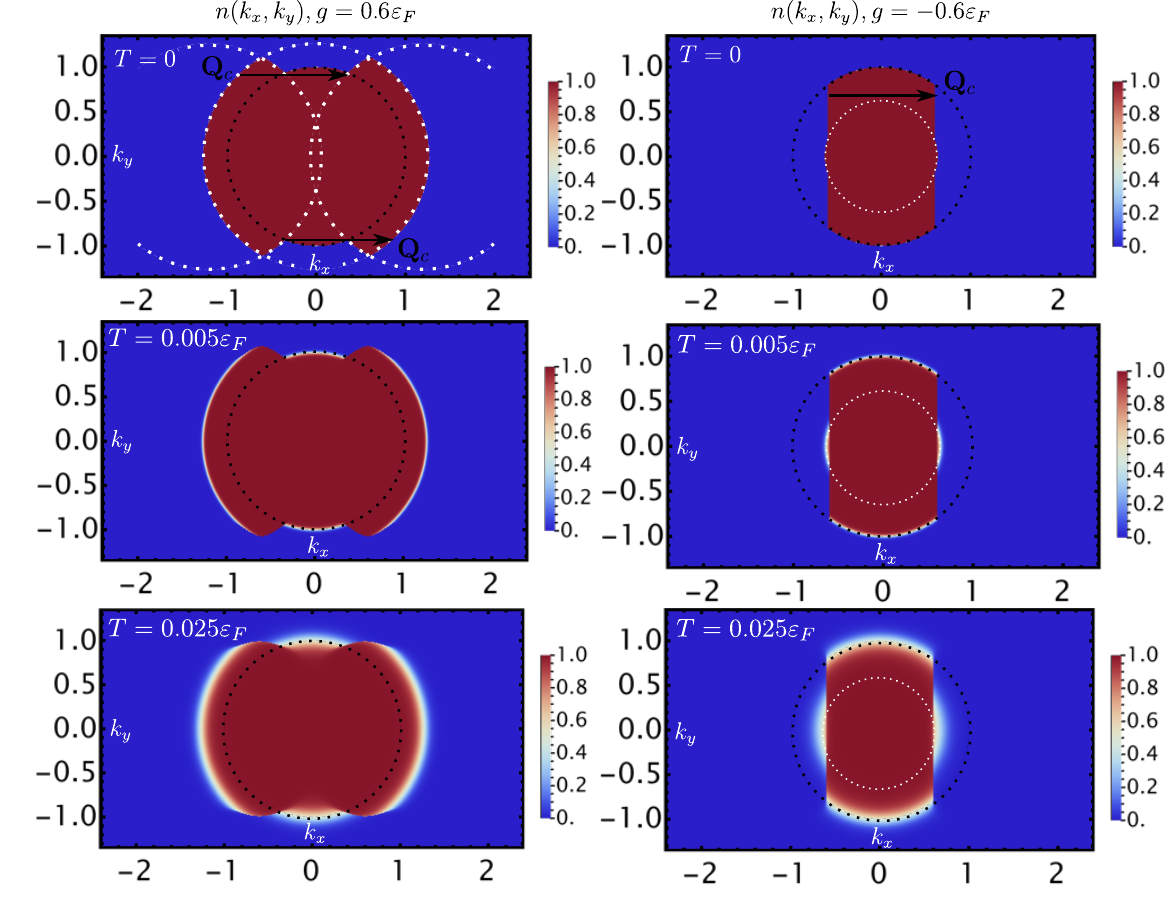}
  \caption{The renormalized momentum occupations reveal a reshaped FS with thermal corrections shown for different temperatures in units of the bare $\eF$. Left: repulsive interactions $g = 0.6 \eF$, right: attractive interactions $g=-0.6$. The dotted black circle indicates the bare FS. In the plot at $T=0$ and $g=0.6 \eF$ the dotted white circles depict the circle $\xi_{\ve k} = g $ and its image under shifts by $\pm \ve Q_c$. As explained in the main text the intersection of these three circles is also occupied in the ground state. The small white dotted circle on the right-hand side depicts also $\xi_{\ve k} = g $, which is the region guaranteed to be occupied for attractive interactions. Outside of this circle occupation is merely found in a vertical strip that is a subset of the bare FS. The black arrows indicate those parts of the renormalized FS that are nested by the cavitiy momentum. We note that thermal broadening of the FS is drastically reduced when nesting occurs, which is a consequence of the cavity-mediated interactions.}
  \label{fig:ren_FS}
\end{figure*}

The solution to~\cref{eq:disp_ren,eq:occ_ren}, which is obtained numerically via iteration, is presented in Fig.~\ref{fig:ren_FS} for various temperatures. At $T=0$, however, the resulting structure of the renormalized FS can be understood by analytical means. The ground state is formed by those quasiparticles with negative $\epsilon^{(n)}_{\ve k^{(\nu)}_l}$ that simultaneously allow to minimize the total energy. For convenience, we consider a grand canonical description with variable particle number and discuss our example with the isotropic bare dispersion and $ Q_c = 1.2 k_F > k_F$ in detail. Other energy-momentum relations or other different values of $Q_c$ can be analyzed analogously.

Repulsive interactions, $g>0$, enhance the tendency to occupy a single-particle state according to Eq.~\eqref{eq:disp_ren}. Therefore, the bare FS that is characterized by $\xi_{\ve k^{(\nu)}_l} < 0$ remains filled in the presence of interactions. In addition, also the larger circle $\xi_{\ve k^{(\nu)}_l} -g < 0$ contains occupied momentum states provided that $\xi_{\ve k^{(\nu)}_{l\pm 1}} -g < 0$, too. Since $Q_c > k_F$ only one of the two states can satisfy this condition. The set of momenta that is filled due the interactions only is thus given by the intersection of the circle $\xi_{\ve k^{(\nu)}_{l\pm 1}} = g $ with its image under shifts by $\pm \ve Q_c$. This is indicated in Fig.~\ref{fig:ren_FS}. 

In contrast, attractive interactions, $g<0$, hinder the occupation of single-particle states. First of all, only the region $\xi_{\ve k^{(\nu)}_l} -g < 0$, which corresponds to a circle smaller than the bare FS, is guaranteed to be filled (see Fig.~\ref{fig:ren_FS}). To minimize the total energy, momenta $\ve k^{(\nu)}_l$ outside of this circle are only occupied if both $\xi_{\ve k^{(\nu)}_l} < 0 $ and  $\xi_{\ve k^{(\nu)}_l} < \xi_{\ve k^{(\nu)}_{l\pm 1}}$. This is the case in the vertical stripes inside of the bare FS. 

\begin{figure}[ht]
\centering
    \includegraphics[width = 0.8 \columnwidth]{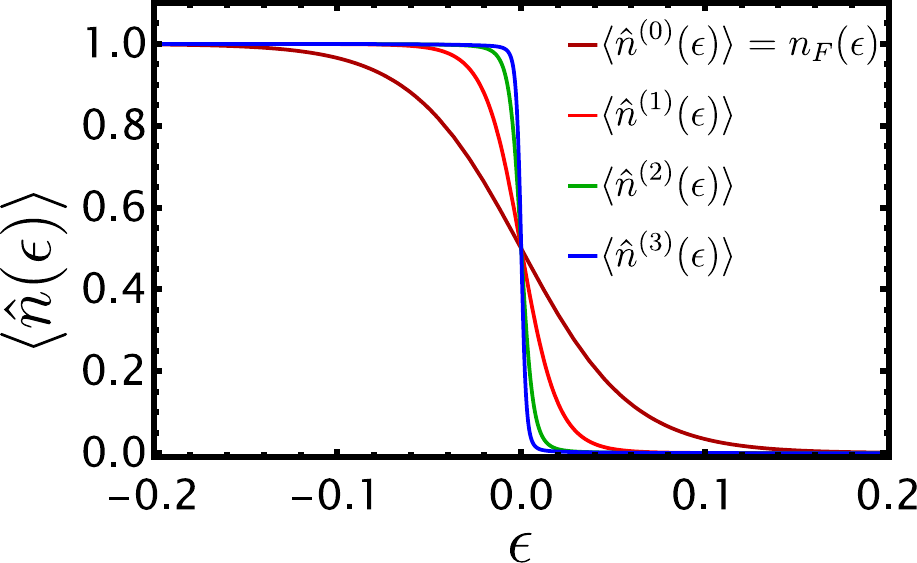}
    \caption{Comparison of the occupation number $\langle \hat n^{(j)}(\epsilon)\rangle$ as function of energy $\epsilon \, \hat = \,  \epsilon^{(\hat n)}_{\ve k^{(\nu)}_l} $ (centered around a common origin) in the case of perfect nesting for different number of self-consistent iterations $j=0,...3$ as introduced in Eq.~\eqref{eq:sc-iteration}. We emphasize that the sharpening of the step around the origin is caused by the interactions and not by the temperature which is fixed to $T=0.06 \eF$. In addtion, we use $g=0.3$ and set $\epsilon_{\ve k^{(\nu)}_{l+1}} =0.8 \epsilon$. The exact value of the ratio is inconsequential apart from a small rescaling of the curves. The contribution from $\ve k^{(\nu)}_{l-1}$ is neglected.}
    \label{fig:avgn}
\end{figure}

Furthermore, Fig.~\ref{fig:ren_FS} allows to make an interesting observation: Some archs of the reshaped FS are perfectly nested with respect to $\pm \ve Q_c$. In particular, these archs show strongly reduced thermal broadening as compared to the parts of the FS without nesting. This effect can be understood from the mean-field equation~\eqref{eq:occ_ren}. In our example, any momentum $\ve k^{(\nu)}_l$ close to the nested archs have only one relevant neigbor that is also close to a nested arch, say $\ve k^{(\nu)}_{l+1}$. Self-consistent iteration of the mean-field equation leads to
\begin{align}\label{eq:sc-iteration}
\begin{split}
    \langle \hat n^{(0)}_{\ve k^{(\nu)}_l} \rangle & = n_F(\xi_{\ve k^{(n)}_l})\\
     \langle \hat n^{(1)}_{\ve k^{(\nu)}_l} \rangle & = n_F(\xi_{\ve k^{(n)}_l}-g \, n_F(\xi_{\ve k^{(\nu)}_{l+1}})) \\
      \langle \hat n^{(2)}_{\ve k^{(\nu)}_l} \rangle & = n_F(\xi_{\ve k^{(n)}_l}-g \, n_F(\xi_{\ve k^{(\nu)}_{l+1}}-g\,  n_F(\xi_{\ve k^{(\nu)}_l})))\, ,\\
     & \!\! \quad \vdots 
\end{split}
\end{align}
where the superindex states the number of insertions. If both momenta are close to the FS the inner Fermi-Dirac distributions vary quickly, which sharpens the behavior of the interacting momentum occupation numbers across the renormalized FS. Effectively this resembles the reduction of the temperature as is exampified in Fig.~\ref{fig:avgn}. Again this effect arises from the special momentum-selective interaction. In a standard FL, like in metals, the chemical potential is shifted to lowest order by the \emph{momentum- integrated} total density, which leads to a much smoother behavior as function of temperature~\footnote{We also note that diagrammatically the shift in the standard case arises from the Hartree diagram. In the cavity scenario the exchange channel corresponds rather to the Fock diagram.}.  

\section{Properties of the symmetry-broken phases of the bare Fermi surface}\label{sec:MF_Ana_inst}
\subsection{The Cooper pairing instability}
We begin by considering only the Cooper channel, that is we set all $\Delta^{(\pm)}_{\ve k^{(\nu)}_l}$ and $X^{(+)}_{\ve k^{(\nu)}_l}, Y^{(+)}_{\ve k^{(\nu)}_l}$ to zero and also neglect the momentum-occupation terms of Eq.~\eqref{eq:Bogoliubov}. In other words, we take only the first mean-field equation in~\eqref{eq:MF_equations} wtih $r=0$ into account. 

Using $\xi_{\ve k^{(iv)}_{-l}} = \xi_{\ve k^{(i)}_{l}}$ and $\xi_{\ve k^{(iii)}_{-l}} = \xi_{\ve k^{(ii)}_{l}}$ the eigenvalues of $\hat H_\mf$ can be obtained in closed form. They are two-fold degenerate (for each sign) and read 
\begin{align}\label{eq:Bog_En_Cooper}
\begin{split}
    \epsilon^{(0)}_{\ve k^{(\nu)}_l} = \pm \sqrt{\xi_{\ve k^{(\nu)}_l}^2 + \left|\Delta^{(0)}_{\ve k^{(\nu)}_{l+1}} ´+ \Delta^{(0)}_{\ve k^{(\nu)}_{l-1}}\right|^2 }\, ,
\end{split}
\end{align}
for $\nu = i,iii$. To eventually solve the mean-field equations in practice, one has to truncate the momentum grids at a value $l_{\max} \geq 0$. The previous formula also applies to $\epsilon^{(0)}_{\ve k^{(i)}_{l_{\max}}}$ and $\epsilon^{(0)}_{\ve k^{(i)}_{-l_{\max}-1}}$ by setting $\Delta^{(0)}_{\ve k^{(i)}_{l_{\max}+1}}$ or, respectively, $\Delta^{(0)}_{\ve k^{(\nu)}_{-l_{\max}-2}}$ to zero, whereas $\epsilon^{(0)}_{\ve k^{(iii)}_{l_{\max}+1}}$ and $\epsilon^{(0)}_{\ve k^{(iii)}_{-l_{\max}}}$ are obtained by setting $\Delta^{(0)}_{\ve k^{(iii)}_{l_{\max}+2}}$ and $\Delta^{(0)}_{\ve k^{(iii)}_{-l_{\max}-1}}$ to zero.

Inserting the two-fold degenerate positive eigenvalues as excitation energies of the Bogoliubov modes into the first mean-field equation of~\eqref{eq:MF_equations} yields
\begin{align}\label{eq:MF_Ana_Cooper}
    \sum_{\ve k^{(\nu)}_{l'}} \left[\delta_{\ve k^{(\nu)}_l, \ve k^{(\nu)}_{l'}} + \frac{g}{2}  \frac{\tanh \left(\frac{1}{2} \beta \epsilon^{(0)}_{\ve k^{(\nu)}_l}\right)}{ \epsilon^{(0)}_{\ve k^{(\nu)}_l} } \underline M_{\ve k^{(\nu)}_l \ve k^{(\nu)}_{l'}}\right]  \Delta^{(0)}_{\ve k^{(\nu)}_{l'}} =0 \, ,
\end{align}
with the ``tight-binding" matrix $\underline M$ from Eq.~\eqref{eq:def_M}. This form of the equation, which is equivalent to $[\mathds 1 + \underline A(\ve \Delta)] \cdot \ve \Delta = 0$, with a matrix $\underline A$ that depends on all gap parameters, is generic and holds irrespective of any truncation of the momentum grid. 

For grids of arbitrary size, the self-consistent equation can only be solved numerically. However, starting out from small grids allows one to obtain analytic insights, which provide an understanding of the critical temperature and the structure of the order parameters in the limit $T \to 0$. To this end, we consider in the following first the case without any shifted copies away from the Fermi surface and then the situation with the four shifted copies closest to the FS. 

\begin{figure}[htb]    
\includegraphics[width=0.99\columnwidth]{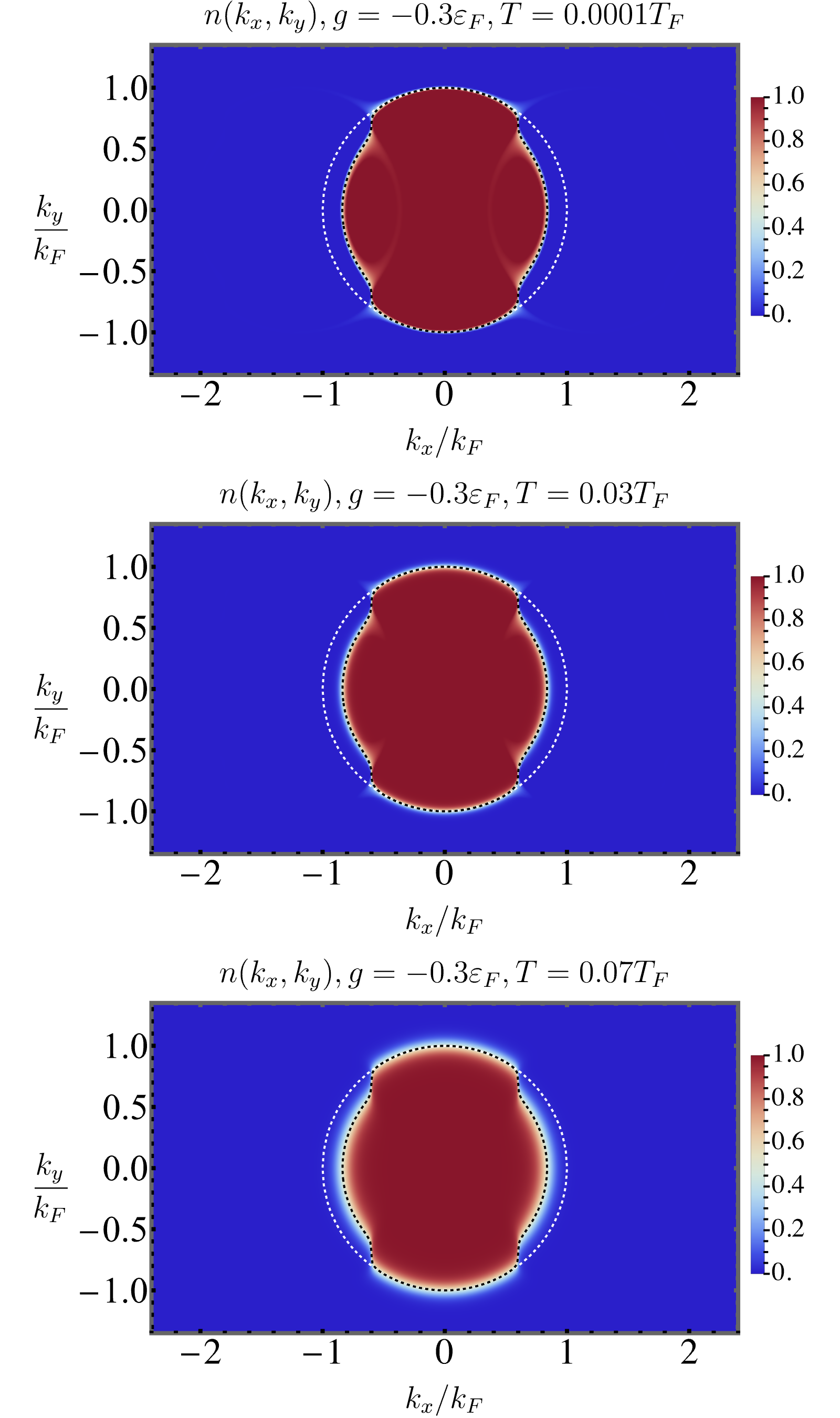}
\caption{Momentum distribution $n_{\ve{k}}$ for different temperatures below $T_c$ for attractive interactions. The black dotted line represents the interacting FS. White dotted line: bare FS.}
\label{fig:n_attractive}
\end{figure}

\begin{figure}[htb]    
\includegraphics[width=0.99\columnwidth]{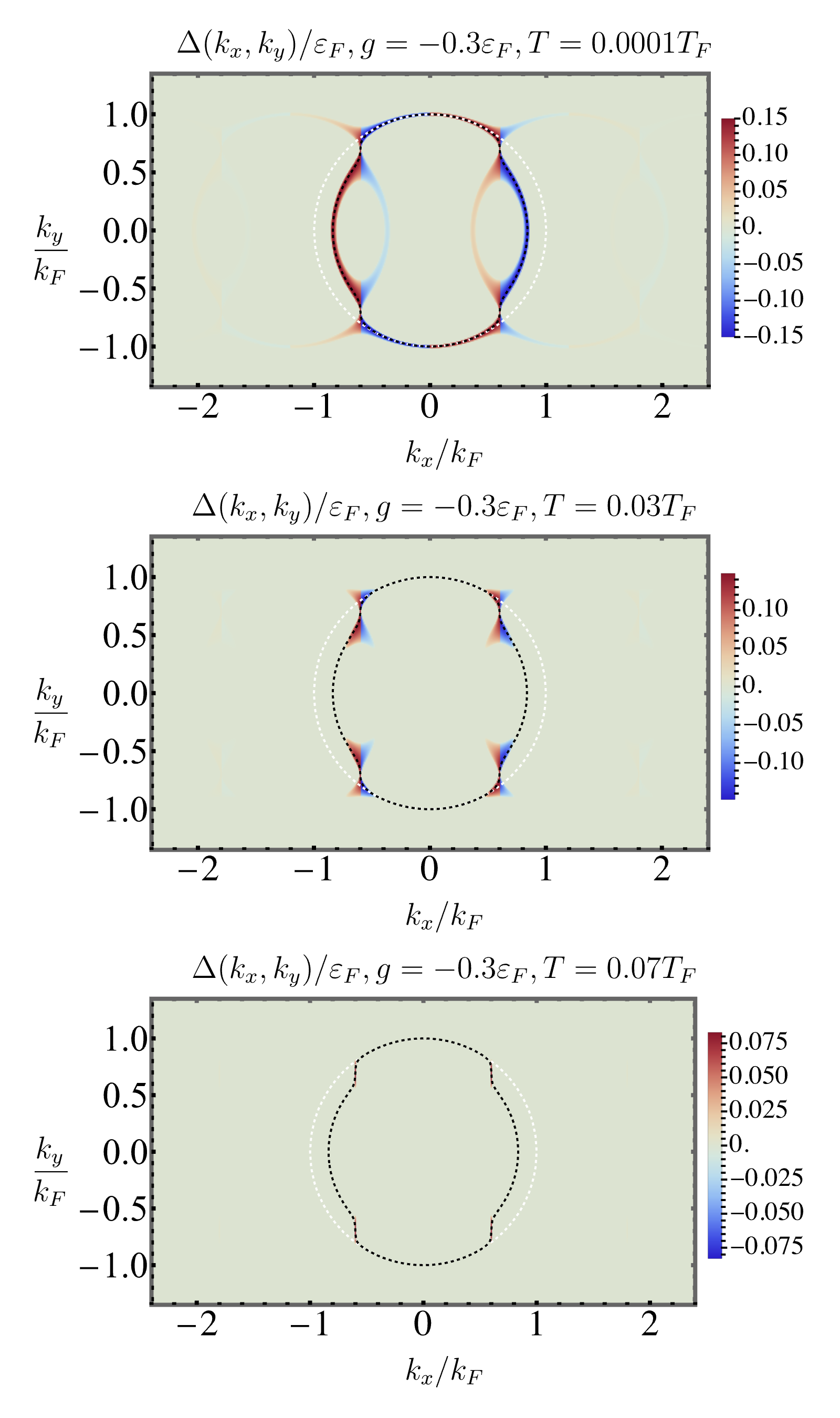}
\caption{Cooper gap $\Delta^{(0)}_{\ve{k}}$ (right) for different temperatures below $T_c$ for attractive interactions.}
\label{fig:delta_attractive}
\end{figure}

Let us focus on the Cooper pairing in momentum groups $(i/iv)$ (red arrows in Fig.~\ref{fig:dkStructure}) corresponding to $\nu = i$ in the above equation. The case $(ii/iii) $ works analogously. Without shifted copies of the hot spots we have the four momenta $\ve k^{(i)}_0 = \khs + \ve \delta \ve^+ k$, $\ve k^{(i)}_{-1} = \khs -\ve Q_c + \ve \delta \ve k^+$, $\ve k^{(iv)}_0 = -\khs - \ve \delta \ve k^+$, and $\ve k^{(iv)}_1 = -\khs + \ve Q_c - \ve \delta \ve k^+$. These are accompanied by two independent order parameters $\Delta^{(0)}_{\ve k^{(i)}_0} = \Delta^{(0)}_{\khs +\ve \delta \ve k^+}$ and $\Delta^{(0)}_{\ve k^{(i)}_{-1}} = \Delta^{(0)}_{\khs -\ve Q_c +\ve \delta \ve k^+}$. The latter have to satisfy the mean-field equations: 
\begin{align}\label{eq:MF_eq_Cooper_HS}
\begin{split}
\Delta^{(0)}_{\khs +\ve \delta \ve k^+} & = - \frac{g}{2} \tanh \left(\frac{1}{2} \beta \epsilon^{(0)}_{\ve k +\ve \delta \ve k^+}\right) \frac{\Delta^{(0)}_{\khs -\ve Q_c +\ve \delta \ve k^+}}{\epsilon^{(0)}_{\khs +\ve \delta \ve k^+}} \\
\Delta^{(0)}_{\khs -\ve Q_c +\ve \delta \ve k^+} & = - \frac{g}{2} \tanh \left(\frac{1}{2} \beta \epsilon^{(0)}_{\ve k -\ve Q_c +\ve \delta \ve k^+}\right) \frac{\Delta^{(0)}_{\khs  +\ve \delta \ve k^+}}{\epsilon^{(0)}_{\khs -\ve Q_c +\ve \delta \ve k^+}}  \, ,
\end{split}
\end{align}
which derive from Eq.~\eqref{eq:MF_Ana_Cooper}. Furthermore, the Bogoliubov energies take only the boundary forms $\epsilon^{(0)}_{\khs + \ve \delta \ve k^+} = \sqrt{\xi_{\khs +\ve \delta \ve k^+}^2 + |\Delta^{(0)}_{\khs - \ve Q_c + \ve \delta \ve k^+}|^2}$ and $\epsilon^{(0)}_{\khs -\ve Q_c + \ve \delta \ve k^+} = \sqrt{\xi_{\khs -\ve Q_c +\ve \delta \ve k^+}^2 + |\Delta^{(0)}_{\khs + \ve \delta \ve k^+}|^2}$ according to the discussion below Eq.~\eqref{eq:Bog_En_Cooper}.
Like in case of the well-known BCS gap equation the above cannot be solved in closed form at arbitrary temperatures but special cases admit an analytic solution. 
The first special case concerns the critical temperature $T_c$. To determine $T_c$ one inserts the two equations~\eqref{eq:MF_eq_Cooper_HS} into each other:
\begin{align}\label{eq:MF_eq_Cooper_HS_fin}
    \frac{\tanh \left( \frac{1}{2} \beta \epsilon^{(0)}_{\khs +\ve \delta \ve k^+} \right)}{\epsilon^{(0)}_{\khs + \ve \delta \ve k^+}} \frac{\tanh\left( \frac{1}{2} \beta \epsilon^{(0)}_{\khs - \ve Q_c + \ve \delta \ve k^+} \right)}{\epsilon^{(0)}_{\khs - \ve Q_c +\ve \delta \ve k^+}} = \frac{4}{g^2} \, .
\end{align}
At $T_c$ this equation admits a solution with vanishing order parameters
\begin{align}\label{eq:MF_eq_Cooper_HS_T_c}
    \left(\frac{g}{2}\right)^2 \frac{\tanh \left( \frac{1}{2} \beta_c \xi_{\khs +\ve \delta \ve k^+} \right)}{\xi_{\khs + \ve \delta \ve k^+}} \frac{\tanh\left( \frac{1}{2} \beta_c \xi_{\khs - \ve Q_c + \ve \delta \ve k^+} \right)}{\xi_{\khs - \ve Q_c +\ve \delta \ve k^+}} = 1 \, .
\end{align}
which, however, seemingly exhibits a momentum dependence. This is resolved by the criterion that $T_c$ is set by the largest temperature for which the system becomes unstable against spontaneous symmetry breaking.
In particular, the left-hand side attains its maximum for the hot spots on the FS, which are reproduced in the limit $\ve \delta \ve k^+ \to 0$ or equivalently $\xi_{\khs +\ve \delta \ve k^+} \to  0$ and $\xi_{\khs - \ve Q_c +\ve \delta \ve k^+} \to  0$. Expanding around vanishing single-particle energies yields
\begin{align}\label{eq:Tc_hs}
    T_c^{\text{hs}} = \frac{|g|}{4} \, ,
\end{align}
where the index hs indicates the fact that only the hot spots have been taken into account. Before including also the shifted copies a few comments are in order:
First of all, we have found, in the simplest approximation, a thermal phase transition for arbitrarily weak coupling that persists in the limit $g \to 0$. This is in contrast to the superradiant transition, which requires a finite attractive interaction strength.
Moreover, the critical temperature depends only on the absolute value of the coupling such that the instability occurs for both attractive and repulsive couplings. This is a consequence of the fact that any two fermions with arbitrary incoming momenta have to undergo an even number of scattering events in order to return to the original incoming momenta. Mathematically, this is reflected in the nearest-neighbor tight-binding structure of the matrix $\underline M$. This holds true even if shifted copies of the hot spots are included, as we show just below. Physically, this means that the instability can be studied for repulsive interactions where the superradiant transition is absent. This is very important for experiments since the collective enhancement of the superradiant transition suppresses the associated quantum critical coupling. Therefore, distinguishing the pairing instability from the superradiant one potentially is hindered for attractive interactions.   
Furthermore, the critical temperature scales linearly with the interaction strength. This is due to the missing momentum integration over the FS as compared to standard BCS theory where it leads to the well-known non-pertubative scaling $T_c \sim \exp(-1/g)$. A similar result has already been found~\cite{Rademaker2016,Gao2020,Chakraborty2021} in a situation with $\ve Q_c=0$. However, it is important to note that the limit $\ve Q_c \to 0$ differs from setting $\ve Q_c = 0$. In the latter case the single-particle momenta do not change in any interaction event such that $\underline M$ becomes diagonal, which results in an instability only for attractive interactions like in the standard BCS theory. \\
Returning to Eq.~\eqref{eq:MF_eq_Cooper_HS_T_c} reveals that also other momenta at finite momentum deviations away from the hot spots become unstable, yet at $\ve \delta\ve k$-dependent temperatures below $T_c$. Fig.~3 in the main text shows that the instability indeed spreads out from the hot spots 
with decreasing $T$.\\
The other special case considers the limit $T \to 0$ of Eq.~\eqref{eq:MF_eq_Cooper_HS}. Inserting the Bogoliubov energies yields
\begin{align}\label{eq:MF_eq_Cooper_HS_T0}
\begin{split}
\Delta^{(0)}_{\khs +\ve \delta \ve k^+} & = - \frac{g}{2} \frac{\Delta^{(0)}_{\khs -\ve Q_c +\ve \delta \ve k^+}}{\sqrt{\xi_{\khs +\ve \delta \ve k^+}^2 + |\Delta^{(0)}_{\khs - \ve Q_c + \ve \delta \ve k^+}|^2}} \\
\Delta^{(0)}_{\khs -\ve Q_c +\ve \delta \ve k^+} & = - \frac{g}{2} \frac{\Delta^{(0)}_{\khs  +\ve \delta \ve k^+}}{\sqrt{\xi_{\khs -\ve Q_c +\ve \delta \ve k^+}^2 + |\Delta^{(0)}_{\khs + \ve \delta \ve k^+}|^2}}  \, ,
\end{split}
\end{align}
because the Bogoliubov energies are positive by definition. 
Most interestingly, for any momentum $\khs + \ve \delta \ve k^+$ on the FS, i.e. $\xi_{\khs + \ve \delta \ve k^+}=0$ we one finds
\begin{align}\label{eq:D0_T0_HS}
    \left|\Delta^{(0),\text{hs}}_{\khs+\ve \delta \ve k^+}\right|_{\xi_{\khs +\ve \delta \ve k^+}=0} = \frac{|g|}{2}\, ,
\end{align}
and the same for $\khs - \ve Q_c + \ve \delta \ve k^+$. We also note that the same result holds also for the reshaped Fermion surface since the evaluation of Eq.~\eqref{eq:D0_T0_HS} is not altered by introducing a renormalized the dispersion $\tilde{\xi}_{\ve k}$. This is confirmed in Figs.~\ref{fig:n_attractive}, \ref{fig:delta_attractive} and Fig.~3 in the main text, which present the numerical solution of the mean-field equations. 

We also point out that Eq.~\eqref{eq:MF_eq_Cooper_HS_T0} implies an algebraic decrease of the gap parameters for increasing momentum deviations $\ve \delta \ve k^+$ away from the Fermi surface. 
The previous analysis has been focused on the hot spots and their environment in momentum space only. Let us now consider how the picture is changed by the presence of shifted copies. Physically, one expects that the latter give rise to corrections since they are away from the FS and that their effects can be described in a perturbative series in the experimentally relevant regime $|g|/\eF \ll 1$. To show this, we include now -- in addition to the hot spots -- the four shifted copies closest to the FS where we focus again on Cooper pairing in the momentum groups (i/iv) (red arrows in Fig.~\ref{fig:dkStructure}). This means we use the eight momenta
$\ve k_1 \equiv \ve k^{(i)}_{-2} = \khs -2\ve Q_c + \ve \delta \ve k^+$, $\ve k_2 \equiv \ve k^{(i)}_{-1} = \khs -\ve Q_c + \ve \delta \ve k^+$, $\ve k_3 \equiv \ve k^{(i)}_{0} = \khs + \ve \delta \ve k^+$, $\ve k_4 = \ve k^{(i)}_1 = \khs + \ve Q_c + \ve \delta \ve k^+$, and $\ve k_5 = \ve k^{(iv)}_{-1} = -\khs -\ve Q_c - \ve \delta \ve k^+$, $\ve k_6 = \ve k^{(iv)}_0 = -\khs - \ve \delta \ve k^+$, $\ve k_7 = \ve k^{(iv)}_{1} = -\khs +\ve Q_c - \ve \delta \ve k^+$, and $\ve k_8 = \ve k^{(iv)}_{2} = -\khs +2\ve Q_c - \ve \delta \ve k^+$, that correspond to four gap parameters, which are arranged in the vector
\begin{align}\label{eq:Cooper_shorthand}
    \ve{\Delta}^{(0)}_{\ve{\delta k}^+}= 
    \left(\begin{array}{c}
    \Delta^{(0)}_{1} \\
    \Delta^{(0)}_{2} \\
    \Delta^{(0)}_{3} \\
    \Delta^{(0)}_{4}
    \end{array}\right)=
    \left(\begin{array}{c}
    \Delta^{(0)}_{\khs-2\ve{Q}_c+\ve{\delta k}^+} \\
    \Delta^{(0)}_{\khs-\ve{Q}_c+\ve{\delta k}^+} \\
    \Delta^{(0)}_{\khs+\ve{\delta k}^+} \\
    \Delta^{(0)}_{\khs+\ve{Q}_c+\ve{\delta k}^+}
    \end{array}\right) \,   .
\end{align}
Within the current truncation we have four momenta from group $(i)$ corresponding to four different Bogoliubov energies: $\epsilon^{(0)}_1 = \epsilon^{(0)}_{\khs - 2 \ve Q_c + \ve \delta \ve k^+}$, 
$ \epsilon^{(0)}_2 = \epsilon^{(0)}_{\khs -  \ve Q_c + \ve \delta \ve k^+}$, 
$ \epsilon^{(0)}_3 = \epsilon^{(0)}_{\khs  + \ve \delta \ve k^+}$, and $\epsilon^{(0)}_4 = \epsilon^{(0)}_{\khs +  \ve Q_c + \ve \delta \ve k^+}$. As mentioned in connection with Eq.~\eqref{eq:Bog_En_Cooper}, these are doubly-degenerate to take into account the momenta for group $(iv)$. These eigenvalues follow again from Eq.~\eqref{eq:Bog_En_Cooper}, which applies now directly to the hot spots $\epsilon^{(0)}_2, \epsilon^{(0)}_3$ while the boundary forms are now attained at the shifted copies $\epsilon^{(0)}_1, \epsilon^{(0)}_4$. 
Recall the structure of the general gap equation~\eqref{eq:MF_Ana_Cooper}, which turns into the matrix equation
\begin{equation}\label{eq:matrix_Delta}
     \left(\mathds 1_{4 \times 4} + \underline{A}_{4 \times 4} \right) \cdot \ve{\Delta}^{(0)}_{\ve{\delta k}^+} = 0
\end{equation}
when the hot spots and their closest shifted copies are taken into account. Most importantly, the matrix $\underline{A}_{4 \times 4}$ is given by
\begin{align}\label{eq:Cooper_mat_example}
 \underline{A}_{4 \times 4} & = 
\frac{g}{2}\left(
\begin{array}{c c c c}
    0 & \Xi(\epsilon^{(0)}_1) & 0 & 0  \\
     \Xi(\epsilon^{(0)}_2) & 0 & \Xi(\epsilon^{(0)}_2) & 0 \\
     0 & \Xi(\epsilon^{(0)}_3) & 0 & \Xi(\epsilon^{(0)}_3) \\
     0 & 0 & \Xi(\epsilon^{(0)}_4) & 0
\end{array}
\right)\, ,
\end{align}
where we have introduced the auxiliary function $\Xi(\epsilon) = \tanh(\beta \epsilon/2)/\epsilon$.
The critical temperature $T_c$ can now be found by imposing the condition
\begin{align}
\label{eq:det_example}
    \begin{split}
     & \det \left(
       \mathds{1}_{4 \times 4} +
            \underline{A}_{4 \times 4} 
        \right) =0\, ,
         \end{split}
\end{align}
The full expression for the determinant in Eq.~\eqref{eq:det_example} is quite cumbersome but can be simplified in the weak-coupling regime $|g|/\eF \ll 1$, which is of relevance for the experiments. Since we are interested in temperatures $T \lesssim |g|$ [cf. Eq.~\eqref{eq:Tc_hs}] we can set $\tanh{(\beta \epsilon^{(0)}_{1,4}/2)} \to 1$. As a result, we obtain
\begin{align}\label{eq:MF_Cooper_det_fin}
\begin{split}
    \Xi(\epsilon^{(0)}_{\khs - \ve Q_c + \ve \delta \ve k^+}) \Xi(\epsilon^{(0)}_{\khs +\ve \delta \ve k^+}) &+
    \frac{\Xi(\epsilon^{(0)}_{\khs -\ve Q_c +\ve \delta \ve k^+})}{\epsilon^{(0)}_{\khs -2 \ve Q_c+\ve \delta \ve k^+}} \\
    & + \frac{\Xi(\epsilon^{(0)}_{\khs +\ve \delta \ve k^+})}{\epsilon^{(0)}_{\khs + \ve Q_c+\ve \delta \ve k^+}}
    \! = \! \frac{4}{g^2} ,
    \end{split}
\end{align}
after neglecting further terms on the left, which are suppressed by a factor $g^2/\eF^2$. Comparing the last result with the corresponding Eq.~\eqref{eq:MF_eq_Cooper_HS} reveals that the second and third contribution on the left-hand side represent the corrections caused by the presence of the shifted copies. Solving for the critical temperature like in the case without shifted copies, we find that the instability occurs again first at the hot spots ($\ve \delta \ve k^+ \to 0$), where we obtain
\begin{align}\label{eq:Cooper_HS_Tc}
    T_c^{(0)} = \frac{|g|}{4} + \frac{g^2}{8 \xi_{\khs + \ve Q_c}} + \mathcal{O}(|g|^3/\eF^2)\, .
\end{align}
    We conclude that the coherent scattering to the shifted copies gives rise to an \emph{enhancement} of the critical temperature both for attractive and repulsive interactions. The correction term $\mathcal{O}(|g|^3/\eF^2)$ contains both the neglected contributions within the present momentum grid truncation but also the effects of the other shifted copies. These are suppressed since the interaction term has to act more repeatedly to scatter a Fermion from the shifted copies further away from the FS to the hot spots~\footnote{Indeed, the calculation with the next four shifted copies gives $T_c^{(0)} = \frac{|g|}{4} + \frac{g^2}{8 \xi_{\khs + \ve Q_c}}+ \frac{g^4}{32 \xi_{\khs +2 \ve Q_c} \xi^2_{\khs + \ve Q_c}}$.}.

We can also determine how the presence of the shifted copies affects the the gap parameter on the FS as compared to Eq.~\eqref{eq:D0_T0_HS} at $T=0$. The second line of Eq.~\eqref{eq:matrix_Delta} reads in the ground state
\begin{align}
    \Delta_2 = - \frac{g}{2} \frac{\Delta_1}{\sqrt{\xi_{\ve k_2}^2  + |\Delta_1 + \Delta_3|^2}} - \frac{g}{2} \frac{\Delta_3}{\sqrt{\xi_{\ve k_2}^2  + |\Delta_1 +\Delta_3|^2}} \, .
\end{align}
Assume that $\ve k_2$ is on the FS, then we have
\begin{align}\label{eq:Delta0_HS_T0}
    \left|\Delta_2\right|_{\xi_{\ve k_2}=0} = \frac{|g|}{2} \, ,
\end{align}
meaning that the closest shifted copies do not at all affect the value of the gap on the Fermi surface as compared to the pure hot spot theory. This result, which is confirmed by the numerical evaluation shown in Figs. 3,4 in the main text, furthermore implies that there is no universal relation between $T_c$ and the value of the gap parameters in the limit $T \to 0$. 

Finally, we note that we do not only find an algebraic variation of the order parameters around the hot spots as function of the momentum deviation $\ve \delta \ve k^+$ as explained below Eq.~\eqref{eq:D0_T0_HS}: In addition, the gap equation in the form~\eqref{eq:MF_Ana_Cooper} establishes an algebraically decreasing recurrence of the gap parameters with respect to the shifted copies. For instance, we have at $T=0$ and $l>0$
\begin{align}\label{eq:Cooper_recurrence}
    |\Delta^{(0)}_{\khs + (l+1)\ve Q_c + \ve \delta \ve k^+}| \approx  \frac{|g|}{2} \frac{|\Delta^{(0)}_{\khs+l \ve Q_c + \ve \delta \ve k^+}|}{\xi_{\khs + (l+1) \ve Q_c + \ve \delta \ve k^+}}\, ,
\end{align}
up to corrections further suppressed in $g$.
These arise from the approximations used: $\epsilon^{(0)}_{\khs + l \ve Q_c + \ve \delta \ve k^+} \approx \xi_{\khs + \ve Q_c + \ve \delta \ve k^+}$ and neglecting the contribution from $\Delta^{(0)}_{\pm \khs +(l+2)\ve Q_c + \ve \delta \ve k^+}$ on the right-hand side. Thermal corrections are exponentially small. The result for negative $l$ is obtained by setting $l+1 \to -(l+1)$ and $l \to - l$. The results for $|\Delta^{(0)}_{\ve k^{(iii)}_l}|=|\Delta^{(0)}_{-\khs +l \ve Q_c + \ve \delta \ve k^-}|$ can be obtained in an analogous fashion.

\subsubsection{The PDW instability}
After the detailed study of the Cooper instability we turn now to the PDW instability. In the absence of the Cooper instabibility, the exciton condensation and without the effects of the FS reshaping the eigenvalues of $\hat H_\mf$ read
\begin{align}\label{eq:Bog_En_PDW}
    \epsilon^{(r)}_{l,(\nu)} = \frac{1}{2} \sqrt{\xi_{\ve k^{(\nu)}_l}^2 + \left|\Delta^{(r)}_{\ve k^{(\nu)}_{l+1}} ´+ \Delta^{(r)}_{\ve k^{(\nu)}_{l-1}}\right|^2 }\, ,
\end{align}
for the combinations $(r,\nu) = (+,i)$ or $(r,\nu) = (-,iii)$ according to the definitions of the PDW order-parameters~\eqref{eq:defDeltaPDW}.
Furthermore, each of the eigenvalues is two-fold degenerate
and we have used $\xi_{\ve k^{(ii)}_{-l-1}} = \xi_{\ve k^{(i)}_{l}}$ and $\xi_{\ve k^{(iv)}_{-l+1}} = \xi_{\ve k^{(iii)}_{l}}$ to simplify the expressions.
We note that $\epsilon^{(+)}_{\ve k^{(i)}_l}$ ($\epsilon^{(-)}_{\ve k^{(iii)}_l}$) from Eq.~\eqref{eq:Bog_En_PDW} maps to the excitation energy $\epsilon^{(0)}_{\ve k^{(i)}_l}$ ($\epsilon^{(-)}_{\ve k^{(iii)}_l}$)  in the Cooper case from Eq.~\eqref{eq:Bog_En_Cooper} via $\Delta^{(+)}_{\ve k^{(i)}_l} \to \Delta^{(0)}_{\ve k^{(i)}_l}$ (
$\Delta^{(-)}_{\ve k^{(iii)}_l} \to \Delta^{(0)}_{\ve k^{(ii)}_l}$). Moreover, inspection of the definitions~\eqref{eq:defDelta0gen} for the Cooper case and~\eqref{eq:defDeltaPDW} for the PDW case, respectively, shows that the single-particle content of $\Delta^{(+)}_{\ve k^{(i)}_l}$ and $\Delta^{(0)}_{\ve k^{(i)}_l}$ ($\Delta^{(-)}_{\ve k^{(iii)}_l}$ and $\Delta^{(0)}_{\ve k^{(iii)}_l}$) is energetically identically since $\xi_{\ve k^{(ii)}_{-l-1}} = \xi_{\ve k^{(iv)}_{-l}}$ ($\xi_{\ve k^{(iv)}_{-l+1}} = \xi_{\ve k^{(ii)}_{-l}}$) (cf. Eq.\eqref{eq:defMomGroups}). 
 After inserting the Bogoliubov energies into the general mean-field equation~\eqref{eq:MF_equations} we obtain
 \begin{align}\label{eq:MF_Ana_PDW}
    \Delta^{(r)}_{\ve k^{(\nu)}_l} = - \frac{g}{2} \sum_{\ve k^{(\nu)}_{l'}} \frac{\tanh \left(\frac{1}{2} \beta \epsilon^{(r)}_{\ve k^{(\nu)}_l}\right)}{ \epsilon^{(r)}_{\ve k^{(\nu)}_l} } \underline M_{\ve k^{(\nu)}_l \ve k^{(\nu)}_{l'}} \Delta^{(r)}_{\ve k^{(\nu)}_{l'}} \, ,
\end{align}
for the combinations $[\nu,r] = [i,+],[iii,-]$. Again we have a structural equivalence with the mean-field equation~\eqref{eq:MF_Ana_Cooper}. Furthermore, we have just argued that the Bogoliubov energies and the content of single-particle energies in the pairing order parameters are the same in the in the PDW and Cooper case. Therefore, the MF equation for $\Delta^{(+)}_{\ve k^{(i)}_l}$ ($\Delta^{(-)}_{\ve k^{(iii)}_l}$) yields the same results as the MF equation for $\Delta^{(0)}_{\ve k^{(i)}_l}$ ($\Delta^{(0)}_{\ve k^{(iii)}_l}$) at all temperatures and coupling strengths.
As a consequence, we can immediately transfer the findings from the analysis of the Cooper instability to the PDW case. In particular, we deduce the critical temperature
\begin{align}\label{eq:PDW_HS_Tc}
    T_c^{(\pm)} = \frac{|g|}{4} + \frac{g^2}{8 \xi_{\khs + \ve Q_c}} + \mathcal{O}(|g|^3/\eF^2)\, ,
\end{align}
from Eq.~\eqref{eq:Cooper_HS_Tc} with identical higher-order corrections. 
By the equivalence of the single-particle energies contained in the $(+)$ and $(-)$ pairs the instability occurs at the same critical temperature. 
Similarly, the gap parameters on the Fermi surface for $T \to 0$ read as
\begin{align}\label{eq:DeltaPDW_HS_T0}
\begin{split}
    |\Delta^{(+)}_{\ve k}| = \frac{|g|}{2} \, ,
\end{split}
\end{align}
and analogously for $|\Delta^{(-)}_{\ve k}|$,
which follows from Eq.~\eqref{eq:Delta0_HS_T0}. Finally, we obtain the algebraic recurrence at the shifted copies from Eq.~\eqref{eq:Cooper_recurrence} for the PDW order-parameters, too:
\begin{align}\label{eq:PDW_recurrence}
    |\Delta^{(+)}_{\khs +(l+1)\ve Q_c + \ve \delta \ve k^+}| \approx - \frac{g}{2} \frac{|\Delta^{(+)}_{\khs + \ve \delta \ve k^+}|}{\xi_{\pm \khs + (l+1) \ve Q_c + \ve \delta \ve k^+}}\, ,
\end{align}
for $l>0$ and analogously $l<0$ and for $|\Delta^{(-)}_{-\khs-(l+1)\ve Q_c - \delta \ve k^-}|$.

We emphasize that the degeneracy between PDW and Cooper instabilities is not caused by considering the two scenarios separately. In Sec.~\ref{sec:MF_Ana_symm} it is shown that $\hat{H}_\mf$ possesses a symmetry corresponding to rotating $\ve k^{(ii)}_l$ to $\ve k^{(iv)}$ that is related to the invariance of the system under $ k_{x,y} \to - k_{x,y}$. This symmetry allows to continuously transform the Cooper order-paramters into PDW order-parameters and vice versa. Consequently, the degeneracy is inherent to the system and is not lifted by taking the coupled mean-field equations into account.

\subsubsection{Exciton condensation}
Next, we repeat the analysis for the XCON in the absence of any pairing and without the reshaped FS. Like in the pairing scenarios we can relate the eigenvalues to the momenta. In the following, we will focus on the determination of $T^{(X)}_c$ for the XCON of type $X^{(+)}_{\ve k_l^{(\nu)}}$ and summarize the very similar results for $Y^{(+)}_{\ve k_l^{(\nu)}}$ below. First, we find the four different expressions
\begin{align}\label{eq:Bogoliubov_XCON}
\begin{split}
    \epsilon^{(X,1)}_{\ve k^{(\nu)}_l} = & \frac{\xi_{\ve k^{(\nu)}_l} + \xi_{\ve k^{(\nu')}_{l+1}}}{2} \\ 
    & + \sqrt{\left(\frac{\xi_{\ve k^{(\nu)}_l} - \xi_{\ve k^{(\nu')}_{l+1}}}{2}\right)^2 + \left|X^{(+)}_{\ve k^{(\nu)}_{l+1}} +X^{(+)}_{\ve k^{(\nu)}_{l-1}} \right|^2} \\
    \epsilon^{(X,2)}_{\ve k^{(\nu)}_l} = &  \frac{\xi_{\ve k^{(\nu)}_l} + \xi_{\ve k^{(\nu')}_{l+1}}}{2} \\ 
    & - \sqrt{\left(\frac{\xi_{\ve k^{(\nu)}_l} - \xi_{\ve k^{(\nu')}_{l+1}}}{2}\right)^2 + \left|X^{(+)}_{\ve k^{(\nu)}_{l+1}} +X^{(+)}_{\ve k^{(\nu)}_{l-1}} \right|^2} \\
    \epsilon^{(X,3)}_{\ve k^{(\nu)}_l} & = -\epsilon^{(X,2)}_{\ve k^{(\nu)}_l}\\
    \epsilon^{(X,4)}_{\ve k^{(\nu)}_l} & = -\epsilon^{(X,1)}_{\ve k^{(\nu)}_l} \, ,
    \end{split}
\end{align}
for the pairs $(\nu,\nu') = (i,iv), (ii,iii)$. In order to determine the concrete forms for the mean-field equation, one has to identify the positive eigenvalues. 
Let us focus on the most relevant scenario with a set of momenta including the vicinity of the hot spots, that is $\ve \delta \ve k^+, \ve \delta \ve k^- \to 0$ and $|X^{(+)}_{\ve k^{(\nu)}} |\ll \eF$, which is certainly satisfied for temperatures close to $T_c^{(X)}$.
In this situation, if $\ve k^{(\nu)}_l$ and $\ve k^{(\nu')}_{l+1}$ are associated with 
a shifted copy away from the FS, the positive eigenvalues are given by $\epsilon^{(X,1)}_{\ve k^{(\nu)}_l}$ and $\epsilon^{(X,2)}_{\ve k^{(\nu)}_l}>0$ since both $\xi_{\ve k^{(\nu)}_l}>0$ and $\xi_{\ve k^{(\nu')}_{l+1}}>0$, which makes the square root subleading.
Regarding the vicinity of the hot spots, 
the most relevant situation occurs for particle-hole pairs.  For instance, 
if $\ve k^{(i)}_0$ is inside (outside) and $\ve k^{(iv)}_{1}$ is outside (inside) of the FS, the result is in both cases $\epsilon^{(X,1)}_{\ve k^{(\nu)}_l}>0$ and $\epsilon^{(X,3)}_{\ve k^{(\nu)}_l}>0$ since now the square roots dominate. 
Inserting these Bogoliubov energies correspondingly into the general mean-field equation~\eqref{eq:MF_equations} yields
\begin{align}\label{eq:MF_XCON_det}
     \sum_{\ve k^{(\nu)}_{l'}} \left(\delta_{\ve k^{(\nu)}_l,\ve k^{(\nu)}_{l'}} - \frac{g}{2} \chi \left(\ve k^{(\nu)}_l\right) \underline M_{\ve k^{(\nu)}_l \ve k^{(\nu)}_{l'}}\right) X^{(+)}_{\ve k^{(\nu)}_{l'}} =0 \,.
\end{align}
where we have arranged the equation in the nontrivial phase in analogy to the pairing case given in Eq.~\eqref{eq:MF_Ana_Cooper}. In addition, we have defined the auxiliary function:
\begin{align}
\chi\left(\ve k^{(\nu)}_l\right) = \dfrac{\tanh(\frac{1}{2}\beta \epsilon^{(X,1)}_{\ve k^{(\nu)}_k}) - \tanh(\frac{1}{2}\beta \epsilon^{(X,2)}_{\ve k^{(\nu)}_k})}{\epsilon^{(X,1)}_{\ve k^{(\nu)}_l} - \epsilon^{(X,2)}_{\ve k^{(\nu)}_l}}\, ,
\end{align}
which applies to both the hot spot and the shifted copies by virtue of the relation $\epsilon^{(X,2)}_{\ve k^{(\nu)}_l} = -\epsilon^{(X,3)}_{\ve k^{(\nu)}_l}$.
Let us again consider the example of groups $(i)/(iv)$ and 
include the hotpspots and the closest shifted copies to the FS like in the case of Cooper pairing as described above Eq.~\eqref{eq:Cooper_recurrence}:
$\ve k_1 \equiv \ve k^{(i)}_{-2} = \khs -2\ve Q_c + \ve \delta \ve k^+$, $\ve k_2 \equiv \ve k^{(i)}_{-1} = \khs -\ve Q_c + \ve \delta \ve k^+$, $\ve k_3 \equiv \ve k^{(i)}_{0} = \khs + \ve \delta \ve k^+$, $\ve k_4 = \ve k^{(i)}_1 = \khs + \ve Q_c + \ve \delta \ve k^+$, and $\ve k_5 = \ve k^{(iv)}_{-1} = -\khs -\ve Q_c - \ve \delta \ve k^+$, $\ve k_6 = \ve k^{(iv)}_0 = -\khs - \ve \delta \ve k^+$, $\ve k_7 = \ve k^{(iv)}_{1} = -\khs +\ve Q_c - \ve \delta \ve k^+$, and $\ve k_8 = \ve k^{(iv)}_{2} = -\khs +2\ve Q_c - \ve \delta \ve k^+$,
To determine $T_c^{(X)}$ it is sufficient to take only the determinant of the matrix in the general equation~\eqref{eq:MF_XCON_det} which becomes in case of our truncation
\begin{align}
\label{eq:det_example_XCON}
    \begin{split}
     & \det \left(
       \mathds{1}_{4 \times 4} -
            \underline{B}_{4 \times 4} 
        \right) =0\, ,
         \end{split}
\end{align}
in analogy to Eq.~\eqref{eq:det_example} from the pairing case, with the matrix
\begin{align}\label{eq:XCON_mat_example}
 \underline{B}_{4 \times 4} & = 
\frac{g}{2}\left(
\begin{array}{c c c c}
    0 & \chi(\ve k_1) & 0 & 0  \\
     \chi(\ve k_2) & 0 & \chi(\ve k_2) & 0 \\
     0 & \chi(\ve k_3) & 0 & \chi(\ve k_3) \\
     0 & 0 & \chi(\ve k_4) & 0
\end{array}
\right)\, .
\end{align}
This is equivalent to 
\begin{align}\label{eq:TX_inter1}
\begin{split}
    \, \chi(\ve k_2) \chi(\ve k_3)
    & + \chi(\ve k_1) \chi(\ve k_2) + \chi(\ve k_3) \chi(\ve k_4) = \frac{4}{g^2}   \, ,
\end{split}
\end{align}
 The first term on the left-hand side contains the contribution from the hot spots only while the other two give the leading corrections from the shifted copies. A term of the form $g^2 \chi(\ve k_1) \chi(\ve k_2) \chi(\ve k_3) \chi(\ve k_4) \sim g^2 \beta^4 e^{-2 \beta \xi_{\khs + \ve Q_c}}$ scales like $e^{-2 \xi_{\khs + \ve Q_c}/|g|}/g^2\ll \beta^2$ in the weak coupling limit $|g| \ll \eF$ and therefore has been neglected. Such terms arise as the square of the asymptotic behavior   
\begin{align}
    \chi \left(\ve k^{(\nu)}_l\right) \simeq
    2\beta e^{-\beta/2 \left(\xi_{\ve k^{(i)}_l} + \xi_{\ve k^{(iv)}_{l+1}} \right) }\, ,
\end{align}
for momenta not in the vicinity of the hot spots, that is $l\neq -1,0$.
Moreover, we consider as usual the limit $\ve \delta \ve k^+ \to 0$ to get them most relevant scenario at the hot spots and set all $X^{(+)}_{\ve k^{(i)}_l} \to 0$ to identify the critical temperature. Using then the result right at the hot spots
\begin{align}
    \chi \left(\ve k^{(\nu)}_l\right) \simeq  \frac{1}{2 T_c^{(X)}} \, ,
\end{align}
we find, as a result from~\eqref{eq:TX_inter1}, the simplified equation for $T^{(X)}_c$:
\begin{align}
    \frac{4}{g^2} = \frac{1}{4 \left(T^{(X)}_c\right)^2} +2\frac{e^{-\xi_{\khs +\ve Q_c}/T^{(X)}_c}}{ \left(T^{(X)}_c\right)^2} \, .
\end{align}
Note that for $ \ve \delta \ve k^+ \to 0$ one has $\xi_{\ve k_1},\xi_{\ve k_4},\xi_{\ve k_5},\xi_{\ve k_8} = \xi_{\khs + \ve Q_c}$, which was used to simplify the expression.
Inverting the previous expression yields the final result for the critical temperature
\begin{align}\label{eq:TCX}
    T_c^{(X)} \simeq \frac{|g|}{4} + |g| e^{-4\xi_{\khs + \ve Q_c}/|g|}\, .
\end{align}
This has several interesting implications: The leading contribution from the hot spots, $|g|/4$, is identical to the pairing cases~\eqref{eq:Tc_hs} and~\eqref{eq:Cooper_HS_Tc} for the Cooper case and Eq.~\eqref{eq:PDW_HS_Tc} for the PDW case. However, the enhancement of $T_c^{(X)}$ arising from coherent processes including the shifted copies is only exponentially small as compared to the pairing scenarios. Therefore, pairing preempts the formation of the exciton condensate. Consequently, we focus on the dominating pairing instabilities in the following and do not study the XCON in greater detail. 

Physically, the different corrections stemming from the shifted copies are connected to the underlying scattering processes. In the pairing cases one has to transfer two particles from the FS to higher-energy states, which is possible as long as the necessary amount of energy is provided by the photon field mediating the interaction. In contrast, in the exchange channel one has to create a particle-hole pair in the vicinity of the shifted copies. However, the probability to create of a hole far outside of the FS in a thermal state is exponentially small. 

To close the chapter we briefly discuss instabilities of the $Y^{(+)}_{\ve k^{(\nu)_l}}$ sector, as introduced in Eq.~\eqref{eq:def:Yldelta}, that occur in momentum groups $(i/ii)$ and $(iv)/(iii)$. In particular, the Bogoliubov energies can be obtained from the previous case~\eqref{eq:Bogoliubov_XCON} via the mapping
\begin{align}
    \epsilon^{Y,1,2,3,4}_{\ve k_l^{(\nu)}} = \epsilon^{X,1,2,3,4}_{\ve k_l^{(\nu)}}(\ve k^{(\nu')}_{l+1} \to \ve k^{(\nu')}_l, X^{(+)}_{\ve k^{(\nu)}_l} \to Y^{(+)}_{\ve k^{(\nu)}_l})\, ,
\end{align}
however, now for the index pairs $(\nu,\nu') = (i,ii),(iv,iii)$.
Similarly, the general mean-field equation is equvialent to Eq.~\eqref{eq:MF_XCON_det} again with the replacement $X^{(+)}_{\ve k^{(\nu)}_l} \to Y^{(+)}_{\ve k^{(\nu)}_l}$. One solves the latter for $T^{(Y)}_c$ along the same lines as in case of the $X^{(+)}_{\ve k^{(\nu)}_l}$, for instance for the instability in the  sectors $(i)/(ii)$. Including again the hot spots and the shifted copies, means here to take the the eight momenta: $\ve k_1 \equiv \ve k^{(i)}_{-2} = \khs -2\ve Q_c + \ve \delta \ve k^+$, $\ve k_2 \equiv \ve k^{(i)}_{-1} = \khs -\ve Q_c + \ve \delta \ve k^+$, $\ve k_3 \equiv \ve k^{(i)}_{0} = \khs + \ve \delta \ve k^+$, $\ve k_4 = \ve k^{(i)}_1 = \khs + \ve Q_c + \ve \delta \ve k^+$ that are identical to the previous case and in addition $\ve k_{5'} = \ve k^{(ii)}_{-2} = \khs -2\ve Q_c + \ve \delta \ve k^-$, $\ve k_{6'} = \ve k^{(ii)}_{-1} = \khs - \ve Q_c - \ve \delta \ve k^-$, $\ve k_{7'} = \ve k^{(ii)}_{0} = \khs + \ve \delta \ve k^-$, and $\ve k_{8'} = \ve k^{(ii)}_{1} = \khs +\ve Q_c +\ve \delta \ve k^-$. Following the same steps 
from Eq.~\eqref{eq:det_example_XCON} to Eq.~\eqref{eq:TCX}, one finally finds the identical critical temperature:
\begin{align}\label{eq:TCY}
    T_c^{(Y)} =
    T_c^{(X)}  \simeq \frac{|g|}{4} + |g| e^{-4\xi_{\khs + \ve Q_c}/|g|}\, ´.
\end{align}
This actually agrees with the physical expectations since the content of single-particle energies in the $Y^{(+)}_{\ve k_l^{(\nu)}}$ is in the limit $\ve \delta \ve k^+, \ve \delta \ve k^- \to 0$ exactly the same as in the calculation for the $X^{(+)}_{\ve k_l^{(\nu)}}$. As a consequence, also the instability to finite expectation values $Y^{(+)}_{\ve k_l^{(\nu)}}$ is suppressed as compared to pairing. 

\subsection{Symmetries of the system and their spontaneous breaking}\label{sec:MF_Ana_symm}
A great deal about the underlying physics of the Fermi gas in the cavity can be learned by analyzing the symmetry properties. As discussed in Sec.~\ref{sec:Hamiltonian_mom_struc}, the momentum reflection symmetries $k_x \to - k_x$ and $k_y \to - k_y$ are present at the level of the full Hamiltonian~\eqref{eq:model} or equivalently~\eqref{eq:modelII}. In the presence of cavity-mediated interactions with nonvanishing $\ve Q_c$ these momentum reflection symmetries are the residuals of the explicitly broken spacial isotropy of the bare Hamiltonian. As a consequence, the classification of possible order parameter solutions in terms of $s,p,d,...$ waves associated with the eigenstates of the angular momentum operator, is only justified in the strict limit $\ve Q_c \to 0$ used in Ref.~\cite{Gao2020}.  

Furthermore, it is very useful to discuss the symmetries that are spontaneously broken in the presence of the different order parameters: The excition condensation order-parameters $X^{(+)}_{\ve k^{(i,ii)}_l}$ in the exchange channel break translation invariance since the finite relative momentum $\ve K(\ve \delta \ve k^+)$ between the two Fermions corresponds to a finite center-of-mass momentum in the language of particle-hole pairs. However, breaking a global $U(1)$ phase rotation symmetry, typically associated with condensation via violation of particle number conservation, is not visible at the level of the fermionic mean-field Hamiltonian. Instead, it requires an effective treatment of the particle-hole pairs in terms of Bosons that proliferate in their ground state. In contrast, the fermion number conservation that is linked in the usual way to the $U(1)$ symmetry 
\begin{align}\label{eq:phi_rot1}
    \hat{c}_{\ve k^{(\nu)}_l} \to e^{i \varphi} \hat{c}_{\ve k^{(\nu)}_l} \qquad 
    \hat{c}^\dagger_{\ve k^{(\nu)}_l} \to e^{-i \varphi} \hat{c}^\dagger_{\ve k^{(\nu)}_l}\qquad \forall\, \nu, l
\end{align}
is broken spontaneously both by the Cooper and by the PDW order parameters in very close analogy to the standard BCS case. For later convenience, we nevertheless give some details here: We first note that each contribution to the full Hamiltonian~\eqref{eq:modelII} is symmetric under the previous transformation. Therefore, also the interaction terms that have been identified as the most relevant for instabilities (\cref{eq:H_Cooper,eq:H_PDW,eq:H_XCON,eq:H_mom_occ}) are invariant before the mean-field decoupling. However, direct application of the transformation~\eqref{eq:phi_rot1} to the gap parameters yields formally 
\begin{align}\label{eq:Delta_phi_rot}
    \Delta^{(r)}_{\ve k^{(\nu)}_l} \to e^{2i \varphi} \Delta^{(r)}_{\ve k^{(\nu)}_l} \qquad 
    \bar \Delta^{(r)}_{\ve k^{(\nu)}_l} \to e^{-2i \varphi} \bar \Delta^{(r)}_{\ve k^{(\nu)}_l}\qquad \forall\, l\, ,
\end{align}
for $r=(0,\pm), \nu = (i,iii)$ showing that the order parameters depend on the choice of $\varphi$ as usual.
Considering the effect of Eq.~\eqref{eq:phi_rot1} on 
$\hat H_\mf$, on the other hand, it boils down to the unitary transformation 
\begin{align}
\left(
\begin{array}{c}
     \ve \hat{\Psi}_{\ve k^{(\nu)}} \\[0.2 cm]
     \ve \hat{\Psi}_{\ve k^{(\nu)}}^\dagger
\end{array}
\right)
\to 
\left(
\begin{array}{c c}
  e^{i\varphi} \mathds 1_{l \times l}   & 0 \\[0.2 cm]
  0   &  e^{-i\varphi} \mathds 1_{l \times l} 
\end{array}
\right)
\left(
\begin{array}{c}
     \ve \hat{\Psi}_{\ve k^{(\nu)}} \\[0.2 cm]
     \ve \hat{\Psi}_{\ve k^{(\nu)}}^\dagger
\end{array}
\right)
\end{align}
acting on the quadratic part of the mean-field Hamiltonian in Eq.~\eqref{eq:HMFfull} for all $\nu$.  Here, the vector $\hat{\ve \Psi}_{\ve k^{(\nu)}} = (\hat{c}_{\ve k^{(\nu)}_{l \in \mathds Z}})$ assembles all the annihilation operators in group $\nu$ for all $l$ (with or without truncation). Since this phase rotation operation is unitary it does not affect the physical properties of $\hat H_\mf$. Simultaneously, the pairing order-paramters $\Delta^{(0)}_{\ve k^{(i,iii)}_l}$ and $\Delta^{(\pm)}_{\ve k^{(i,iii)}_l}$ are not invariant under the phase rotation~\eqref{eq:phi_rot1} as already illustrated in Eq.~\eqref{eq:Delta_phi_rot}. As expected, we thus deal indeed with the mechanism of spontaneous symmetry breaking. 

However, this gives rise to an interesting question:
Since both Cooper and PDW order parameters break the same symmetry, one would typically expect competing instabilities. In Sec.~\ref{sec:MF_Ana_inst}, in contrast, it has been found that both instabilities are degenerate, yet on the basis of a decoupled mean-field theory. Therefore, it is natural to ask whether this signals an accidental coincidence introduced by the approximation and if so, which instability preempts the other one. In the case of the Fermi gas subject to cavity-mediated interactions, however, the degeneracy turns out to be physical because there is an additional $SO(2)$ rotation symmetry in momentum space present. The latter exhibits spontaneous symmetry breaking simultaneously with the global $U(1)$ from Eq.~\eqref{eq:phi_rot1}. Thereby, it determines the relative admixtures of Cooper and PDW states in the symmetry-broken phase below $T_c$. The special symmetry transformation under consideration rotates momenta of groups $(ii)$ and $(iv)$ with identical $k_x$ components into each other in the sense
\begin{align}\label{eq:alpha_rot}
\begin{split}
  \left(
    \begin{array}{c}
         \hat{c}_{\ve k^{(ii)}_l}  \\
         \hat{c}_{\ve k^{(iv)}_{l+1}}
    \end{array}
    \right) 
    & \to 
    \underline R(\alpha) 
    \left(
    \begin{array}{c}
         \hat{c}_{\ve k^{(ii)}_l}  \\
         \hat{c}_{\ve k^{(iv)}_{l+1}}
    \end{array}
    \right) \qquad \forall \, l\\
    & = \left(
    \begin{array}{cc}
        \cos(\alpha) & - \sin(\alpha)  \\
         \sin(\alpha) & \cos(\alpha) 
    \end{array}
    \right)
    \left(
    \begin{array}{c}
         \hat{c}_{\ve k^{(ii)}_l}  \\
         \hat{c}_{\ve k^{(iv)}_{l+1}}
    \end{array}
    \right)\, ,
    \end{split}
\end{align}
see Fig.~\ref{fig:dkStructure} and Eq.\eqref{eq:defMomGroups} for the labeling conventions. 
An equivalent representation of the symmetry is obtained by rotating groups $(i)$ and $(iii)$ into each other,
which is related to the above by performing an additional $k_x \to -k_x$ operation. 
Most interestingly, this transformation acts like a rotation in the space of the pairing-order parameters
\begin{align}\label{eq:alpha_rot_pairing}
\begin{split}
    \left(
    \begin{array}{c}
        \Delta^{(0)}_{\ve k^{(i)}_l}  \\[0.35cm]
        \Delta^{(+)}_{\ve k^{(i)}_l}
    \end{array}
    \right)
    & \rightarrow 
    \underline R^{-1}(\alpha) 
    \left(
    \begin{array}{c}
        \Delta^{(0)}_{\ve k^{(i)}_l}  \\[0.35cm]
        \Delta^{(+)}_{\ve k^{(i)}_l}
    \end{array}
    \right) \\[0.5cm]
 \left(
    \begin{array}{c}
        \Delta^{(0)}_{\ve k^{(iii)}_l}  \\[0.35cm]
        \Delta^{(-)}_{\ve k^{(iii)}_l}
    \end{array}
    \right)
    & \rightarrow 
     \underline R(\alpha) 
    \left(
    \begin{array}{c}
        \Delta^{(0)}_{\ve k^{(iii)}_l}  \\[0.35cm]
        \Delta^{(-)}_{\ve k^{(iii)}_l}
    \end{array}
    \right) \, ,  
\end{split}
\end{align}
as well as in the case of the exciton condensation
\begin{align}\label{eq:alpha_rot_XCON}
\begin{split}
    \left(
    \begin{array}{c}
        Y^{(+)}_{\ve k^{(i)}_l}  \\[0.35cm]
        X^{(+)}_{\ve k^{(i)}_l}
    \end{array}
    \right)
    & \rightarrow 
    \underline R(\alpha) 
    \left(
    \begin{array}{c}
        Y^{(+)}_{\ve k^{(i)}_l}  \\[0.35cm]
        X^{(+)}_{\ve k^{(i)}_l}
    \end{array}
    \right) \\[0.5cm]
 \left(
    \begin{array}{c}
        Y^{(+)}_{\ve k^{(iv)}_l}  \\[0.35cm]
        X^{(+)}_{\ve k^{(ii)}_l}
    \end{array}
    \right)
    & \rightarrow 
     \underline R^{-1}(\alpha) 
    \left(
    \begin{array}{c}
        Y^{(+)}_{\ve k^{(iv)}_l}  \\[0.35cm]
        X^{(+)}_{\ve k^{(ii)}_l}
    \end{array}
    \right) \, .
\end{split}
\end{align}
This order-parameter rotation also represent the physically most accessible interpretation of the symmetry. 
Next, we show that Eq.~\eqref{eq:alpha_rot} indeed represents a symmetry of the Hamiltonian. To this end, we analyze the consequences of the $\alpha$ rotation along the same lines as in the previous case of the $\varphi$ phase rotation.
First of all, we note that due to the particular form of the interaction the full Hamiltonian~\eqref{eq:modelII} actually turns out to be invariant under the transformation~\eqref{eq:alpha_rot}. 
This can be seen by rewriting the full interaction part
\begin{align}
\label{eq:sym_proof_1}
\begin{split}
& \frac{g}{2} \sum_{\substack{\nu,\nu' = i,...,iv\\
    l,l' \in \mathds Z; s= \pm 1}}
    \sum_{\substack{\delta k_x,\delta k_y \\ \delta k'_x,\delta k'_y}} \hat{c}^\dagger_{k^{(\nu)}_{l+s}}\hat{c}^\dagger_{k^{(\nu')}_{l'-s}} \hat{c}_{k^{(\nu')}_{l'}} \hat{c}_{k^{(\nu)}_l} = \\
    & \frac{g}{2} \sum_s
    \left[\sum_{\delta k_x, \delta k_y} \sum_{\nu ,l} \hat{c}^\dagger_{\ve k^{(\nu)}_{l+s}} \hat{c}_{\ve k^{(\nu)}_{l}} \right]
    \left[\sum_{\delta k'_x, \delta k'_y} \sum_{\nu' ,l'} \hat{c}^\dagger_{\ve k^{(\nu')}_{l'-s}} \hat{c}_{\ve k^{(\nu')}_{l'}} \right] \, .
\end{split}
\end{align}
Now it follows from simply inserting the definition of the transformation~\eqref{eq:alpha_rot} in combination with suitable index shifts $l \to l \pm 1 $ that the expression 
\begin{align}\label{eq:sym_proof_2}
    \sum_{\delta k_x, \delta k_y} \sum_{\nu ,l} \hat{c}^\dagger_{\ve k^{(\nu)}_{l+s}} \hat{c}_{\ve k^{(\nu)}_{l}}
\end{align}
is invariant under the action of $\underline R(\alpha)$ for arbitrary values of $l,\nu $ and in particular for arbitrary integer values of $s$. As a result, the interaction Hamiltonian from above is invariant under the transformation~\eqref{eq:alpha_rot}. Moreover, using the last result for $s = 0$ and the fact that $\xi_{\ve k^{(ii)}_l} = \xi_{\ve k^{(iv)}_{l+1}}$ we conclude that the kinetic part of Eq.~\eqref{eq:modelII} is also not affected by the momentum rotation. Consequently, Eq.~\eqref{eq:alpha_rot} indeed represents a symmetry of the full Hamiltonian. 
Next, we have to show that $\underline R(\alpha)$ acts merely like a unitary transformation on the mean-field Hamiltonian $\hat H_\mf$ from Eq.~\eqref{eq:HMFfull}. To see this, we note that the latter is equivalent to the Nambu Hamiltonian
\begin{align}\label{eq:H_MF_Nambu}
    \hat{H}_\mf = \frac{1}{2}\sum_{\substack{\delta k_x,\delta k_y\\l \in \mathds Z}}
    \left(
    \begin{array}{c}
         \hat{\Psi}^\dagger_l  \\
         \hat{\Psi}_l
    \end{array}
    \right)^T
    \underline h_\mf
    \left(
    \begin{array}{c}
         \hat{\Psi}_l  \\
         \hat{\Psi}^\dagger_l
    \end{array}
    \right) + \text{const} \, ,
\end{align}
where we have defined
\begin{align}
    \hat{\Psi}_l = (\hat{c}_{\ve k^{(i)}_l}, \hat{c}_{\ve k^{(i)}_{-l-1}}, \hat{c}_{\ve k^{(ii)}_l}, \hat{c}_{\ve k^{(ii)}_{-l-1}}, \hat{c}_{\ve k^{(iii)}_{l+1}},\hat{c}_{\ve k^{(iii)}_{-l}},\hat{c}_{\ve k^{(iv)}_{l+1}},\hat{c}_{\ve k^{(iv)}_{-l}}).
\end{align}
The matrix elements of $\underline h_\mf$ can be read off from the different contributions of $\hat H_\mf$. 
For instance, the diagonal elements for arbitrary $\nu$ are given by
\begin{align}
    \left(\underline h_\mf\right)_{\hat{c}^\dagger_{\ve k^{(\nu)}_l} \hat{c}_{\ve k^{(\nu)}_{l'}}} = \frac{\delta_{l,l'}}{2}\left(\xi_{\ve k^{(\nu)}_l} -g \sum_{s=\pm 1} \langle \hat{n}_{\ve k^{(\nu)}_{l+s}} \rangle\right) \, ,
\end{align}
which contains both the kinetic energy and the contribution from $\hat H_{\hat n}$ from Eq.~\eqref{eq:H_mom_occ}. The additional factor of $1/2$ is needed to avoid double-counting with $l$. Similarly, we find from $\hat H^{(X)}_{\text{XCON}}$ (cf. Eq.~\eqref{eq:H_XCON_X}) 
\begin{align}
\begin{split}
    \left(\underline h_\mf\right)_{\hat{c}^\dagger_{\ve k^{(iv)}_{l'}} \hat{c}_{\ve k^{(i)}_{l}}} & \!\! = - \left(\underline h_\mf\right)_{\hat{c}_{\ve k^{(i)}_{l}} \hat{c}^\dagger_{\ve k^{(iv)}_{l'}}}\!\! = \frac{\delta_{l+1,l'}}{2}\sum_{s=\pm 1} \!\! \bar{X}^{(+)}_{\ve k^{(i)}_{l+s}} \\
     \left(\underline h_\mf\right)_{\hat{c}^\dagger_{\ve k^{(iii)}_{l'}} \hat{c}_{\ve k^{(ii)}_{l}}} & \!\! = - \left(\underline h_\mf\right)_{\hat{c}_{\ve k^{(ii)}_{l}} \hat{c}^\dagger_{\ve k^{(iii)}_{l'}}} \!\! = \frac{\delta_{l+1,l'}}{2}\sum_{s=\pm 1} \!\! \bar{X}^{(+)}_{\ve k^{(ii)}_{l+s}},
\end{split}
\end{align}
whereas we obtain from $\hat H^{(Y)}_{\text{XCON}}$ (cf. Eq.~\eqref{eq:H_XCON_Y})
\begin{align}
\begin{split}
    \left(\underline h_\mf\right)_{\hat{c}^\dagger_{\ve k^{(ii)}_{l'}} \hat{c}_{\ve k^{(i)}_{l}}} & \!\! = - \left(\underline h_\mf\right)_{\hat{c}_{\ve k^{(i)}_{l}} \hat{c}^\dagger_{\ve k^{(ii)}_{l'}}}\!\! = \frac{\delta_{l,l'}}{2}\sum_{s=\pm 1} \!\! \bar{Y}^{(+)}_{\ve k^{(i)}_{l+s}} \\
     \left(\underline h_\mf\right)_{\hat{c}^\dagger_{\ve k^{(iii)}_{l'}} \hat{c}_{\ve k^{(iv)}_{l}}} & \!\! = - \left(\underline h_\mf\right)_{\hat{c}_{\ve k^{(iv)}_{l}} \hat{c}^\dagger_{\ve k^{(iii)}_{l'}}} \!\! = \frac{\delta_{l,l'}}{2}\sum_{s=\pm 1} \!\! \bar{Y}^{(+)}_{\ve k^{(iv)}_{l+s}}.
\end{split}
\end{align}
In analogy, $\hat H_{\text{Cooper}}$ (cf. Eq.~\eqref{eq:H_Cooper}) yields the matrix elements
\begin{align}
\begin{split}
    \left(\underline h_\mf\right)_{\hat{c}_{\ve k^{(iv)}_{l'}} \hat{c}_{\ve k^{(i)}_{l}}} & \!\! = - \left(\underline h_\mf\right)_{\hat{c}_{\ve k^{(i)}_{l}} \hat{c}_{\ve k^{(iv)}_{l'}}}\!\! = \frac{\delta_{-l,l'}}{2}\sum_{s=\pm 1} \!\! \bar{\Delta}^{(0)}_{\ve k^{(i)}_{l+s}} \\
     \left(\underline h_\mf\right)_{\hat{c}_{\ve k^{(ii)}_{l'}} \hat{c}_{\ve k^{(iii)}_{l}}} & \!\! = - \left(\underline h_\mf\right)_{\hat{c}_{\ve k^{(iii)}_{l}} \hat{c}^\dagger_{\ve k^{(ii)}_{l'}}} \!\! = \frac{\delta_{-l,l'}}{2}\sum_{s=\pm 1} \!\! \bar{\Delta}^{(0)}_{\ve k^{(iii)}_{l+s}},
\end{split}
\end{align}
and finally $\hat H_{\text{PDW}}$ (cf. Eq.~\eqref{eq:H_PDW}) gives rise to
\begin{align}
\begin{split}
    \left(\underline h_\mf\right)_{\hat{c}_{\ve k^{(ii)}_{l'}} \hat{c}_{\ve k^{(i)}_{l}}} & \!\! = - \left(\underline h_\mf\right)_{\hat{c}_{\ve k^{(i)}_{l}} \hat{c}_{\ve k^{(ii)}_{l'}}}\!\! = \frac{\delta_{-l-1,l'}}{2}\sum_{s=\pm 1} \!\! \bar{\Delta}^{(+)}_{\ve k^{(i)}_{l+s}} \\
     \left(\underline h_\mf\right)_{\hat{c}_{\ve k^{(iv)}_{l'}} \hat{c}_{\ve k^{(iii)}_{l}}} & \!\! = - \left(\underline h_\mf\right)_{\hat{c}_{\ve k^{(iii)}_{l}} \hat{c}^\dagger_{\ve k^{(iv)}_{l'}}} \!\! = \frac{\delta_{-l+1,l'}}{2}\!\!\!\!\sum_{s=\pm 1} \!\! \bar{\Delta}^{(-)}_{\ve k^{(iii)}_{l+s}}.
\end{split}
\end{align}
The remaining nontrivial elements of $\underline h_\mf$ are obtained by virtue of $\underline h_\mf^\dagger = \underline h_\mf$ whereas $\underline h_\mf$   vanishes for all other combinations of indices. After establishing the mean-field Hamiltonian in the Nambu structure of Eq.~\eqref{eq:H_MF_Nambu}, we one observes that the corresponding representation of $R(\alpha)$ from Eq.~\eqref{eq:alpha_rot} in the basis $(\hat{\Psi}_l,\hat{\Psi}_l^\dagger)$ reads
\begin{align}
    \left(
    \begin{array}{c}
         \hat{\Psi}_l  \\
         \hat{\Psi}_l^\dagger
    \end{array}
    \right)
    \rightarrow 
    \left(
    \begin{array}{cc}
    \underline{\tilde{R}}(\alpha)     & 0 \\
     0    & \underline{\tilde{R}}(\alpha)  
    \end{array}
    \right)
    \left(
    \begin{array}{c}
         \hat{\Psi}_l  \\
         \hat{\Psi}_l^\dagger
    \end{array}
    \right) \, ,
\end{align}
where
\begin{align}
    \underline{\tilde R}(\alpha) = 
    \left(
    \begin{array}{cccc}
       \mathds 1_{2 \times 2}  & 0 & 0 & 0    \\
       0  & \cos(\alpha) \mathds 1_{2 \times 2} & 0 & - \sin(\alpha) \mathds 1_{2 \times 2} \\
       0  & 0 & \mathds 1_{2 \times 2} & 0 \\
       0 & \sin(\alpha) \mathds 1_{2 \times 2} & 0 & \cos(\alpha) \mathds 1_{2 \times 2}
    \end{array}
    \right)\, .
\end{align}
From the latter definition one directly shows that $\underline{\tilde R}(\alpha)^\dagger \cdot \underline{\tilde R}(\alpha) = \mathds 1_{8 \times 8 }$. Consequently, $\underline{\tilde R}(\alpha)$ represents indeed a unitary transformation of $\hat H_\mf$, which does not affect the physical properties of the Hamiltonian. On the other hand, however, $\underline R(\alpha)$ transforms the order parameters in a nontrivial fashion as is shown in Eqs.~\eqref{eq:alpha_rot_pairing} and~\eqref{eq:alpha_rot_XCON}. As a result, we find spontaneous symmetry breaking of this special momentum rotation when finite order parameters form, like in case of the standard $U(1)$ symmetry breaking associated with pairing (see Eq.~\eqref{eq:phi_rot1}), as discussed above. The fact the the transition into a state with pairs preempts the XCON instabilities is not affected by the presence of this additional symmetry $\underline R(\alpha)$. However, we conclude from its presence that the statement $T_c^{(0)} = T_c^{(\pm)}$ (cf. Eqs.~\eqref{eq:Cooper_HS_Tc} and~\eqref{eq:PDW_HS_Tc}) is not caused accidentally by the invoked approximations. In contrast, both the $U(1)$ phase symmetry and the momentum rotation symmetry are spontaneously broken simultaneously below the critical temperature. While spontaneously attained value of $\varphi$ describes the overall phase of the pairs (cf.~\eqref{eq:Delta_phi_rot}) the value of $\alpha$ determines the fractions of Cooper pairs and pairs with finite center-of-mass momentum, as is seen in Eq.~\eqref{eq:alpha_rot_pairing}. Moreover, this statement holds also beyond mean-field theory since the identified symmetries leave also the exact Hamiltonian invariant.

\section{Pairing instabilities in the standing-wave cavity}
\label{sec:instabilities_SWcavity}
\subsection{Physical picture and mean-field theory}
Finally, let us consider a standing-wave cavity, which represents the most common experimental platform for studying the coupling between an optical resonator and an ultracold Fermi gas ~\cite{ZhangWu2021,HelsonBrantut2023,ZwettlerBrantut2026}.
The most prominent difference between a ring cavity and a standing-wave cavity is the absence of momentum conservation in the latter due to reflections of the photons on the mirrors back into the cavity, see Fig.~1(a) of the main text for a pictorial representation. 
As it turns out, however, the results for the pairing instabilities in both setups are qualitatively similar, since the derivations for the ring cavity can be applied with little modifications also in the absence of momentum conservation. Establishing the physics of pairing in the standing-wave cavity is the main focus of this section. We will neglect instabilities in the exchange channel, since they are suppressed in the same way as in the ring cavity
[cf.~discussion below \cref{eq:TCX}]. In addition, we also exclude the effects of reshaping the Fermi surface,  since they essentially affect pairing only via a renormalized dispersion. 

The Hamiltonian for the standing-wave cavity reads \cite{Piazza2014PRL,Mivehar2021rev,HelsonBrantut2023}:
\begin{align}\label{eq:modelSW}
  \hat H_{\sw} =\sum_{\ve k} \xi_{\ve k}
  \hat c^\dagger_{\ve k} \hat c^{}_{\ve k}   +\frac{g}{2} \sum_{\substack{\ve p, \ve k\\ s,s'= \pm 1}}  \hat{c}^\dagger_{\ve p+s\ve Q_c} \hat{c}^\dagger_{\ve k+s'\ve Q_c}  \hat{c}^{}_{\ve k} \hat{c}^{}_{\ve p}\, .
\end{align}
As already mentioned, the crucial difference with respect to the Hamiltonian~\eqref{eq:model} arises from the possibility that the momenta $\ve p , \ve k$ change independently during any scattering process, so that the total momentum of the two interacting Fermions may change by $\pm 2 \ve Q_c$. Consequently, a generic pair
\begin{align}\label{eq:DeltaSW}
    \Delta_{\ve p, \ve k} \equiv g \langle  \hat{c}_{\ve k} \hat{c}_{\ve p} \rangle \, ,
\end{align}
which carries the total momentum $\ve P = \ve p + \ve k$ is both coupled via momentum-conserving interaction processes to $\Delta_{\ve p \pm \ve Q_c, \ve k \mp \ve Q_c}$, like in case of the ring cavity, but also via non-momentum-conserving processes to $\Delta_{\ve p \pm \ve Q_c, \ve k \pm \ve Q_c}$ with c.o.m.~momentum $\ve P \pm 2 \ve Q_c $. Performing a general mean-field decoupling of $\hat H_\sw$ in the pairing sector using the order parameters $\Delta_{\ve p, \ve k}$ yields 
\begin{align}
\begin{split}
    & \hat H_\swmf = \frac{1}{2} \sum_{\ve k} \xi_{\ve k}- \frac{1}{g} \operatorname{Tr}(\underline M \bar{\underline \Delta} \underline M \underline \Delta)\\
    & +\frac{1}{2} \sum_{\ve k, \ve p}
    \left(
    \begin{array}{c}
           \hat{c}^\dagger_{\ve k}  \\
           \hat{c}_{\ve k} 
    \end{array}
    \right)^{T}
    \left(
    \begin{array}{cc}
        \xi_{\ve k} \delta_{\ve k, \ve p} & (\underline M  \underline \Delta  \underline M)_{\ve k, \ve p}  \\
        (\underline M \bar{\underline \Delta}  \underline M)_{\ve k, \ve p} & -\xi_{\ve k} \delta_{\ve k, \ve p}
    \end{array}
    \right) 
    \left(
    \begin{array}{c}
          \hat{c}^\dagger_{\ve p}  \\
          \hat{c}_{ \ve p} 
    \end{array}
    \right)\, .
    \end{split}
    \label{eq:H_swmf}
\end{align}
Note that the underlined expressions correspond to matrix multiplications in momentum of the order parameter matrix $(\underline \Delta)_{\ve p, \ve k} \equiv \Delta_{\ve p, \ve k}$ and the matrix of allowed momentum transfers $(\underline M)_{\ve k, \ve k'} = \delta_{\ve k', \ve k + \ve Q_c} + \delta_{\ve k', \ve k - \ve Q_c}$, which takes into account the momentum change of Fermions due to the interaction with cavity photons, equivalently to Eq.~\eqref{eq:def_M}.
We can now follow the same procedure as in the ring-cavity case: we diagonalize the Hamiltonian \eqref{eq:H_swmf} into a Bogoliubov form with positive eigenenergies $\epsilon_\alpha$ [cf.~\eqref{eq:Bogoliubov}], we derive the mean-field free energy [cf.~\eqref{eq:FreeEnMF}] and we set its derivative with respect the the mean-field parameters to zero to derive the mean-field equation
\begin{align}
   \Delta_{\ve k,\ve p} = - \frac{g}{2} \sum_{\alpha} \tanh\!\left(\frac{\beta \epsilon_\alpha}{2} \right)\!\! \sum_{\ve k', \ve p'} (\underline M)^{-1}_{\ve k ,\ve k'} \frac{\delta \epsilon_\alpha} {\delta \bar \Delta_{\ve k', \ve p'}} (\underline M)^{-1}_{\ve p' ,\ve p} \,,
   \label{eq:gapeq_SWcavity}
\end{align}
where, following the same notation as above, $\epsilon_\alpha$ are the eigenvalues of $\hat H_\swmf$.
Next, we identify again the most dominant pairing instabilities by focusing on the hot spots of~\cref{fig:hotspots}. Since the Hamiltonian $\hat H_\sw $ from Eq.~\eqref{eq:modelSW} contains the same momentum-conserving terms as the Hamiltonian of the ring cavity~\eqref{eq:model}, both the decoupling in Cooper and PDW pairs, given in Eqs.~\eqref{eq:HS_a} and~\eqref{eq:HS_b}, respectively, yield resonant processes on the FS.
\begin{figure}[htbp]
    \centering
    \includegraphics[width = .5 \columnwidth]{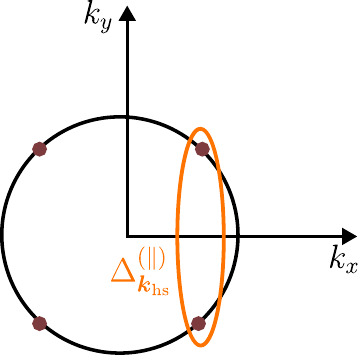}
    \caption{Additional pairing instability on the hot spots (dots) for the standing-wave-cavity case, to be considered together with the two pairing instabilities of \cref{fig:inst_hot_spots}.}
    \label{fig:Delta_par}
\end{figure}
Furthermore, the non-momentum-conserving contributions of the interaction allow for an additional finite c.o.m.~momentum pair related to scattering processes with all Fermions on the FS (cf.~Fig~\ref{fig:Delta_par} for a graphical representation). The corresponding mean-field decoupling is given by 
\begin{align}
\begin{split}
& g \hat{c}^\dagger_{\khs - \ve Q_c} \hat{c}^\dagger_{-\khs} \hat{c}_{-\khs+\ve Q_c} \hat{c}_{\khs} \\
& \simeq   \Delta^{(\parallel)}_{\khs} \hat{c}^\dagger_{\khs -\ve Q_c} \hat{c}^\dagger_{-\khs} + \bar\Delta^{(\parallel)}_{\khs-\ve Q_c} \hat{c}_{-\khs + \ve Q_c} \hat{c}_{\khs } \\
& \quad -\frac{\bar\Delta^{(\parallel)}_{\khs-\ve Q_c} \Delta^{(\parallel)}_{\khs}}{g}\, .
\end{split}
\end{align}
Note that the total momentum of the pair $\Delta^{(\parallel)}_{\khs} = g \langle c_{-\khs+\ve Q_c} c_{\khs} \rangle $ ($\Delta^{(\parallel)}_{\khs - \ve Q_c}= g \langle c_{-\khs} c_{\khs-\ve Q_c} \rangle $) is given by $\ve P = \ve Q_c$ ($\ve P = -\ve Q_c$), i.e.~it is parallel to the cavity axis. This remains true even when the shifted copies are included since the interactions only change the total momentum by $\pm 2\ve Q_c$ or leave it invariant. Therefore, this new pair with non-zero total momentum will be denoted with the symbol $\parallel$.

\subsection{Analytic results for $T_c$}

We now have all the ingredients to derive the weak-coupling analytical expression for $T_c$ in the standing-wave cavity for the three different kinds of pairs.
We will first provide the derivation for Cooper pairs, and then show an analogous procedure can be performed for the other two kinds of pairs.
As presented in \cref{fig:dkStructure}, we include the first shifted copies of the hot spots in the momentum grid, like in case of the ring cavity outlined in \cref{sec:MF_Ana_inst}, but for simplicity we consider only infinitesimal momentum deviations $|\ve\delta \ve{k}^{+}|\to0^+$ to focus on the instabilities at the hot spots that determine $T_c$.
Similarly to \cref{eq:Cooper_shorthand}, we can organize all the possible pairs $\Delta_{\ve{k},\ve{p}}$ in the Cooper sector in the vector structure
\begin{align}
    \label{eq:vector_Delta_Cooper_SWcavity}
    \ve{\Delta}^{(0)}= 
    \left(\begin{array}{c}
    \Delta^{(0)}_{1} \\
    \Delta^{(0)}_{2} \\
    \Delta^{(0)}_{3} \\
    \Delta^{(0)}_{4} \\  
    \Delta^{(0)}_{5} \\
    \Delta^{(0)}_{6} \\
    \Delta^{(0)}_{7} \\
    \Delta^{(0)}_{8}
    \end{array}\right)=
    \left(\begin{array}{c}
    \Delta^{}_{\khs-2\ve{Q}_c,-\khs+2\ve{Q}_c} \\
    \Delta^{}_{\khs-2\ve{Q}_c,-\khs} \\
    \Delta^{}_{\khs-\ve{Q}_c,-\khs+\ve{Q}_c} \\
    \Delta^{}_{\khs,-\khs+2\ve{Q}_c} \\
    \Delta^{}_{\khs-\ve{Q}_c,-\khs+\ve{Q}_c} \\
    \Delta^{}_{\khs,-\khs} \\
    \Delta^{}_{\khs+\ve{Q}_c,-\khs+\ve{Q}_c} \\
    \Delta^{}_{\khs+\ve{Q}_c,-\khs-\ve{Q}_c}
    \end{array}\right) \,.
\end{align}
With respect to \cref{eq:Cooper_shorthand}, we have now four additional momentum pairs ($\Delta^{(0)}_{2,4,5,7}$) participating in the gap equation. These are pairs with total momentum $\ve P=\pm 2 \ve Q_c$, which are coupled in the standing-wave cavity to the $\ve{P}=0$ pairs via non-momentum-conserving interaction terms. As none of these additional pairs has both particles on the hot spots, we anticipate from the calculation in the ring cavity that led to the result~\eqref{eq:Cooper_HS_Tc} that they will influence only the subleading ($\sim g^2$) behavior of $T_c$.

Representing \cref{eq:gapeq_SWcavity} in the basis~\eqref{eq:vector_Delta_Cooper_SWcavity}, one obtains an equation of the form
\begin{equation}\label{eq:matrix_Delta_Cooper_SWcavity}
     \left(\mathds 1_{8 \times 8} + \underline{A}_{8 \times 8} \right) \cdot \ve{\Delta}^{(0)} = 0 \,,
\end{equation}
which admits a non-zero solution for $\ve \Delta^{(0)}$ when
\begin{align}
\label{eq:det_Delta_Cooper_SWcavity}
    \begin{split}
     & \det \left(
       \mathds{1}_{8 \times 8} +
            \underline{A}_{8 \times 8} 
        \right) =0\, .
         \end{split}
\end{align}
This equation, together with taking $\Delta_{\ve k, \ve p}\to 0$ inside $\underline{A}_{8 \times 8}$, corresponds to the condition for the critical temperature $T_c$.
The matrix elements of $\underline{A}_{8 \times 8}$ can be obtained by evaluating the derivative of the eigenvalues $\epsilon_\alpha$ with respect to each possible pair $\bar\Delta_{\ve k',\ve p'}$ in \eqref{eq:gapeq_SWcavity}. At the critical temperature $T_c$, they reduce to the standard expression for the particle-particle bubble in the Bethe-Salpeter equation for the pairing susceptibility \cite{Fetter&Walecka}, that is
\begin{equation}
\label{eq:Xi_kp}
    \Xi_{\ve k, \ve p}=g \frac{1-n_F(\xi_{\ve k})-n_F(\xi_{\ve p})}{\xi_{\ve k}+\xi_{\ve p}} \, ,   
\end{equation}
where $n_F(\xi)=1/(e^{-\xi/T}+1)$ is the Fermi-Dirac distribution, $\xi_{\ve k}$ is the dispersion in the normal phase at the momentum $\ve k$ and $(\ve k, \ve p)$ are the momenta of the pairs $\Delta_{\ve k ,\ve p}$ of Eq.~\eqref{eq:matrix_Delta_Cooper_SWcavity} on which the matrix $\underline{A}_{8 \times 8}$ acts.
The non-zero matrix elements of $\underline{A}_{8 \times 8}$ are determined via the matrices of allowed momentum transfers $(\underline{M})_{\ve k, \ve k'}$ and $(\underline{M})_{\ve p, \ve p'}$ in \cref{eq:gapeq_SWcavity}.
By direct inspection, one can see that on our momentum grid, \cref{eq:Xi_kp} can only take two values: $\Xi_{\rm on}=g/(4T_c)$ if both momenta $\ve{p}$ and $\ve{k}$ are at hot spots on the FS, or $\Xi_{\rm off}\approx g/(2 \xi_{\khs+\ve Q_c})$ (assuming $T_c \ll \xi_{\khs + \ve Q_c} \sim \eF$) if at least one momentum corresponds to a shifted copy of the hot spots. The matrix $\underline{A}_{8 \times 8}$ attains then the form
\begin{align}\label{eq:Cooper_mat_SWcavity}
 \underline{A}_{8 \times 8} & = 
\left(
\begin{array}{c c c c c c c c}
    0 & 0 & \Xi_{\rm on} & 0 & 0 & 0 & 0 & 0 \\
     0 & 0 & \Xi_{\rm on} & 0 & \Xi_{\rm off} & 0 & 0 & 0 \\
     \Xi_{\rm off} & \Xi_{\rm off} & 0 & \Xi_{\rm off} & 0 & \Xi_{\rm on} & 0 & 0 \\
     0 & 0 & \Xi_{\rm on} & 0 & 0 & 0 & \Xi_{\rm off} & 0 \\
     0 & \Xi_{\rm off} & 0 & 0 & 0 & \Xi_{\rm on} & 0 & 0 \\
     0 & 0 & \Xi_{\rm on} & 0 & \Xi_{\rm off} & 0 & \Xi_{\rm off} & \Xi_{\rm off} \\
     0 & 0 & 0 & \Xi_{\rm off} & 0 & \Xi_{\rm on} & 0 & 0 \\
     0 & 0 & 0 & 0 & 0 & \Xi_{\rm on} & 0 & 0
\end{array}
\right)\, .
\end{align}
Calculating the determinant in \cref{eq:det_Delta_Cooper_SWcavity}, isolating $T_c$ and expanding for small $g$, one obtains
\begin{align}\label{eq:Cooper_Tc_SWcavity}
    T_c^{(0)} = \frac{|g|}{4} + \frac{3 g^2}{8 \xi_{\khs + \ve Q_c}} + \mathcal{O}(|g|^3/\eF^2)\, ,
\end{align}
which differs from the ring cavity case \eqref{eq:Cooper_HS_Tc} only by the numerical coefficient of the $g^2$ term, which arises again by the inclusion of the Fermions at the shifted copies of the hot spots. 
This means, in particular, that the coupling of the Cooper pairs at the hot spots to the additional pairs $\Delta^{(0)}_{2,4,5,7}$ via the non-momentum conserving interactions \emph{enhances} the critical temperature with respect to the ring cavity case, where the only correction arises from $\Delta^{(0)}_{1,8}$.

An analogous procedure can be performed for the $\Delta^{(\pm)}$ and $\Delta^{(\parallel)}$ pairs. In these cases, the momentum pairs can be reorganized in the following vector structures \footnote{Note that in the $\Delta^{(\pm)}$ case, pairs in the form $\Delta_{\ve k, \ve p}$ and $\Delta_{\ve p, \ve k}$ are to be considered independent, since we suppose to still have an infinitesimal deviation from the hot spots $\delta \ve k^{\pm}$ which allows to distinguish between them.}:
\begin{align}
    \label{eq:vector_Delta_+_SWcavity}
    \ve{\Delta}^{(+)}= 
    \left(\begin{array}{c}
    \Delta^{(+)}_{1} \\
    \Delta^{(+)}_{2} \\
    \Delta^{(+)}_{3} \\
    \Delta^{(+)}_{4} \\  
    \Delta^{(+)}_{5} \\
    \Delta^{(+)}_{6} \\
    \Delta^{(+)}_{7} \\
    \Delta^{(+)}_{8}
    \end{array}\right)=
    \left(\begin{array}{c}
    \Delta^{}_{\khs-2\ve{Q}_c,\khs+\ve{Q}_c} \\
    \Delta^{}_{\khs-2\ve{Q}_c,\khs-\ve{Q}_c} \\
    \Delta^{}_{\khs-\ve{Q}_c,\khs} \\
    \Delta^{}_{\khs,\khs+\ve{Q}_c} \\
    \Delta^{}_{\khs-\ve{Q}_c,\khs-2\ve{Q}_c} \\
    \Delta^{}_{\khs,\khs-\ve{Q}_c} \\
    \Delta^{}_{\khs+\ve{Q}_c,\khs} \\
    \Delta^{}_{\khs+\ve{Q}_c,\khs-2\ve{Q}_c}
    \end{array}\right) \,,
\end{align}
\begin{align}
    \label{eq:vector_Delta_-_SWcavity}
    \ve{\Delta}^{(-)}= 
    \left(\begin{array}{c}
    \Delta^{(-)}_{1} \\
    \Delta^{(-)}_{2} \\
    \Delta^{(-)}_{3} \\
    \Delta^{(-)}_{4} \\  
    \Delta^{(-)}_{5} \\
    \Delta^{(-)}_{6} \\
    \Delta^{(-)}_{7} \\
    \Delta^{(-)}_{8}
    \end{array}\right)=
    \left(\begin{array}{c}
    \Delta^{}_{-\khs-\ve{Q}_c,-\khs+2\ve{Q}_c} \\
    \Delta^{}_{-\khs-\ve{Q}_c,-\khs} \\
    \Delta^{}_{-\khs,-\khs+\ve{Q}_c} \\
    \Delta^{}_{-\khs+\ve{Q}_c,-\khs+2\ve{Q}_c} \\
    \Delta^{}_{-\khs,-\khs-\ve{Q}_c} \\
    \Delta^{}_{-\khs+\ve{Q}_c,-\khs} \\
    \Delta^{}_{-\khs+2\ve{Q}_c,-\khs+\ve{Q}_c} \\
    \Delta^{}_{-\khs+2\ve{Q}_c,-\khs-\ve{Q}_c}
    \end{array}\right) \,,
\end{align}
\begin{align}
    \label{eq:vector_Delta_parallel_SWcavity}
    \ve{\Delta}^{(\parallel)}= 
    \left(\begin{array}{c}
    \Delta^{(\parallel)}_{1} \\
    \Delta^{(\parallel)}_{2} \\
    \Delta^{(\parallel)}_{3} \\
    \Delta^{(\parallel)}_{4} \\  
    \Delta^{(\parallel)}_{5} \\
    \Delta^{(\parallel)}_{6} \\
    \Delta^{(\parallel)}_{7} \\
    \Delta^{(\parallel)}_{8}
    \end{array}\right)=
    \left(\begin{array}{c}
    \Delta^{}_{\khs-2\ve{Q}_c,-\khs-\ve{Q}_c} \\
    \Delta^{}_{\khs-2\ve{Q}_c,-\khs+\ve{Q}_c} \\
    \Delta^{}_{\khs-\ve{Q}_c,-\khs} \\
    \Delta^{}_{\khs,-\khs-\ve{Q}_c} \\
    \Delta^{}_{\khs-\ve{Q}_c,-\khs+2\ve{Q}_c} \\
    \Delta^{}_{\khs,-\khs+\ve{Q}_c} \\
    \Delta^{}_{\khs+\ve{Q}_c,-\khs} \\
    \Delta^{}_{\khs+\ve{Q}_c,-\khs+2\ve{Q}_c}
    \end{array}\right) \,.
\end{align}
With these vector structures, 
one finds that the matrix $\underline{A}_{8 \times 8}$ in the gap equations always assumes the exact same structure of \cref{eq:Cooper_mat_SWcavity}. This thus means that the critical temperatures are given by the same expression
\begin{align}\label{eq:pm_parallel_Tc_SWcavity}
    T_c^{(\pm)}=T_c^{(\parallel)} = \frac{|g|}{4} + \frac{3 g^2}{8 \xi_{\khs + \ve Q_c}} + \mathcal{O}(|g|^3/\eF^2)\, ,
\end{align}
confirming the degeneracy of the instabilities for the $\Delta^{(0)}$,$\Delta^{(\pm)}$ and $\Delta^{(\parallel)}$  pairs in the standing-wave cavity case, analogously to the degeneracy between $\Delta^{(0)}$ and $\Delta^{(\pm)}$ instabilities in the ring cavity case.

\subsection{Symmetries}
The degeneracy of the critical temperatures immediately gives rise to the question whether the latter is an artifact of the mean-field theory or whether it is protected by symmetry like in case of the ring cavity. Indeed, the $\text{SO}(2)$ transformation $\underline R(\alpha)$ in Eq.~\eqref{eq:alpha_rot} represents also a symmetry of the full standing-wave-cavity Hamiltonian~\eqref{eq:modelSW}. Moreover, the system exhibits an additional, closely related symmetry:
\begin{align}\label{eq:alpha_rot_x}
\begin{split}
  \left(
    \begin{array}{c}
         \hat{c}_{\ve k^{(iii)}_l}  \\
         \hat{c}_{\ve k^{(iv)}_{-l+1}}
    \end{array}
    \right) 
    & \to 
    \underline R(\alpha) 
    \left(
    \begin{array}{c}
         \hat{c}_{\ve k^{(iii)}_l}  \\
         \hat{c}_{\ve k^{(iv)}_{-l+1}}
    \end{array}
    \right) \, ,
    \end{split}
\end{align}
with $\underline R(\alpha)$ from Eq.~\eqref{eq:alpha_rot} and the momentum notation of Eq.~\eqref{eq:defMomGroups}. Physically, it rotates fermionic states with momentum component $k_x$ into $-k_x$, if \mbox{$k_y<0$}. For instance, at the level of pairing order parameters, $g \langle c_{\ve k^{(iii)}_1} c_{\ve k^{(i)}_0 } \rangle$, which generalizes the pair $\Delta^{(\parallel)}_{\khs}$ depicted in Fig.~\ref{fig:Delta_par} to finite momentum deviation from the hot spots, is connected to the Cooper pair with vanishing center-of-mass momentum $g \langle c_{\ve k^{(iv)}_0 } c_{\ve k^{(i)}_0 } \rangle$. As this example shows, Eq.~\eqref{eq:alpha_rot_x} connects pairs bearing different c.o.m.~momenta. Likewise, the symmetry~\eqref{eq:alpha_rot} does not only transform the $\Delta^{(0)}$ and $\Delta^{(\pm)}$ pairs into each other, but also implies relations between the pairs that arise from the non-momentum conserving interactions in the standing-wave cavity. 

On a formal level, one shows that \cref{eq:alpha_rot,eq:alpha_rot_x} indeed leave the Hamiltonian invariant by repeating the procedure leading from Eq.~\eqref{eq:sym_proof_1} to Eq.~\eqref{eq:sym_proof_2} in Section~\ref{sec:MF_Ana_symm}.
In contrast to the ring-cavity case, the index $-s$ is replaced by an index $s'$ that is summed over independently from $s$ to take care of the non-momentum-conserving interactions. This is a necessary condition for ~\cref{eq:alpha_rot_x} to represent a symmetry of the Hamiltonian, such that only the standing-wave cavity exhibits the additional symmetry. This is also illustrated by the given example $g \langle c_{\ve k^{(iii)}_1} c_{\ve k^{(i)}_0 } \rangle\leftrightarrow g \langle c_{\ve k^{(iv)}_0 } c_{\ve k^{(i)}_0 } \rangle$, since the former pair does not represent a leading instability in the ring cavity. In total, we expect the dominant instabilities $\Delta^{(0)}$, $\Delta^{(\pm)}$ and $\Delta^{(\parallel)}$ of the standing-wave cavity to be degenerate due to the peculiar $\text{SO}(2)$ symmetries of the Hamiltonian \cref{eq:modelSW}.

\subsection{Numerics}
To confirm the previous analytic results and to obtain further insights, we additionally solve the mean-field equation~\eqref{eq:gapeq_SWcavity} numerically on the momentum grid~\eqref{eq:defMomGroups} including the closest shifted copies from the hot spots and finite momentum deviations $\ve \delta \ve k^{\pm}$ as in the case of the ring cavity covered above and in the main text. 
Due to the presence of the additional relevant kind of pairing $\Delta^{(\parallel)}_{\ve k, \ve p}$ and the fact that the total momentum of the pairs is not conserved, $\hat H_\mf$ in Eq.~\eqref{eq:H_swmf} is a dense matrix in the anomalous sectors. Therefore, we directly solve the mean-field equation for order parameters $\Delta_{\ve p,\ve k}$ with arbitrary momentum indices.

First of all, we note that below $T_c$ the solution separates into two sets of order parameters $\Delta^{(>)}_{\ve p, \ve k}$ and $\Delta^{(<)}_{\ve p, \ve k}$. 
The first one contains the $\Delta^{(0)}$ pairs $g\langle \hat c_{\ve k^{(iv)}_0} \hat c_{\ve k_0^{(i)} }\rangle$, $g\langle \hat c_{\ve k^{(ii)}_0} \hat c_{\ve k_0^{(iii)} }\rangle$ and the pairs obtained from the latter via the two $\text{SO(2)}$ symmetries: the $\Delta^{(+)}$ pair $g\langle \hat c_{\ve k^{(ii)}_{-1}} \hat c_{\ve k_0^{(i)} }\rangle$, the $\Delta^{(-)}$ pair $g\langle \hat c_{\ve k^{(iv)}_{1}} \hat c_{\ve k_0^{(iii)}}\rangle$ 
and the $\Delta^{(\parallel)}$ pairs $g\langle \hat c_{\ve k^{(iii)}_1} \hat c_{\ve k^{(i)}_0} \rangle$, $g\langle \hat c_{\ve k^{(ii)}_0} \hat c_{\ve k^{(iv)}_1} \rangle$, all of which represent dominant instabilities when taking the limit $\ve \delta \ve k^{\pm} \to 0$. Furthermore, $\Delta^{(>)}_{\ve p, \ve k}$ includes all pairs reached from the previous ones by the standing-wave-cavity interaction of Eq.~\eqref{eq:modelSW}, either by respecting momentum conservation or by violating it by $\pm 2 \ve Q_c$. In contrast, $\Delta^{(<)}_{\ve p, \ve k}$ contains the remaining pairs, like $g\langle \hat c_{\ve k^{(ii)}_0} \hat c_{\ve k^{(i)}_0} \rangle$. 
These two sectors do not mix as any single interaction only allows for a change of the c.o.m.~momentum of a pair $\ve P$ by $ \ve 0$ or $\pm 2 \ve Q_c$. 
For any finite $\ve \delta \ve k^\pm$ the numerics shows that $\Delta^{(<)}_{\ve p, \ve k}$ is suppressed against $\Delta^{(>)}_{\ve p, \ve k}$.

\begin{figure}[tb]
\includegraphics[width=0.9\columnwidth]{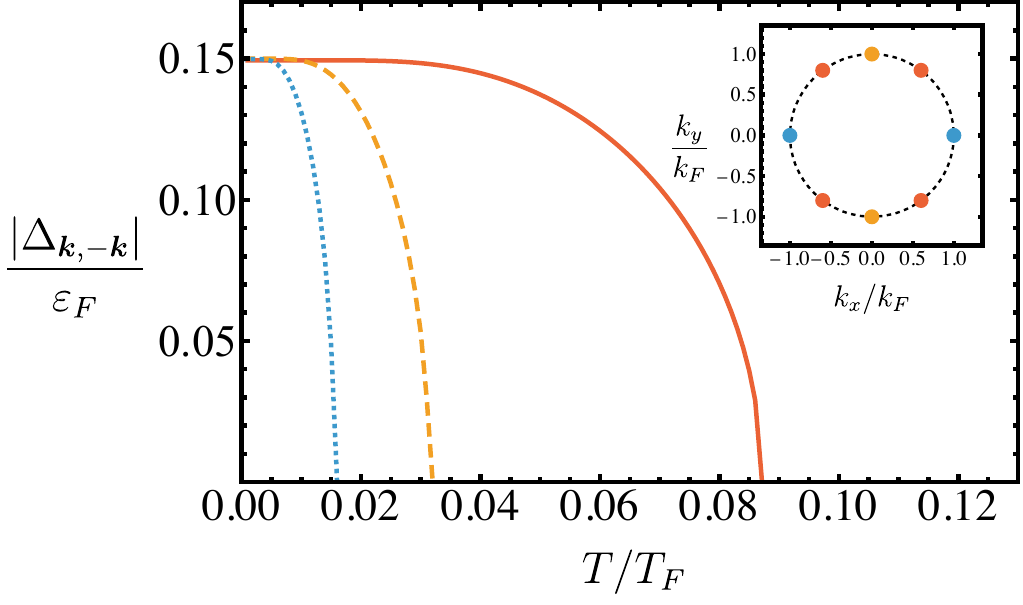}
\caption{$|\Delta_{\ve{k},-\ve{k}}|$ as a function of temperature $T$ at specific $\ve{k}$-points on the FS, indicated in the inset (hot spots: continuous line, upper and lower $\ve{k}$-points: dashed line, left and right $\ve{k}$-points: dotted line). The coupling is fixed to $g=0.3\varepsilon_F$.
}
\label{fig:results_momenta}
\end{figure}

Fig.~\ref{fig:results_momenta} represents the pairing gap $|\Delta_{\ve k, -\ve k}|$ at different momenta $\ve k$ on the FS (with neglected reshaping), in analogy to Fig.~4(a) in the main text. This choice represents the largest member of the $\Delta^{(0)}$ class of order parameters. Identical numerical results are found also for analogous $\Delta^{(\pm)}$ and $\Delta^{(\parallel)}$ pairs, as expected from their degeneracy.
The hot spots indeed exhibit the instability at the highest temperature, which agrees with $T_c$ in Eq.~\eqref{eq:Cooper_Tc_SWcavity}. As in the ring cavity case, the FS becomes fully gapped at low enough temperatures. Furthermore, at $T=0$, the numerics find a common value for the gap on the FS of approximately $|g|/2$, 
which indicates a connection to Eq.~\eqref{eq:D0_T0_HS}.
To further confirm the degeneracy of $\Delta^{(0)}$, $\Delta^{(\pm)}$, and $\Delta^{(\parallel)}$, we have checked that convergence of the self-consistent code is also possible in arbitrary superpositions of the three kinds of pairs.



\bibliography{Comp_inst_bibliography}